\documentclass[default]{aastex701}

\usepackage{graphicx}

\shorttitle{}
\shortauthors{Hoai et al.}
\graphicspath{{./}{}}
\nolinenumbers
\begin{document}

\title{On the shape of the ascending branch of the light curves of Long Period Variables}

\author[0000-0002-3816-4735]{Do Thi Hoai}
\affiliation{Vietnam National Space Center (VNSC), Vietnam Academy of Science and Technology (VAST),\\
18 Hoang Quoc Viet, Nghia Do, Ha Noi, Vietnam}
\email[show]{dthoai@vnsc.org.vn (DTH)}

\author[0000-0002-0311-0809]{Pham Tuyet Nhung}
\affiliation{Vietnam National Space Center (VNSC), Vietnam Academy of Science and Technology (VAST),\\
  18 Hoang Quoc Viet, Nghia Do, Ha Noi, Vietnam}
\email[show]{pttnhung@vnsc.org.vn (PTN)}

\author[0000-0002-8979-6898]{Pierre Darriulat}
\affiliation{Vietnam National Space Center (VNSC), Vietnam Academy of Science and Technology (VAST),\\
18 Hoang Quoc Viet, Nghia Do, Ha Noi, Vietnam}
\email[show]{darriulat@vnsc.org.vn (PD)}



\begin{abstract}

  The present article is written in the wake of a recently published study of the relation between the light curves of Long Period Variables and their evolution along the Asymptotic Giant Branch (AGB). It introduces a pair of new parameters that describe the shape of the ascending branches of such light curves. This parameterization reveals strong correlations with other parameters describing the evolution of the star on the AGB: periods, regularity and amplitudes of the oscillations, effective temperatures, mass loss rates, carbon and oxygen isotopic ratios, colour indices. It sheds new light on the occasional presence of humps on the ascending branches and on the distinction between stars that have not experienced strong Third Dredge Up (TDU) events and those that have as well as between oxygen-rich and carbon-rich stars. One of the new parameters, referred to as $q$, is particularly efficient at tracking the evolution of the star along the AGB. Globally, a simple picture can be drawn in the plane spanned by the period and by $q$. Yet, many details remain unexplained when looking at this picture in finer details. Overall, the results presented in the article help significantly with clarifying the complex set of observations that have been made in this domain and should inspire new considerations on their relation with the underlying physical mechanisms at stake inside the stars.
  
\end{abstract}



\section{Introduction} \label{sec1}

Light curves of Long Period Variables have been measured over more than a century by many observers and reduced to an easily accessible form by the American Association of Variable Stars Observers (AAVSO\footnote{https://www.aavso.org/}). Much information is contained in such data, but their interpretation in terms of the physical properties of the pulsating stars is not straightforward. An abundant literature is dedicated to efforts that have been made in this direction; references can be found in two articles that we published recently: \citet{Hoai2025}, hereafter referred to as Paper I, and \citet{Nhung2025}, hereafter referred to as Paper II. In the wake of these papers, the present work aims at improving our understanding of the relation between the physics governing the evolution of pulsating stars along the Asymptotic Giant Branch (AGB) and the properties of their light curves; it addresses questions that had been left unanswered by Papers I and II. As was done in Paper II, the procedure adopted consists in selecting groups of stars with known stellar parameters giving a good idea of their evolution in the Hertzsprung-Russell diagram and examining how their light curves evolve in the space of parameters that describe their shapes and main features. Specifically, Paper II was using the width of the light pulse to characterize the shape of a light curve; the present work uses instead the shape of the ascending branch of the light cycle, which had not been considered in Papers I and II. We recall that in principle, when a star pulsates, the way it does so depends more on the physics at stake in its interior (essentially governed by the properties of the convective envelope) than on what happens on its photosphere and in its atmosphere. But what happens on its photosphere and in its atmosphere has a strong impact on the appearance of the light curve. As a result, while there are good reasons to hope that a study of the light curves might be an efficient tool to help with understanding the inner star dynamics, this is far from being proven. To which extent the continuity of the evolution of the star on the AGB is associated with some continuity of the shape of the light curve must be investigated and cannot be taken as granted. While very significant correlations have been revealed by earlier studies, and in particular by Paper II, demonstrating the validity of such an approach, many questions remain unanswered.

The parameters that Paper II uses to describe a light curve are the mean period $P$, the mean amplitude $A$, the mean phase at minimum light $\varphi_{\rm{min}}$ (offering a measure of the mean asymmetry between the durations of the ascending and descending branches of a same cycle), a width parameter $W$ measuring the mean duration of the light pulse relative to the period and a parameter $\Delta$ measuring the irregularity of the pulsation regime. The parameters used to evaluate the evolution of the star along the AGB are a mean colour index $C$, which offers an evaluation of the abundance of dust emitting in the infrared, and the spectral type and associated index, which offers an estimate of the effective temperature. In addition, when available, we take in due consideration the values taken by the gas mass-loss rate and the $^{12}$C/$^{13}$C isotopic ratio. We refer the reader to Paper II for a precise definition of these parameters. They leave open the detailed shapes of the ascending and descending branches of the mean light cycle. In Paper I, we have shown that the mean profiles of the descending branches, when normalized to same amplitude and same duration, vary only little from curve to curve, leaving little hope to learn significant information from their study. In contrast, the normalized profiles of the ascending branches are known to display a broad range of different shapes, in particular to occasionally host abrupt changes of slope that may take the form of humps. Accordingly, the present work focuses on seeking possible relations between the shape of the ascending branch and the stellar parameters. 

\begin{figure*}
  \centering
  \includegraphics[width=6cm]{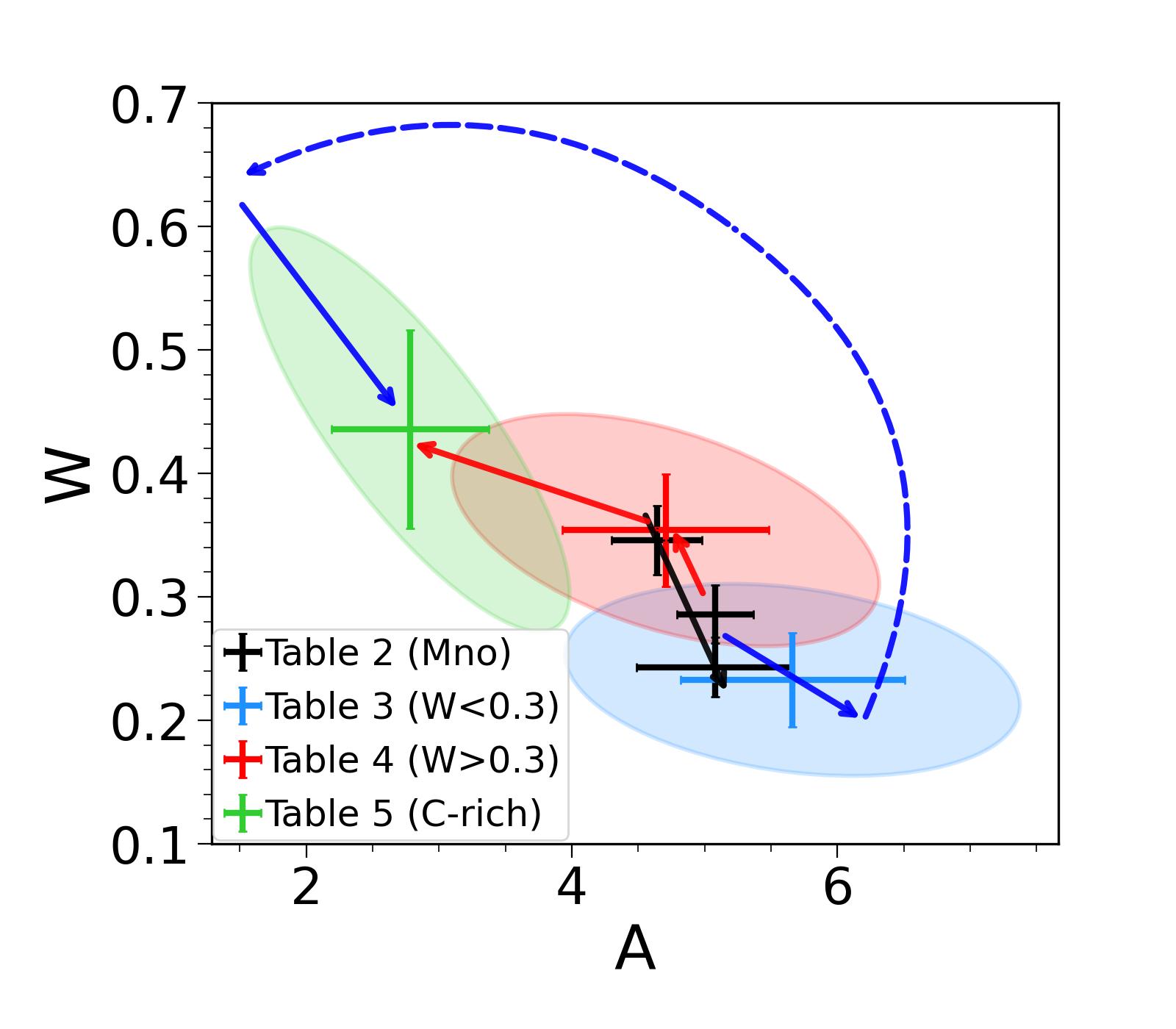}
  \includegraphics[width=6cm]{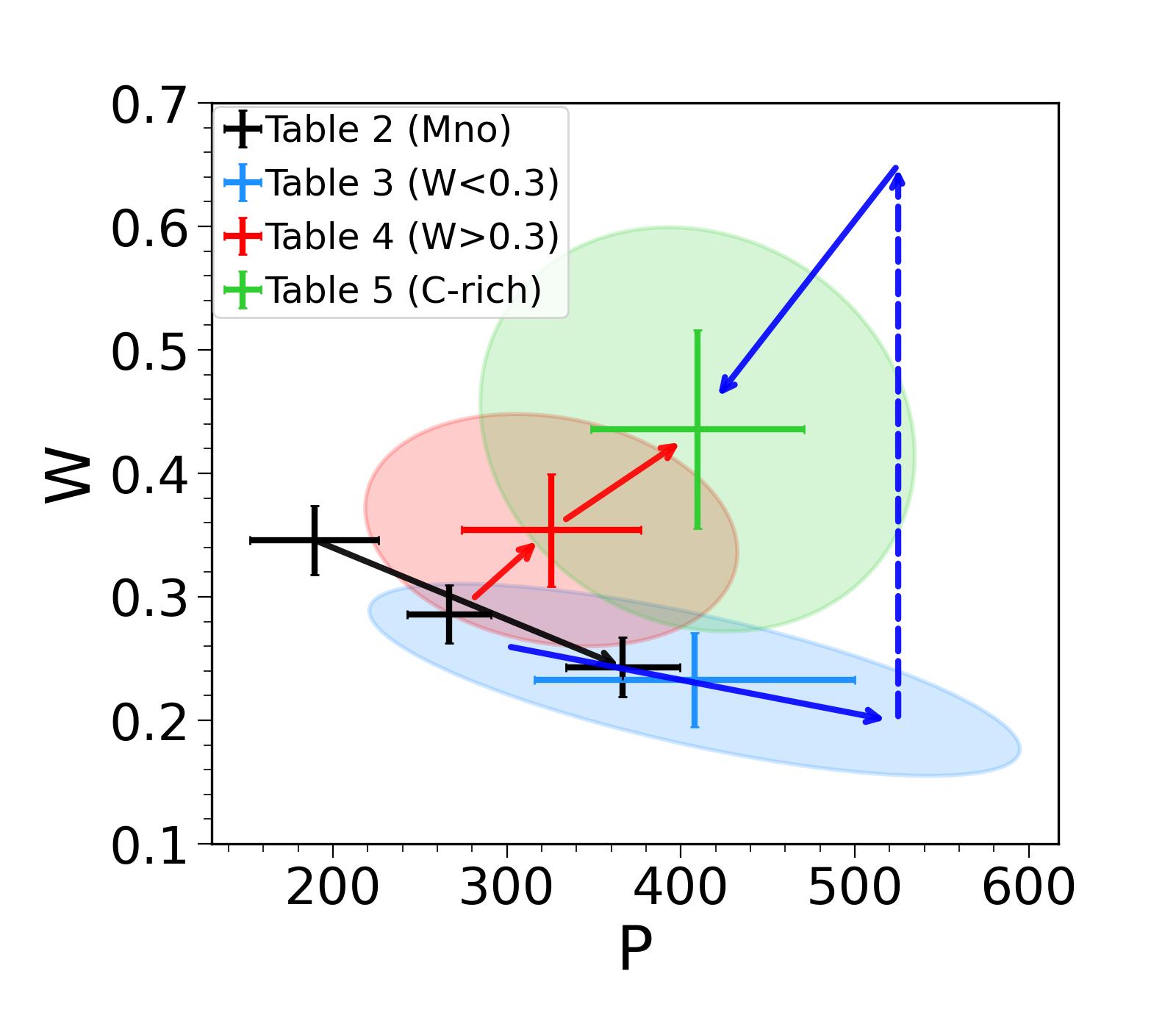}
  \caption{Schematic evolution of the light curves in parameter space ($W$ vs $A$ projection in the left panel and $W$ vs $P$ projection in the right panel) as depicted in Paper II. Crosses show mean and rms values of the samples listed in the inserts, where Table 2 stands for Mno spectral types split in three groups of temperature increasing with decreasing $W$, Table 3 and Table 4 for other M spectral types and S spectral types evolving toward lower, and respectively larger, values of $W$, and Table 5 for carbon-rich stars. Coloured ellipses approximate the regions covered by these samples. Arrows show the suggested evolution as described in the text: black for Mno stars, blue and red for stars experiencing strong TDU events, the former stopping pulsating regularly while possibly transiting to the carbon-rich stars of the sample.}
 \label{fig1}
\end{figure*}

The main results of Papers I and II can be summarized as follows:

While on the early part of the AGB, stars expand, cool down and tend to pulsate with increasing periods; when their period exceeds 100 days, they enter the sample considered in the present work. At that time, they are oxygen-rich with an M spectral type and have not yet experienced sufficient TDU events for the presence of technetium in their spectrum to be detectable. As the temperature of the star decreases and as its size increases, the light curve evolves as illustrated in Figure \ref{fig1} (black arrow): the period increases, the width parameter $W$ decreases, the phase at minimal light increases, weak humps may appear on the ascending branch close to minimal light, the pulsations become less regular, the colour indices increase and the ratio between the oscillation amplitudes in the infrared and in the visible increases. The appearance of technetium in the spectrum, corresponding to the transition from Mno to Myes spectral types, namely the occurrence of TDU events strong enough to have a detectable impact, occurs at different stages along the TP-AGB, with accordingly different values of $W$. It is expected to never occur for stars of low initial masses, which remain as Mno spectral type for their whole AGB life.

The presence of stars of Myes, and even S spectral types having curves covering a broad range of $W$ values suggests that stars may leave the Mno spectral type at very different stages of their evolution along the AGB. Some curves (blue arrow in Figure \ref{fig1}) seem to keep evolving in the same direction after the transition to the Myes and S spectral types: $W$ keeps decreasing, the periods and amplitudes of the oscillations keep increasing and they become less regular. This is consistent with the expectation that thermal pulses modify only progressively the internal state of the star. Other curves (red arrow in Figure \ref{fig1}) seem instead to transit to the Myes and S spectral types very early with the width parameter $W$ increasing rather than decreasing. Their location in the parameter space suggests that they evolve towards the carbon-rich stars of the sample. In contrast, the absence in the parameter space of carbon-rich curves close to those evolving along the blue arrow suggests that some of these curves may leave the domain of stable pulsation and stop pulsating regularly when transiting from oxygen-rich to carbon-rich (dotted arrows in Figure \ref{fig1}).

However, such suggestions, albeit sensible and plausible, are far from being proven. In particular, a major difficulty that has been recurrently met during the study is the lack of a clear picture of the physics governing the transition from a stable to an unstable regime of pulsation, namely of the nature of the instability region(s) in the Hertzsprung-Russell diagram. 

\section{Sample of light curves studied in the present work}
The present work requires a particularly severe selection of curves having data of sufficient quality for a detailed study of the profile of their ascending branches, which implies retaining only 86 curves from the Paper II sample of 116 curves, restricted, in the case of M spectral types, to cases where the presence of technetium in the spectrum has been searched for and reliably defined: either positively, in which case we talk of Myes spectral types, or negatively, in which case we talk of Mno spectral types. The presence of technetium is a precious indicator of the evolution of the star along the AGB, giving evidence for the occurrence of strong enough third-dredge-up (TDU) events. Unfortunately, only a few stars of M spectral type have been searched for it, and when they have, the difficulty of the measurement has often caused the result to be uncertain. In addition, we select from the \citet{MerchanBenitez2023} list 7 curves of stars of M spectral type having a sufficiently high density and quality of observations and a period between 400 and 450 days; the reason for this addition is the relative scarcity of such stars in Papers I and II. As a result, the present sample includes 93 stars having well measured light curves, of which 25 are of type Mno, 10 of type Myes, 19 of type M with unknown technetium content, 18 of type S and 21 of type C.

In order to ease the discussion, we find it convenient to study separately curves sharing a same spectral type, as was done in Papers I and II. This implies to tentatively assign the 19 M-type curves of stars having an unknown technetium content to either the Mno or Myes group, which we discuss below, of course keeping in mind that these are just guesses and should not be confused with proven assignments. In doing so, we anticipate on the arguments presented in Section 3, in particular the definition of the $p$ and $q$ parameters, but we think that these tentative assignments must be disposed of as early as possible in order to ease the presentation of the article.

RT Cyg is a star of M2 spectral type and short period, 190 days, which shares similar values of its curve parameters with the warmer members of the Mno family, which form a compact group in the parameter space. This leaves little doubt on its proper assignment to the Mno group, as had been done in Paper II, and in particular to the sub-group of its warmer members.

R Dra, S CrB and R UMa, of respectively M4.5, M6.5 and M5 spectral types, share with a group of 5 Mno stars a distinctively different shape of their ascending branches; we refer to these as Mno outliers and study them in detail in Section 4. This suggests strongly that they also are of Mno type. However, if it were not for this particularity, we might as well have assigned S CrB and R UMa to the Myes family. In contrast, R Dra is marginally outlier in the $p$ vs $q$ plane (see Section 4) but its other parameters, in particular the period, favour an Mno assignment.

T Cas and U UMi, both of M6 spectral type, have large W values, respectively 0.39 and 0.37, which strongly suggest an evolution toward curves of broader light maxima. T Cas has a large $^{12}$C/$^{13}$C isotopic ratio of 33. We accordingly assign them tentatively to the Myes category. They have negative $q$ values and are very different from the Myes curves having positive $q$ values.

This leaves 13 M-type curves of unknown technetium content. They are RU Aur, V Cam, X Cep, R Tel, R CVn and V Cas, together with the seven stars added for having a period between 400 and 450 days: RW And, Y Cas, SX Cyg, R Oct, S Pic, Y Vel and Z Vel. The first four are curves having long periods, which could be retained from the Paper II sample as having a sufficient quality and density of observations. Most of these curves have parameters that match those of Mno spectral types. Only two of them, RW And and S Pic, display slightly larger oscillation amplitudes typical of Myes curves. We assign them accordingly to the Myes spectral type and assign all the others to the Mno spectral type. We note, however, that the validity of such an Mno assignment is not certain. In particular long period stars such as RU Aur, V Cam, X Cep and R Tel are being assigned an Mno spectral type on the basis of an assumption that may well be wrong: that having reached such an evolved state while remaining oxygen-rich, these stars will never experience a TDU event, most likely because of having too low an initial mass.

With such tentative assignments, the sample of curves includes 40 Mno, 14 Myes, 18 S and 21 C spectral-types. They are listed in Table \ref{tab1} in alphabetic order. As illustrated in Figure \ref{fig1}, stars sharing similar properties are often observed to have light curves sharing similar values of their parameters, therefore confined to well defined regions of the parameter space. This is the basis of the argumentation used to seek possible relations between the evolutions of stars and of their light curves. 

\startlongtable
\begin{deluxetable*}{lcl ccc ccc ccc c}
\tablenum{1}
\tablecaption{List of the selected stars and of the parameters of their light curves. Column 1 lists the star names; column 2 the period (measured in days); column 3 the spectral type as quoted in the SIMBAD astronomical database \citep[][except for RW And for which we retain the Merch\'{a}n Ben\'{i}tez assignment]{Wenger2000} and the type Mno, Myes, S or C as defined in Section 2 (an asterisk signals a tentative type assignment made as described in the text); column 4 specifies the presence or absence of technetium in the spectrum; column 5 lists the  $^{12}$C/$^{13}$C isotopic ratio when available; columns 6 to 10 list the Paper II curve parameters as defined in the text; columns 11 to 12 list the values of the parameters describing the shape of the ascending branch as described in the text. \label{tab1} }
\colnumbers
\tablewidth{0pt}
\tablehead{
 \colhead{Name}&\colhead{$P$}&\colhead{Spectral type}&\colhead{Tc}&\colhead{$^{12}$C/$^{13}$C}&\colhead{$C$}&\colhead{$A$}&\colhead{$W$}&\colhead{$\Delta$}&\colhead{$\varphi_{\rm{min}}$}&
 \colhead{$p$$\times$10$^3$}&\colhead{$q$$\times$10$^3$}&\colhead{$\chi^2$$\times$10$^4$}\\
  \colhead{}&\colhead{(day)}&\colhead{}&\colhead{}&\colhead{}&\colhead{(mag)}&\colhead{(mag)}&\colhead{}&\colhead{(mag)}&\colhead{}& \colhead{}&\colhead{}&\colhead{}
}
\startdata
R And&	410&	S5/S&	yes&	40&	2.6&	7.4&	0.22&	0.76&	0.61&	$-$135&	45&	65\\
RW And&	430&	M5/Myes*&	$-$&	$-$&	2.4&	6&	0.2&	0.78&	0.57&	$-$391&	157&	261\\
W And&	398&	S7/S&	yes&	$-$&	1.8&	6.3&	0.22&	0.79&	0.58&	$-$555&	79&	52\\
X And&	345&	S2.5/S&	$-$&	$-$&	2.2&	5.5&	0.27&	0.51&	0.62&	88&	39&	54\\
R Aql&	274&	M6.5/Mno&	no&	8&	2.3&	4.7&	0.31&	0.33&	0.53&	$-$403&	$-$12&	8\\
W Aql&	483&	S6/S&	yes&	26&	3.2&	5.2&	0.26&	0.77&	0.62&	$-$438&	57&	174\\
T Aqr&	202&	M2/Mno&	no&	$-$&	1.4&	5&	0.34&	0.4&	0.5&	$-$316&	$-$49&	48\\
R Aur&	457&	M6.5/Myes&	yes&	33&	2.1&	5.9&	0.21&	0.5&	0.46&	$-$995&	0&	120\\
RU Aur&	470&	M9/Mno*&	$-$&	$-$&	3.6&	5&	0.21&	0.73&	0.59&	$-$723&	192&	300\\
R Boo&	224&	M4/Mno&	no&	$-$&	1.7&	5.1&	0.31&	0.35&	0.52&	$-$186&	$-$18&	16\\
R Cam&	270&	S2/S&	$-$&	$-$&	0.4&	4.5&	0.41&	0.28&	0.52&	$-$474&	$-$139&	35\\
S Cam&	330&	C6/C&	$-$&	14&	1.1&	1.9&	0.56&	0.15&	0.49&	$-$657&	$-$146&	72\\
T Cam&	376&	S5/S&	yes&	31&	0.7&	5.4&	0.4&	0.18&	0.49&	$-$676&	$-$169&	17\\
V Cam&	521&	M7/Mno*&	$-$&	$-$&	3.6&	5.9&	0.17&	0.72&	0.61&	$-$946&	190&	450\\
R Cap&	342&	C/C&	$-$&	$-$&	3&	3.5&	0.35&	0.44&	0.57&	$-$212&	38&	231\\
R Cas&	432&	M6.5/Mno&	no&	12&	2.1&	6&	0.21&	0.84&	0.61&	$-$472&	94&	73\\
S Cas&	613&	S3/S&	$-$&	32&	3.6&	5.6&	0.2&	0.74&	0.56&	$-$772&	88&	218\\
T Cas&	445&	M7/Myes*&	prob&	33&	1.9&	4.3&	0.39&	0.27&	0.43&	$-$919&	$-$95&	43\\
U Cas&	277&	S4.5/S&	$-$&	$-$&	1.2&	6.2&	0.25&	0.38&	0.53&	$-$75&	28&	44\\
V Cas&	229&	M5/Mno*&	$-$&	$-$&	1.6&	4.8&	0.28&	0.28&	0.51&	$-$355&	14&	21\\
W Cas&	408&	C9/C&	yes&	25&	1&	2.8&	0.45&	0.21&	0.52&	$-$592&	$-$79&	35\\
Y Cas&	413&	M7.5/Mno*&	$-$&	$-$&	3&	4.5&	0.26&	0.42&	0.57&	$-$501&	105&	73\\
RV Cen&	445&	C3/C&	$-$&	$-$&	1.6&	2.4&	0.51&	0.24&	0.43&	$-$1022&	$-$99&	234\\
AX Cep&	396&	C/C&	$-$&	$-$&	2.7&	2.9&	0.39&	0.56&	0.55&	$-$237&	26&	211\\
PQ Cep&	443&	C6/C&	$-$&	$-$&	2.2&	2.6&	0.46&	0.46&	0.47&	$-$763&	$-$41&	329\\
S Cep&	484&	C7/C&	yes&	224&	2.2&	2.4&	0.4&	0.43&	0.46&	$-$887&	33&	303\\
T Cep&	338&	M6/Myes&	yes&	33&	1.2&	4.1&	0.33&	0.29&	0.47&	$-$707&	$-$49&	31\\
X Cep&	540&	M4/Mno*&	$-$&	$-$&	3.2&	6&	0.16&	0.68&	0.58&	$-$1286&	217&	1526\\
S CMi&	333&	M7/Myes&	yes&	18&	1.9&	4.9&	0.32&	0.33&	0.49&	$-$583&	$-$60&	20\\
R Cnc&	362&	M6.5/Mno&	no&	17&	1.9&	4.4&	0.25&	0.33&	0.53&	$-$670&	61&	48\\
V Cnc&	272&	S0/S&	yes&	$-$&	1.4&	5&	0.33&	0.3&	0.55&	$-$375&	$-$69&	25\\
T Col&	226&	M4/Mno&	no&	$-$&	1.7&	4.3&	0.33&	0.37&	0.51&	$-$284&	$-$28&	19\\
S CrB&	360&	M6.5/Mno*&	poss&	$-$&	2.8&	5.5&	0.26&	0.52&	0.64&	$-$51&	81&	133\\
V CrB&	358&	C6/C&	yes&	$-$&	2.1&	3&	0.37&	0.52&	0.58&	$-$5&	33&	105\\
BH Cru&	513&	C8/C&	yes&	8&	1.5&	2.9&	0.47&	0.33&	0.5&	$-$910&	$-$49&	66\\
R CVn&	329&	M6.5/Mno*&	poss&	$-$&	2.1&	4.4&	0.3&	0.41&	0.51&	$-$483&	$-$2&	12\\
chi Cyg&	408&	S6/S&	yes&	36&	2.1&	8.3&	0.18&	0.64&	0.56&	$-$581&	19&	24\\
FF Cyg&	328&	S5.5/S&	$-$&	$-$&	1&	4.5&	0.34&	0.22&	0.5&	$-$422&	$-$93&	172\\
LX Cyg&	581&	C/C&	yes&	32&	1.7&	3.4&	0.46&	0.41&	0.59&	$-$463&	42&	267\\
R Cyg&	427&	S4/S&	yes&	29&	2.9&	6.2&	0.24&	1.02&	0.62&	$-$56&	36&	47\\
RS Cyg&	419&	C5.5/C&	$-$&	$-$&	1.2&	1.6&	0.66&	0.12&	nan&	$-$757&	$-$127&	295\\
RT Cyg&	190&	M2/Mno*&	dbfl&	$-$&	1.5&	4.7&	0.38&	0.54&	0.54&	$-$195&	$-$57&	21\\
SX Cyg&	410&	M7/Mno*&	$-$&	$-$&	2.3&	5.2&	0.2&	0.63&	0.58&	$-$654&	156&	141\\
U Cyg&	463&	C9/C&	yes&	14&	2.1&	3.1&	0.39&	0.45&	0.51&	$-$627&	$-$19&	79\\
V Cyg&	420&	C7/C&	$-$&	20&	2.8&	3.4&	0.38&	0.49&	0.56&	152&	29&	122\\
Z Del&	305&	S4.5/S&	yes&	$-$&	1.4&	5.5&	0.3&	0.45&	0.51&	$-$342&	$-$12&	112\\
R Dra&	247&	M4.5/Mno*&	dbfl&	$-$&	2.1&	5.1&	0.3&	0.46&	0.55&	$-$66&	17&	25\\
T Dra&	422&	C6/C&	yes&	24&	3&	3&	0.37&	0.34&	0.56&	$-$87&	46&	147\\
T Eri&	253&	M5/Mno&	no&	$-$&	1.9&	4.8&	0.31&	0.44&	0.51&	$-$289&	$-$27&	7\\
W Eri&	376&	M7/Mno&	no&	$-$&	2.4&	5.4&	0.22&	0.5&	0.59&	$-$247&	118&	116\\
R For&	387&	C4/C&	$-$&	$-$&	2.7&	3&	0.35&	0.53&	0.51&	$-$348&	52&	140\\
R Gem&	361&	S3.5/S&	yes&	22&	2&	6.2&	0.31&	0.52&	0.64&	211&	$-$35&	23\\
T Gem&	287&	S2/S&	yes&	$-$&	0.5&	5.4&	0.38&	0.21&	0.5&	$-$329&	$-$139&	31\\
VX Gem&	379&	C9/C&	yes&	9&	1.6&	4&	0.36&	0.32&	0.59&	$-$584&	35&	62\\
RU Her&	486&	M6/Myes&	yes&	25&	2.3&	5.9&	0.18&	0.74&	0.54&	$-$1064&	122&	273\\
S Her&	307&	M4/Myes&	yes&	$-$&	1.1&	5.4&	0.31&	0.33&	0.49&	$-$564&	$-$77&	13\\
T Her&	165&	M2.5/Mno&	no&	$-$&	1.5&	5&	0.34&	0.44&	0.52&	$-$195&	$-$46&	15\\
U Her&	404&	M6.5/Mno&	no&	19&	2.5&	4.9&	0.21&	0.4&	0.58&	$-$717&	121&	33\\
R Hor&	404&	M5/Myes&	yes&	$-$&	2.1&	7.3&	0.23&	0.53&	0.59&	$-$211&	106&	186\\
R Hya&	386&	M6/Myes&	yes&	26&	0.9&	3.7&	0.36&	0.38&	0.49&	$-$589&	$-$35&	84\\
RU Hya&	332&	M6.5/Mno&	no&	$-$&	3.3&	5.4&	0.24&	0.59&	0.6&	41&	101&	203\\
R Leo&	312&	M7/Mno&	no&	10&	1.2&	4.3&	0.26&	0.4&	0.56&	$-$533&	65&	26\\
R Lep&	441&	C7/C&	$-$&	34&	2.5&	2.5&	0.42&	0.59&	0.47&	$-$811&	12&	192\\
R Lyn&	378&	S4/S&	yes&	$-$&	1.4&	5.7&	0.26&	0.44&	0.56&	$-$442&	$-$36&	26\\
R Oct&	405&	M5.5/Mno*&	$-$&	$-$&	2&	4.6&	0.18&	0.64&	0.56&	$-$861&	134&	190\\
U Oct&	303&	M6/Mno&	no&	$-$&	1.8&	5.3&	0.27&	0.4&	0.53&	$-$316&	16&	12\\
R Oph&	303&	M5/Mno&	no&	$-$&	2.1&	5.6&	0.27&	0.43&	0.56&	$-$407&	11&	31\\
RY Oph&	150&	M3/Mno&	no&	$-$&	1.7&	4.7&	0.32&	0.47&	0.52&	$-$229&	$-$21&	14\\
R Ori&	377&	SC5/C&	$-$&	$-$&	1.8&	3.5&	0.32&	0.27&	0.55&	$-$428&	11&	48\\
R Peg&	377&	M6/Mno&	no&	10&	2.3&	5&	0.24&	0.49&	0.56&	$-$582&	72&	25\\
RZ Peg&	437&	SC5/C&	yes&	9&	2.3&	4.1&	0.33&	0.25&	0.56&	$-$372&	$-$12&	42\\
W Peg&	345&	M6.5/Mno&	no&	17&	2&	4.2&	0.28&	0.37&	0.52&	$-$619&	13&	40\\
X Peg&	201&	M2.5/Mno&	no&	$-$&	1.3&	4&	0.39&	0.34&	0.5&	$-$341&	$-$85&	30\\
Z Peg&	326&	M7/Myes&	yes&	$-$&	1.7&	4.7&	0.3&	0.33&	0.5&	$-$580&	$-$14&	130\\
S Pic&	425&	M7/Myes*&	$-$&	$-$&	2.9&	6&	0.21&	0.66&	0.62&	$-$317&	128&	86\\
R Ser&	355&	M5/Myes&	yes&	14&	2&	6.5&	0.24&	0.58&	0.57&	$-$323&	61&	55\\
R Sgr&	269&	M4/Mno&	no&	$-$&	1.5&	4.9&	0.31&	0.44&	0.53&	$-$211&	$-$15&	17\\
T Sgr&	391&	S4.5/S&	yes&	$-$&	1.5&	4.2&	0.27&	0.37&	0.54&	$-$424&	10&	207\\
R Tel&	462&	M5/Mno*&	$-$&	$-$&	2.5&	5.2&	0.16&	0.68&	0.53&	$-$1095&	131&	1090\\
R Tri&	266&	M3.5/Mno&	no&	$-$&	1.7&	5.4&	0.25&	0.39&	0.52&	$-$398&	8&	20\\
R UMa&	301&	M5/Mno*&	$-$&	$-$&	2.9&	5.5&	0.27&	0.43&	0.61&	42&	73&	187\\
RR UMa&	231&	M4/Mno&	no&	$-$&	1.6&	4.8&	0.29&	0.44&	0.55&	$-$2&	53&	75\\
S UMa&	226&	S0.5/S&	yes&	$-$&	1.2&	3.7&	0.46&	0.09&	0.5&	$-$510&	$-$147&	9\\
T UMa&	256&	M4/Mno&	no&	$-$&	2.4&	5.3&	0.31&	0.62&	0.58&	187&	23&	42\\
U UMi&	324&	M6/Myes*&	$-$&	$-$&	1.6&	3.3&	0.37&	0.24&	0.5&	$-$281&	$-$34&	32\\
Y Vel&	442&	M9/Mno*&	$-$&	$-$&	2.4&	4.3&	0.17&	0.71&	0.58&	$-$891&	227&	311\\
Z Vel&	417&	M9/Mno*&	$-$&	$-$&	2.1&	4&	0.25&	0.44&	0.47&	$-$819&	47&	175\\
R Vir&	146&	M3.5/Mno&	no&	$-$&	1.8&	4.3&	0.37&	0.34&	0.5&	$-$308&	$-$65&	39\\
RS Vir&	354&	M8/Mno&	no&	$-$&	3.2&	5.5&	0.28&	0.53&	0.63&	284&	66&	196\\
RU Vir&	437&	C8/C&	yes&	$-$&	3.7&	3&	0.37&	0.72&	0.56&	$-$433&	43&	249\\
S Vir&	377&	M6/Myes&	yes&	$-$&	2&	5.7&	0.24&	0.36&	0.55&	$-$453&	55&	18\\
R Vol&	451&	C/C&	$-$&	$-$&	3.4&	2.9&	0.31&	0.63&	0.5&	$-$429&	87&	76\\
R Vul&	137&	M3/Mno&	no&	$-$&	1.8&	4.5&	0.37&	0.45&	0.51&	$-$278&	$-$61&	12\\
\enddata
\end{deluxetable*}
\clearpage

\section{Normalized profiles of the ascending branches}
\subsection{Definition}
\begin{figure*}
  \centering
  \includegraphics[width=7cm,trim={0cm 0 0cm 0},clip]{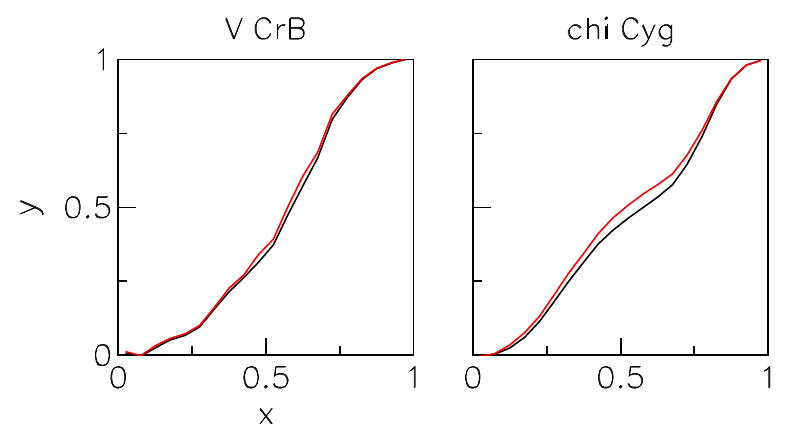}
  \caption{Mean normalized ascending branches of the light curves of V CrB (left) and of chi Cyg (right). Averaging is made either on magnitude (black) or on luminosity (red).}
\label{fig2}
\end{figure*}

For each of the curves listed in Table \ref{tab1} we produce the mean normalised profile of the ascending branches calculated on cycles for which both a light minimum and the following light maximum can be reliably defined. We call $x$ the ratio between the time from minimum and the duration of the ascending branch; $x$ varies from 0 to 1 in 20 bins. We call $y$ the ratio of the difference of magnitude between the values measured at $x$ and at minimum light to the difference of magnitude between the values measured at maximum and minimum light of the relevant cycle; $y$ varies from 0 to 1. Note that this definition of a normalized profile is somewhat arbitrary; other choices may be considered. In particular one might argue that using luminosity rather than magnitude should offer a more direct relation to the physical properties of the pulsating star. We carefully studied such issues and, as we are only interested in the evolution of the light curve, we found that in practice it makes little difference. An illustration is given in Figure \ref{fig2}, using as examples two representative cases of cycle profile: a carbon-rich star, V CrB and a very large amplitude curve, that of chi Cyg.

\subsection{Deviation from a sine wave profile: a two-parameter description}
\begin{figure*}
  \centering
  \includegraphics[width=3.15cm,trim={0.3cm 3 0.3cm 0},clip]{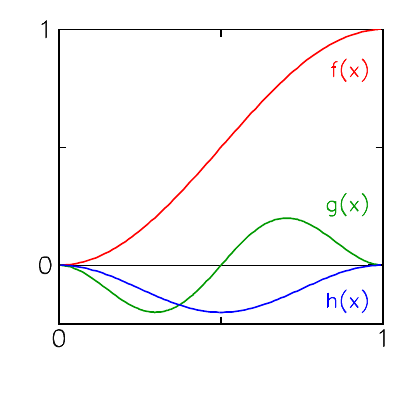}
  \includegraphics[width=3.15cm,trim={0.3cm 3 0.3cm 0},clip]{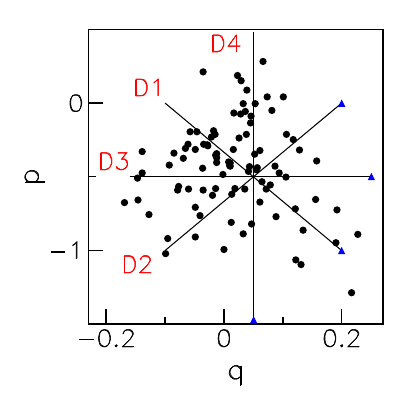}
  \includegraphics[width=11cm,trim={0cm 3 0cm 0},clip]{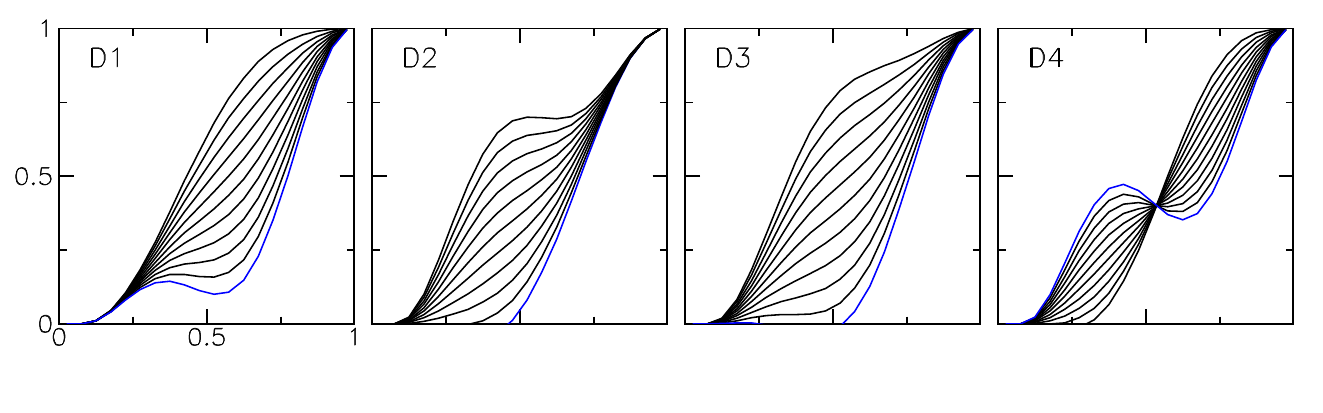}
  
  \caption{From left to right: Dependence on $x$ of the three terms of the fit function $S(x)$; for clarity, $h(x)$ has been scaled down by a factor 10. Definition of four directions in the $p$ vs $q$ plane used to illustrate the evolution of the ascending branch profiles when spanning across it, which is done in the four rightmost panels.}
  
\label{fig3}
\end{figure*}

With the aim of revealing simple trends in the evolution of the shapes of the ascending branches of the light curves, we try to characterize each of them as simply as possible using a two-parameter parameterisation.
As we shall see in Subsection 3.4, the curves of stars entering the thermally-pulsing AGB (TP-AGB) have ascending branches displaying a nearly sinusoidal profile, which we write as $f(x)$=$0.5(1-\cos\pi x$). To a good approximation, we may expect such curves to be the precursors of most of the curves in the sample. To describe the evolution of the light curves when the star evolves along the AGB, it is therefore convenient to parameterize their deviation from such a sine wave curve. We do it in the form of two terms, one antisymmetric and the other symmetric in $x$. Specifically, we write these terms as $g(x)$=$x(x-1)\sin(2\pi x$) and $h(x)$=$\cos(2\pi x)-1$, respectively (left panel of Figure \ref{fig3}). The role of the factor $x(x-1)$ is to make the derivative of $h(x)$ cancel at $x=0$ and $x=1$, as it must do. We checked that other similar forms, such as $h(x)$=$x(x-1)\sin(\pi x)$, $h(x)$=$x^2(x-1)^2$ or $g(x)$=$x^2(x-1)^2(x-0.5)$ give the same results. We then fit each of the normalized profiles of the 93 ascending branches to a form $S(x)$=$f(x)$+$p\times g(x)$+$q\times h(x)$ by finding the best-fit values of the parameters $p$ and $q$ which minimize $\chi^2$=$\sum[y(x)-S(x)]^2$, where the sum extends over the 20 bins of $x$. The parameters $p$ and $q$ measure the deviation of $S(x)$ from $f(x)$ at $x$=0.5, $p$ for its derivative and $q$ for its value: $q$=$0.5[0.5-S(0.5)]$ and $p$=$(2/\pi)$d$S$(0.5)/d$x$$-$1. Figure \ref{fig3} illustrates the evolution of the ascending branch profiles when spanning across the $p$ vs $q$ plane. 

\begin{figure*}
  \centering
  \includegraphics[width=0.8\textwidth,trim={0cm 0 0cm 0},clip]{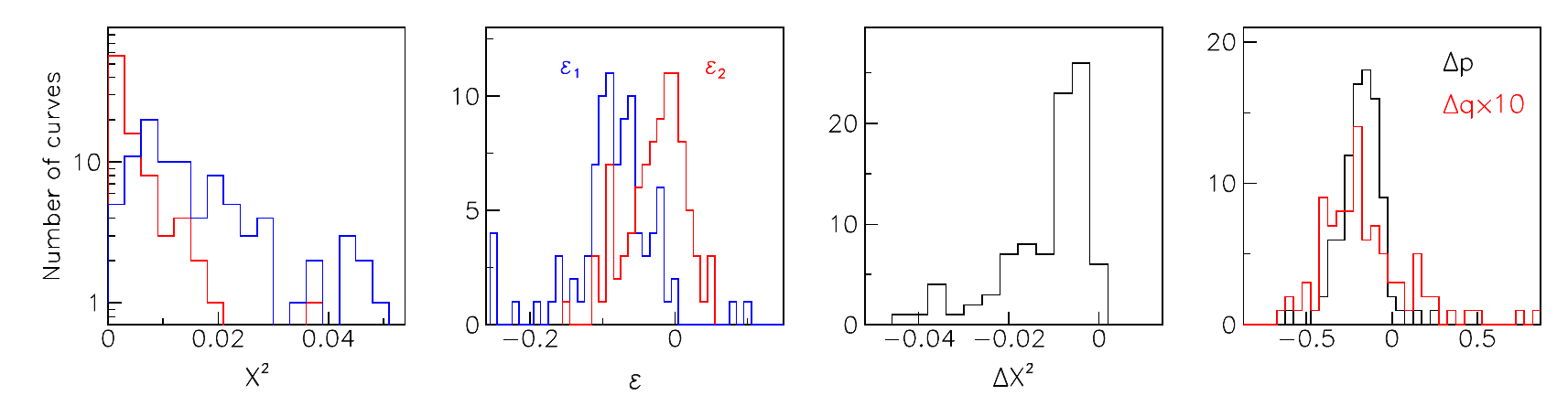}
  \caption{Left: $\chi^2$ distributions (in blue for $\varepsilon_1$=$\varepsilon_2$=0, in red for  $\varepsilon_1$ and  $\varepsilon_2$ free to vary).  Centre-left: distributions of the best-fit values of  $\varepsilon_1$ and  $\varepsilon_2$. Centre-right: distribution of the improvement in $\chi^2$ due to the freedom given to  $\varepsilon_1$ and  $\varepsilon_2$ to vary. Right: distribution of the resulting shifts in $p$ and $q$, $\Delta{p}$ and $\Delta{q}$. }
  \label{fig4}
\end{figure*}

The fits are generally of good quality but we noted some small but significant deviations in the close vicinity of the minima and maxima. As the $x$ values of their locations are not defined to better than $\sim$0.02 on average, we repeated the fits by allowing for small shifts in $x$, $\varepsilon_1$ and $\varepsilon_2$, of the minimum and maximum, respectively. Features of the new best fits are illustrated in Figure \ref{fig4}, showing the distributions of $\chi^2$, $\varepsilon_1$ and $\varepsilon_2$. The mean value of $\chi^2$ is 0.0032, corresponding to an rms deviation of only 0.013, which is quite remarkable for such a low number of parameters. The distributions of $\varepsilon_1$ and $\varepsilon_2$ peak at $\varepsilon_1$$\sim$$-$0.09 and $\varepsilon_2$$\sim$$-$0.01, with full widths at half maximum of the order of 0.05; they result in shifts in $p$ and $q$, $\Delta{p}$ and $\Delta{q}$, having distributions centred at $\sim$$-$0.2 and $-$0.02, respectively, displaying significant tails. We studied in great detail the differences between the results of the two fits and found strong evidence for their consistency in terms of the conclusions which they lead to. While significant changes of the values of parameters $p$ and $q$ were sometimes observed, the relative location of the various curves in parameter space, with respect to each other, was not significantly altered; so was therefore their interpretation in terms of evolution of the stars along the AGB. In particular, we were concerned that the unexpected evidence for outliers in the $p$ vs $q$ distribution of the Mno curves, discussed in Section 4, might be an artefact of the two-parameter form that we adopted. Such is not the case, both fits give very consistent results, giving confidence in the robustness of the conclusions presented in the present article. 

\begin{figure*}
  \centering
  \includegraphics[width=0.9\textwidth,trim={0cm 0 0cm 0},clip]{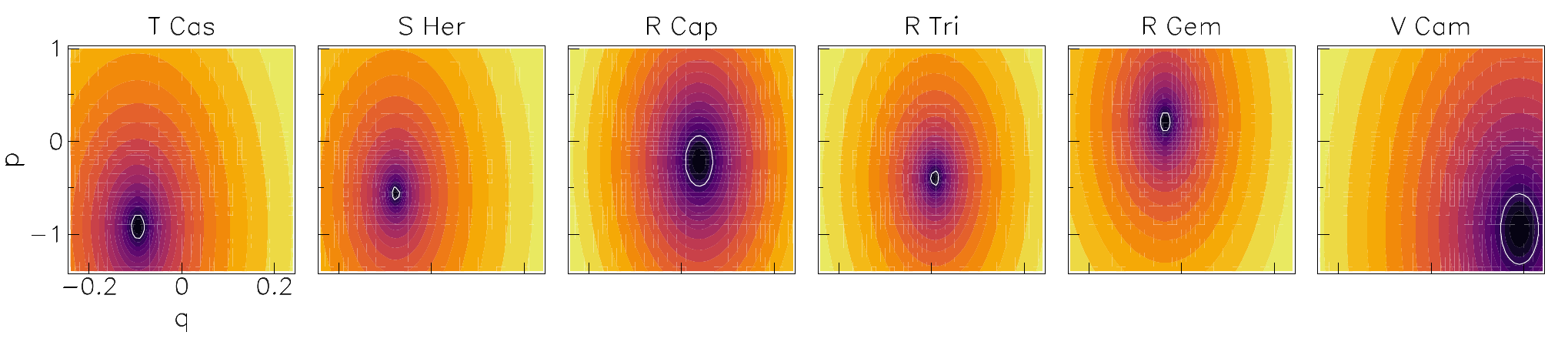}
  \caption{Six examples of $\chi^2$ maps. The level of the white contour is twice that of the minimal value. }
  \label{fig5}
\end{figure*}

Yet, we conservatively adopted a fit procedure that is intermediate between the two extreme cases mentioned above, $\varepsilon_1$=$\varepsilon_2$=0 and both $\varepsilon_1$ and $\varepsilon_2$ free to vary.  Using the former would mean ignoring the significant improvement obtained by allowing small $x$ shifts of the minima and maxima. Using the latter would require a better control over the uncertainties attached to each individual measurement in order to have confidence in the reliability of the different $\varepsilon_1$ and $\varepsilon_2$ values, obtained for each curve, at the level of a percent. The adopted procedure consists in using common values of $\varepsilon_1$ and $\varepsilon_2$ for all curves, $-$0.09 and $-$0.01, respectively. In practice, it means to use as model a form obtained by rescaling $S(x)$ between $x$=$\varepsilon_1$ and $x$=1$-$$\varepsilon_2$ but it ignores the different measurement errors affecting the location of the minima and maxima of each individual curve. Again, we have checked that the results presented in the remaining of the article would be unchanged if we had used instead any of the other two forms discussed above.

\begin{figure*}
  \centering
  \includegraphics[width=0.85\textwidth,trim={0cm 0 0cm 0},clip]{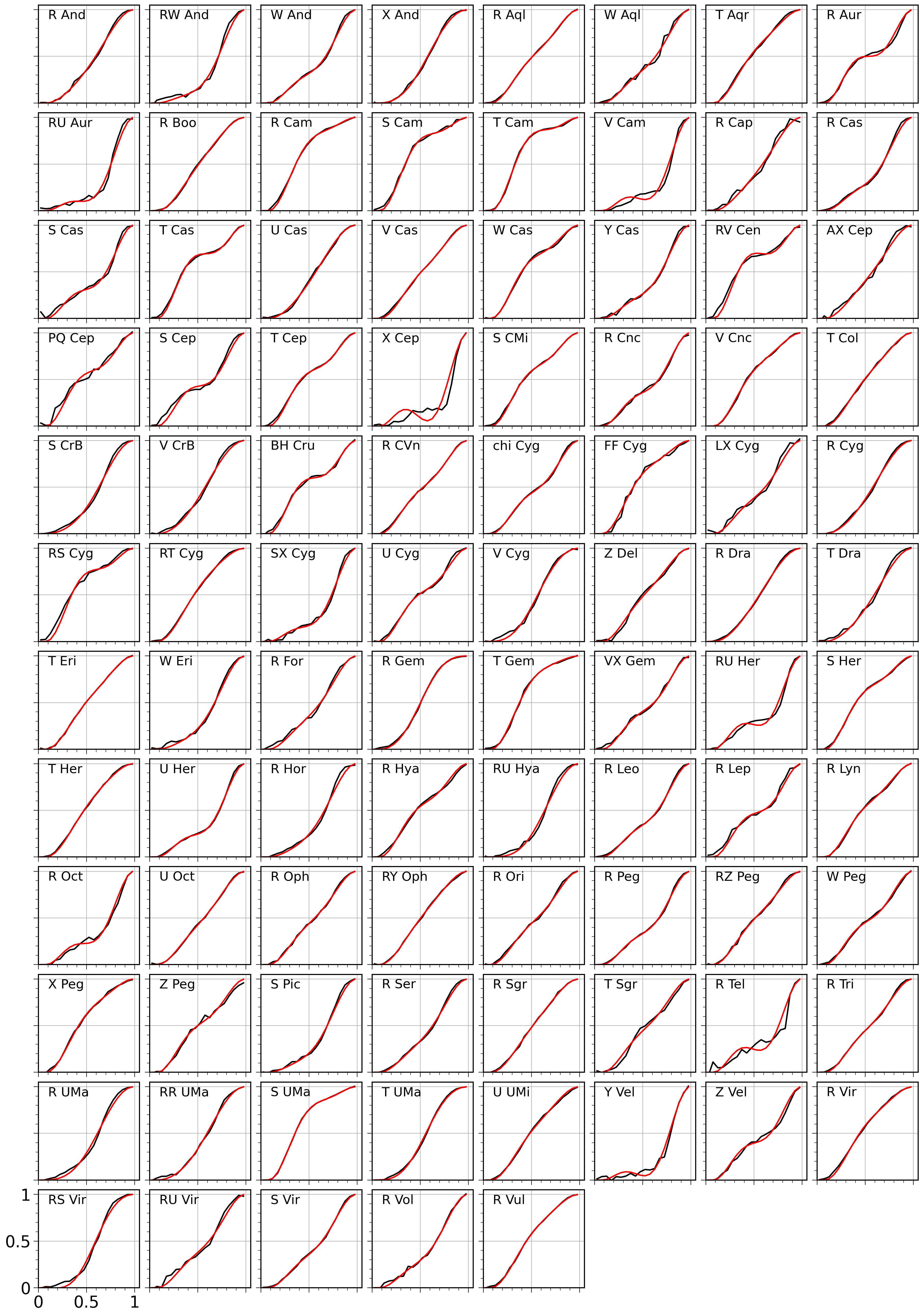}
  \caption{Normalized profiles of the ascending branches of the 93 curves of the sample. The red curves show the best-fit results to the two-parameter form. }
  \label{fig6}
\end{figure*}

The best-fit results are illustrated in Figures \ref{fig5} and \ref{fig6} and listed in Table \ref{tab1}. Figure \ref{fig5} displays six typical $\chi^2$ maps in the $p$ vs $q$ plane. On average, we estimate uncertainties of $\pm$0.1 in $p$ and $\pm$0.02 in $q$. Figure \ref{fig6} displays the 93 ascending branch profiles together with the results of the best fits. 

\begin{figure*}
  \centering
  \includegraphics[width=0.365\textwidth,trim={0cm 0cm 1.3cm 0},clip]{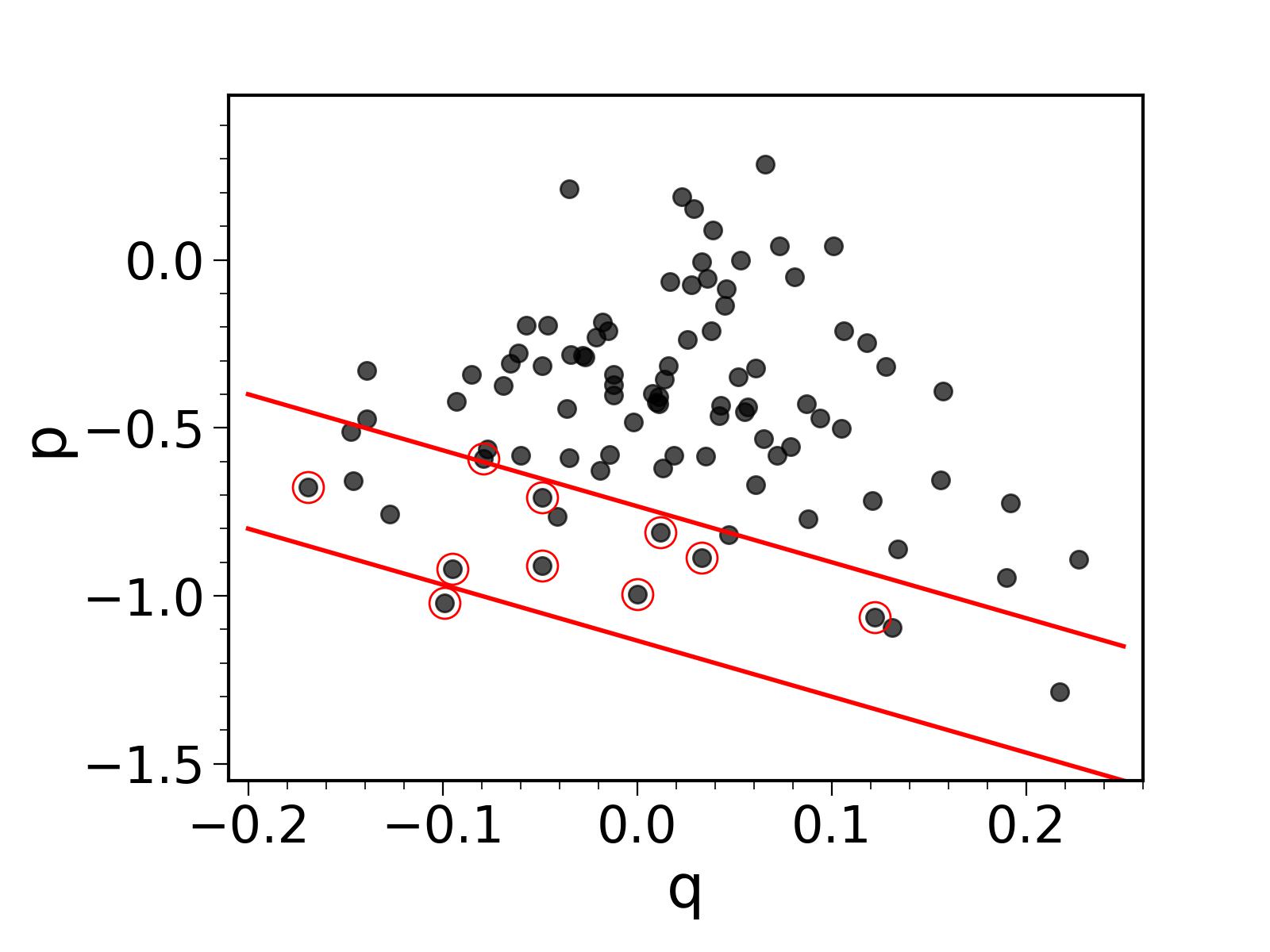}
  \includegraphics[width=0.6\textwidth,trim={0cm -1.4cm 0cm 0},clip]{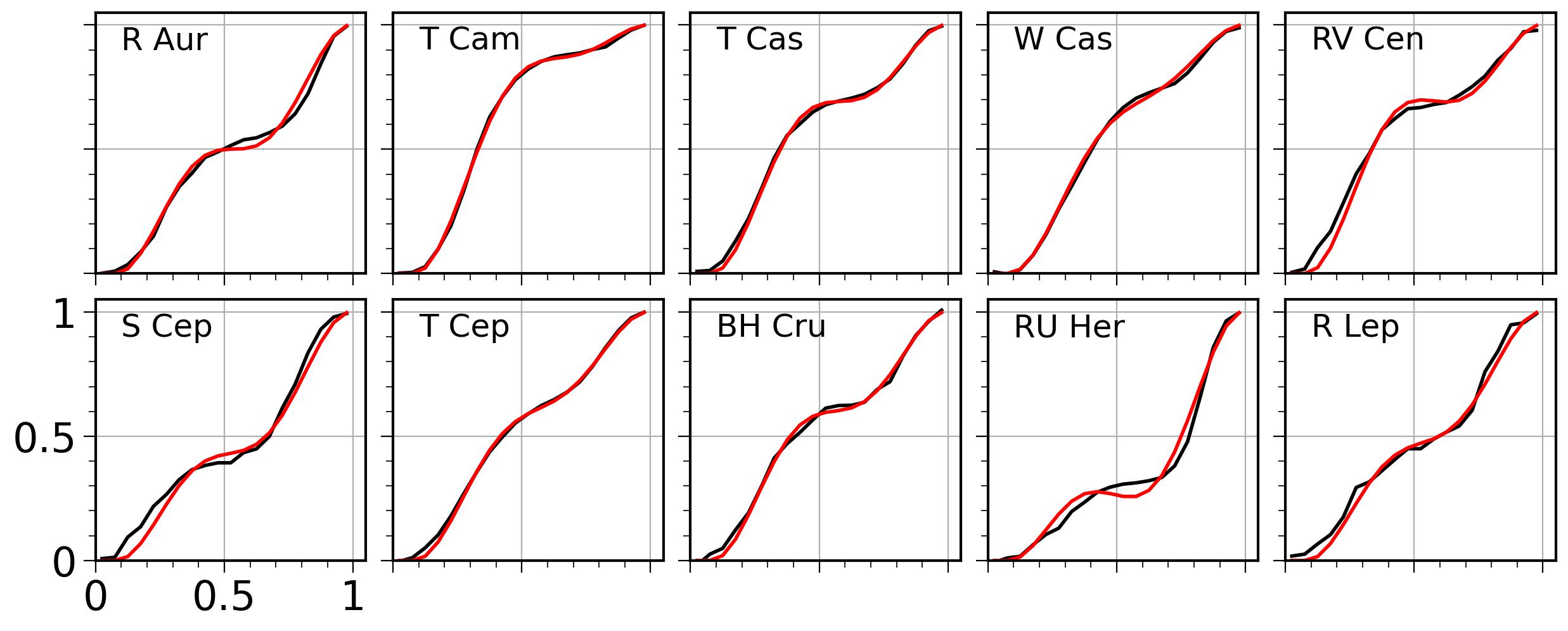}
  \caption{Left: location in the $p$ vs $q$ plane of 10 curves selected for displaying a clear hump (circled in red). They are seen to populate the lower part of the region covered by the 93 curves of the whole sample. Right: observed (black) and best fit (red) normalized profiles of the ascending branches of these 10 curves.}
  \label{fig7}
\end{figure*}

A remarkable result of the parameterization is the ability of the fit to describe accurately profiles displaying humps. An illustration is given in Figure \ref{fig7}, which shows ten profiles of ascending branches selected for displaying a clear hump. The mean $\chi^2$ is only 0.023 (rms deviation of 3.4\%) and the parameters cover a well-defined band in the $p$ vs $q$ plane. This is an important result, which describes the occurrence of humps as an expected episode in the evolution of the shape of the pulse profile, as had been already the case for the Hertzsprung progression in Cepheids, using in that case a two-mode description.

\subsection{Evolution in the parameter space}
\begin{figure*}
  \centering
  \includegraphics[width=4.5cm,trim={0cm -0.8cm 0cm 0},clip]{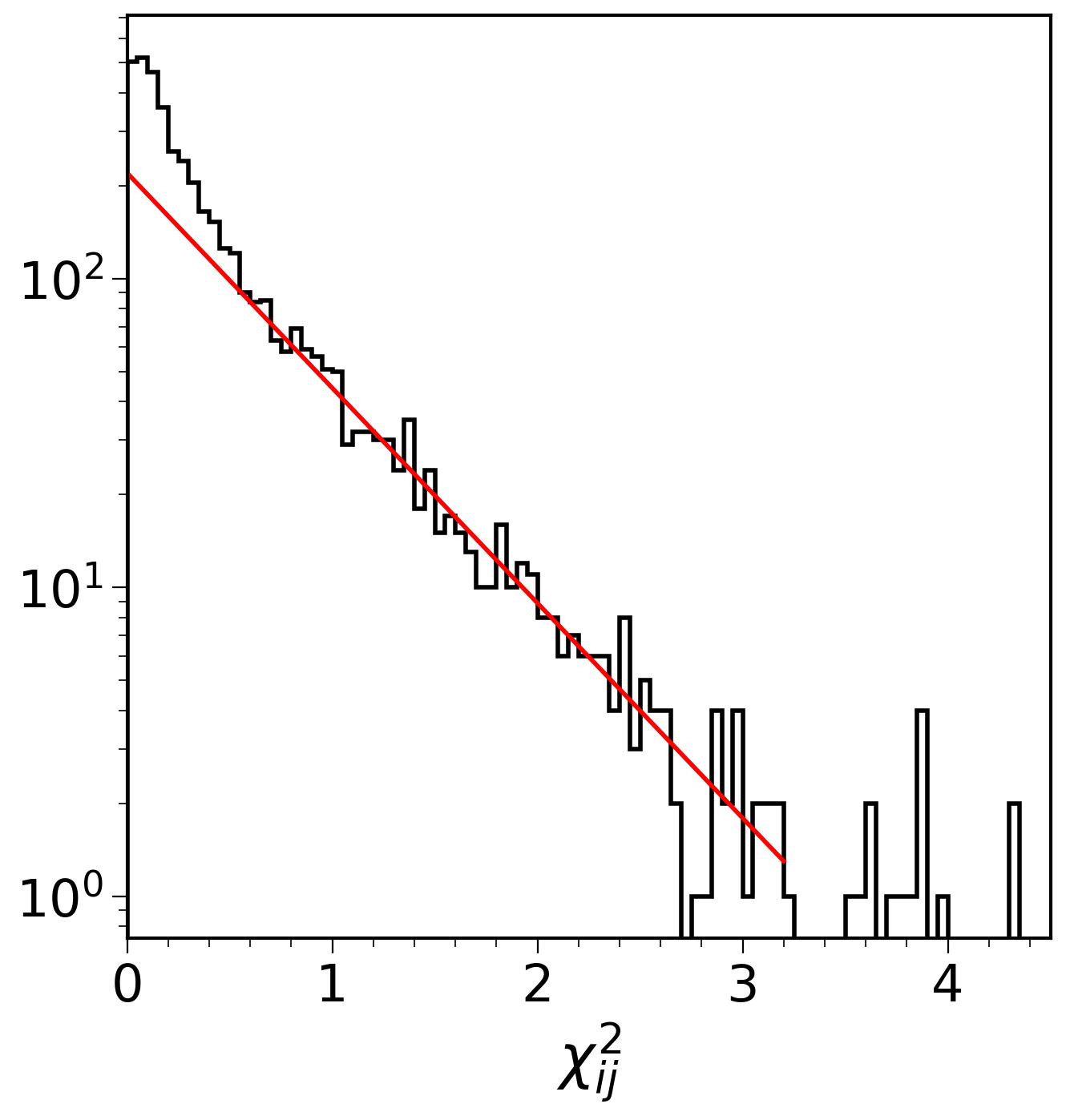}
  \includegraphics[width=5.4cm,trim={0cm 0 0cm 1cm},clip]{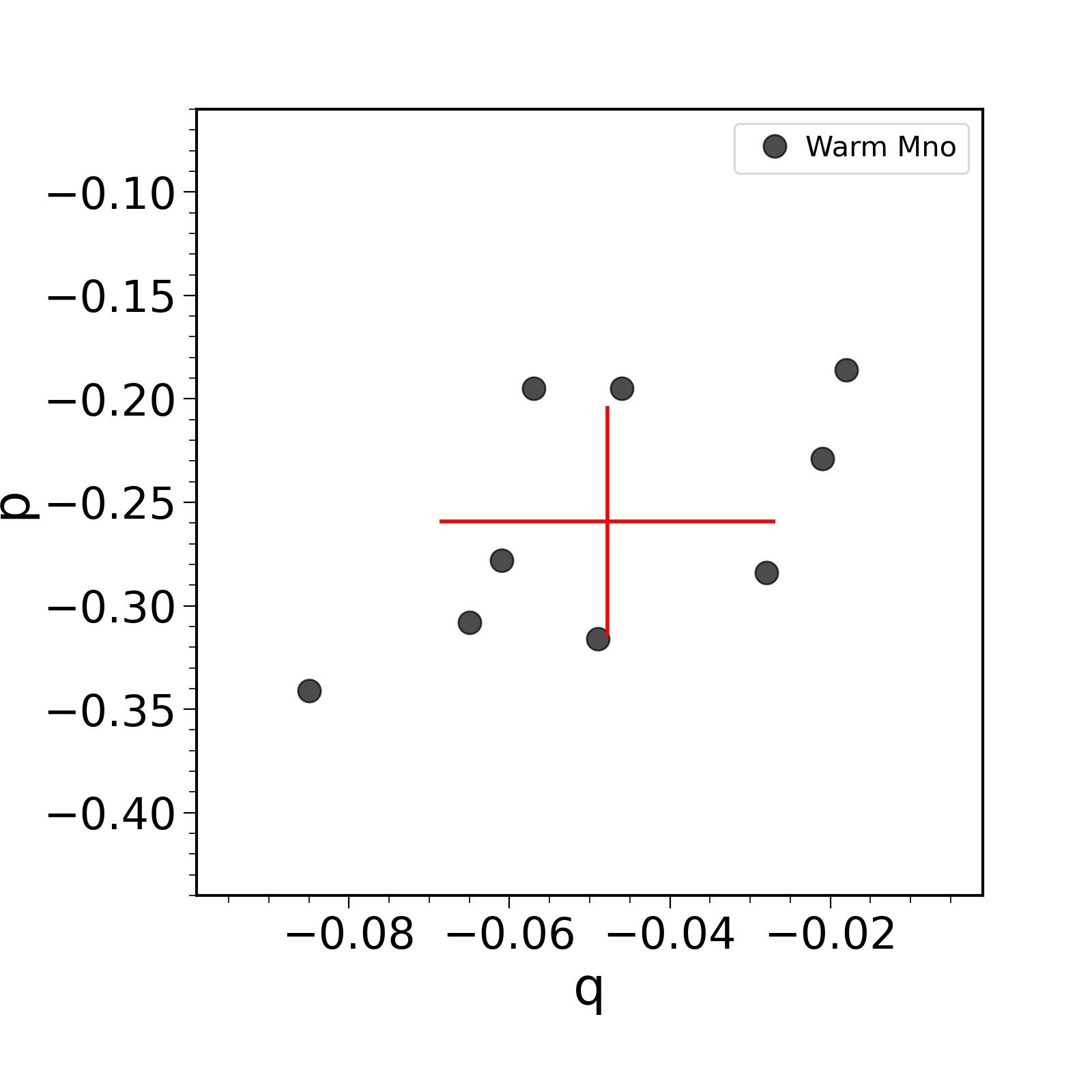}
  \includegraphics[width=4.7cm,trim={0cm -0.3cm 0cm 0},clip]{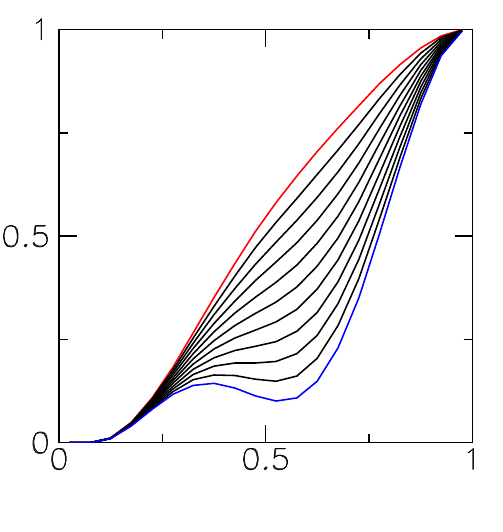}

  \caption{Left: $\chi_{\rm{ij}}^2$ distribution of the 4278 pairs of curves. Centre: $p$ vs $q$ distribution of the sample of warm Mno stars (Table \ref{tab2}). Right: evolution of the normalized ascending branch profiles along a straight line in parameter space linking the state of the warmer Mno stars (red line, Table \ref{tab2}) to that of the cooler Mno stars (blue line, [$p$,$q$]=[$-$1.0,+0.2]).}
  \label{fig8}
\end{figure*}

For each of the 93$\times$92/2=4278 pairs ($i,j$) of curves we calculate $\chi^2_{ij}$=$\sum[y_i(x_k)-y_j(x_k)]^2$ where the sum runs from $k$=1 to $k$=20.  Its distribution is shown in the left panel of Figure \ref{fig8}. The strong peaking at low values of $\chi^2$ reveals the presence of groups of curves having very similar shapes. Remarkably, but not unexpectedly, these groups are closely related to curves and stars sharing similar values of all their parameters. A particularly clear example is illustrated in the central panel of Figure \ref{fig8}. It includes the warmer members of the stars of Mno spectral type, listed in Table \ref{tab2}. Their confinement in a compact region of the $p$ vs $q$ plane goes together with significant confinement in the other curve and star parameters. This is a remarkable result: one would expect stars that differ by their initial mass, usually the same as their Main Sequence mass, and by their initial metallicity to follow well defined trajectories in the space of stellar parameters, their precise location on such trajectories depending only on their age. In the present case, even assuming that the whole sample is of solar metallicity, being dominated by stars in the solar neighbourhood, one would therefore expect the warmer Mno stars of the sample to be located in different regions of the stellar parameters, depending on their initial mass. One might then expect that their light curves be also located in different regions of the curve parameter space, in particular in the $p$ vs $q$ plane. But such is not the case. Essentially, this may imply that all precursors of the stars of the sample are entering the parameter space at a single location, independently of their initial mass or that the warm Mno stars of the present sample have initial masses confined to a narrow region. 

Related to such arguments, is the assumption that light curves of oxygen-rich stars evolve progressively and continuously along a trajectory in the parameter space when the star evolves along the AGB branch. Such an assumption relies on the fact that TDU events modify only progressively the properties of the stars, as amply suggested by model simulations \citep{Herwig2005}. This, of course, excludes short episodes in the vicinity of TDU events, but assumes that after a TDU event the parameters describing the star and its light curve take values that differ only modestly from the values that they had before the TDU event. We tend therefore to assume that the warmer Mno stars of the sample are the precursors of most of the stars in the sample, which evolve in the parameter space in different ways depending essentially on the values of their initial masses. But this is only an assumption; one might instead imagine that some stars enter the parameter space in other regions, at a more advanced state of evolution on the AGB than the warmer Mno stars do, when their regime of pulsation becomes regular enough for them to be accepted as Mira variables. 

\begin{deluxetable*}{lcl ccc ccc cc}
\tablenum{2}
\tablecaption{  Parameters of the warmer stars of Mno spectral type and of their curves. When the spectral types listed in the SIMBAD data 
base and in the \citet{MerchanBenitez2023} list differ, both are mentioned. The last line lists the mean values of the parameters and 
the rms deviations with respect to the mean. \label{tab2}}
\tablewidth{0pt}
\tablehead{
 \colhead{Name}&\colhead{$P$}&\colhead{Spectral type}&\colhead{$C$}&\colhead{$A$}&\colhead{$W$}&\colhead{$\Delta$}&\colhead{$\varphi_{\rm{min}}$}&
 \colhead{$p$}&\colhead{$q$}\\
  \colhead{}&\colhead{(day)}&\colhead{}&\colhead{(mag)}&\colhead{(mag)}&\colhead{}&\colhead{(mag)}&\colhead{}& \colhead{}&\colhead{}&\colhead{}
}
\startdata
T Aqr&	202&	M2&	1.4&	5&	0.34&	0.4&	0.5&	$-$0.316&	$-$0.049\\
R Boo&	224&	M4/M3&	1.7&	5.1&	0.31&	0.35&	0.52&	$-$0.186&	$-$0.018\\
T Col&	226&	M4/M3&	1.7&	4.3&	0.33&	0.37&	0.51&	$-$0.284&	$-$0.028\\
RT Cyg&	190&	M2&	1.5&	4.7&	0.38&	0.54&	0.54&	$-$0.195&	$-$0.057\\
T Her&	165&	M2.5&	1.5&	5&	0.34&	0.44&	0.52&	$-$0.195&	$-$0.046\\
RY Oph&	150&	M4/M3&	1.7&	4.7&	0.32&	0.47&	0.52&	$-$0.229&	$-$0.021\\
X Peg&	201&	M2.5&	1.3&	4&	0.39&	0.34&	0.5&	$-$0.341&	$-$0.085\\
R Vir&	146&	M3.5&	1.8&	4.3&	0.37&	0.34&	0.5&	$-$0.308&	$-$0.065\\
R Vul&	137&	M3&	1.8&	4.5&	0.37&	0.45&	0.51&	$-$0.278&	$-$0.061\\
\hline
Mean/rms&	182/32&	M3.1/M2.8&	1.6/0.2&	4.6/0.4&	0.35/0.03&	0.41/0.07&	0.51/0.01&	$-$0.26/0.05&	$-$0.048/0.021\\
\enddata
\end{deluxetable*}

Evolution from state 1 to state 2 along a straight line in the parameter space implies a similar evolution of its projection on any plane associated with a pair of parameters. For each parameter $\psi$ describing the curves and the stars, taking values $\psi_1$ and $\psi_2$ for state 1 and state 2, respectively, the parameter $\psi_\lambda$=$(\psi_1+\psi_2)/2$+$\lambda(\psi_2-\psi_1)/2$, with $\lambda$=$-$1 in state 1 and $\lambda$=$+$1 in state 2, describes the state of the star along such a straight line. As an example, we show in the right panel of Figure \ref{fig8} the evolution of the profile of the ascending branch linking the state of the warmer Mno stars described in Table \ref{tab2} to a state corresponding approximately to the cooler Mno stars of the sample, ($p,q$)=($-$1.0,$+$0.2). As we shall see in the next section, this corresponds to a region of the $p$ vs $q$ plane hosting a majority of Mno curves.

\begin{figure*}
  \centering
  \includegraphics[width=15cm,trim={0cm 3cm 0cm 2cm},clip]{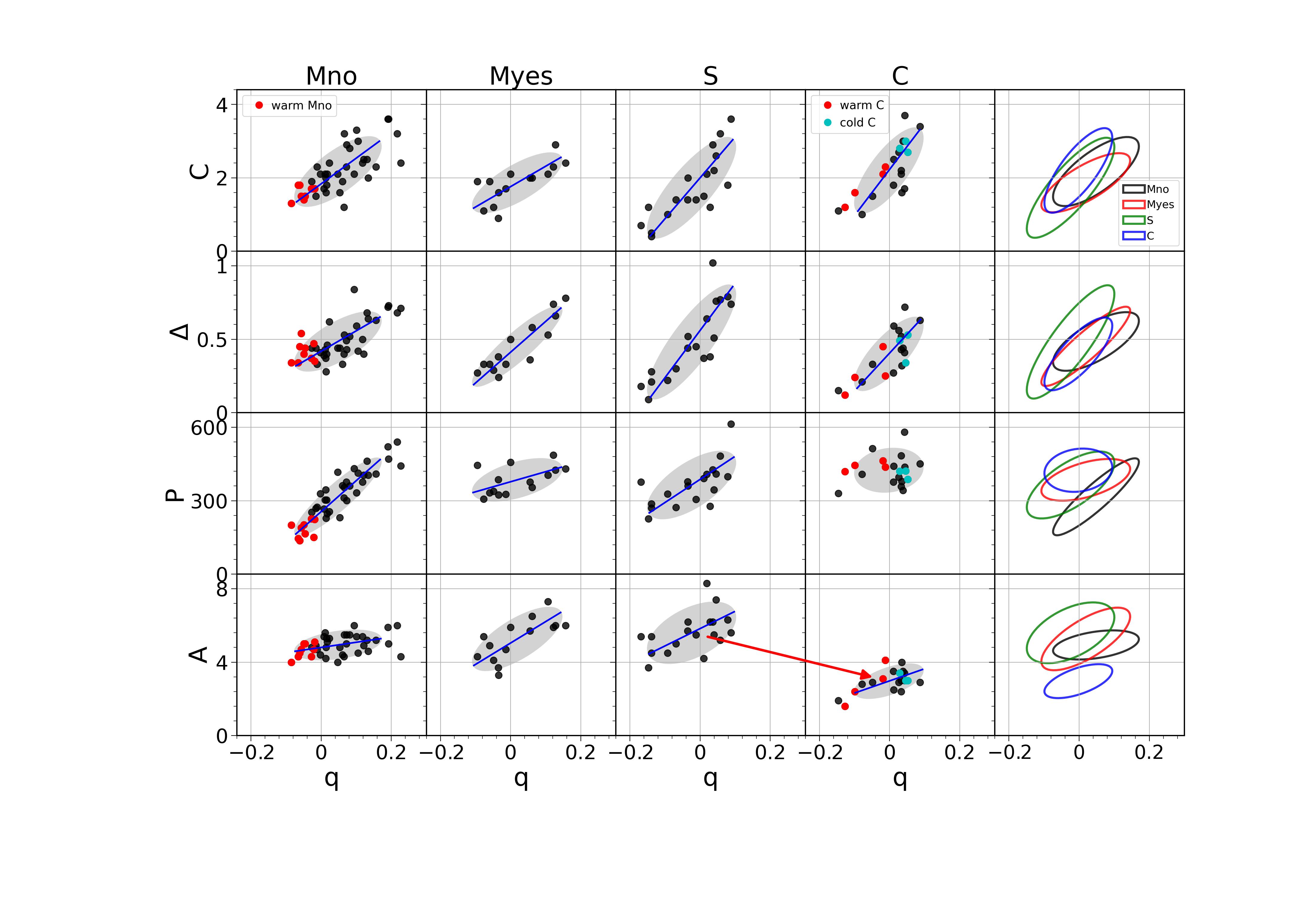}
  \includegraphics[width=15cm,trim={0cm 2.5cm 0cm 1.5cm},clip]{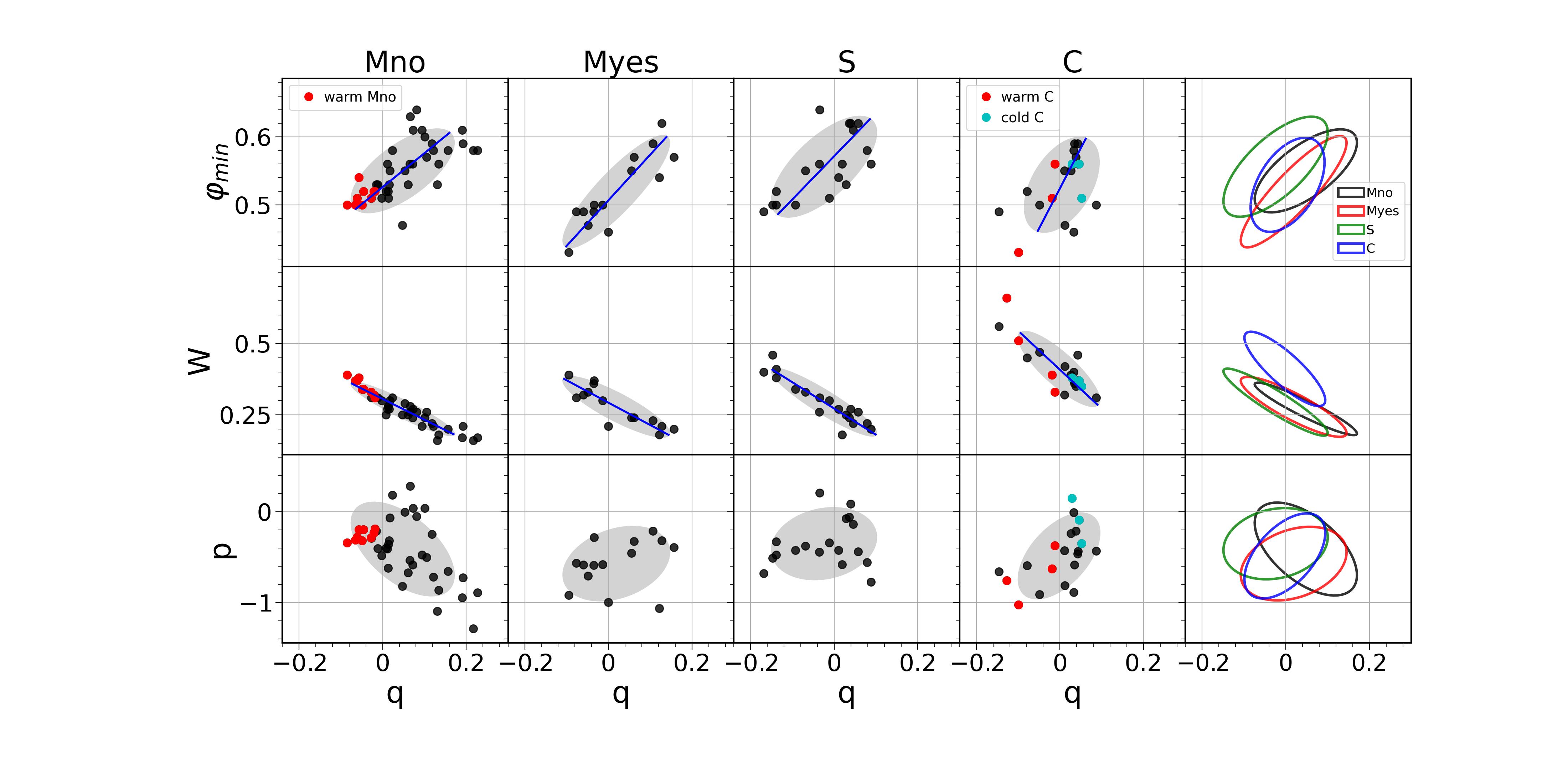}

  \caption{ Dependence on $q$ of the shape-insensitive (upper set of panels) and -sensitive (lower set of panels) parameters of the curves listed in Table \ref{tab1} according to spectral type (columns). The grey ellipses are for a deviation of $\pm$1.5 $\sigma$ and their contours are summarized in the right column with a colour code defined in the upper right panel of each set. Green and red full circles are used to distinguish between the cooler and warmer carbon-rich stars. In cases where a clear correlation is seen, straight lines show the results of linear fits. }
  \label{fig9}
\end{figure*}  

Figure \ref{fig9} displays the evolution of the curve parameters as a function of $q$ for each spectral type separately. The left panels are for the shape-sensitive parameters, $p$, $W$ and $\varphi_{\rm{min}}$, for which a correlation with $q$ is expected. The lower panels are instead for the shape-insensitive parameters, $P$, $A$, $C$ and $\Delta$, for which a correlation with $q$ reveals a non-trivial relation between the evolution of the star on the AGB and the shape of its light curve. We list below a few general comments that Figure \ref{fig9} suggests:

- the Mno stars have shorter periods than all other stars, consistent with the idea that these are precursors of all stars in the sample;
  
- in most planes, clear correlations with $q$ are observed, particularly strong for $P$, $C$, $\Delta$ and $W$, revealing the pertinence of using this shape parameter to describe the evolution of the star; the location of the warmer Mno stars on the low $q$ side and of the cooler carbon stars (see Section 6) on the high $q$ side of their respective spectral types strengthens this remark;
  
- a major drop of the amplitude of oscillation is observed in the transition from S to C spectral types. This is probably, at last in part, an effect of the different natures of the molecules in the photosphere, C-bearing in the former case and oxides in the latter. Other similar, but smaller jumps are observed in a few other cases. Yet, clear similarities between the $q$-distributions of the parameters in adjacent spectral types strongly suggest that such transitions imply only modest shifts of the $q$ parameter and that the light curves evolve in continuity with the evolution of the stars on the AGB.

\section{Stars of Mno spectral types} 
\begin{figure*}
  \centering
  \includegraphics[width=4.9cm,trim={0cm 0.5cm 2cm 1.3cm},clip]{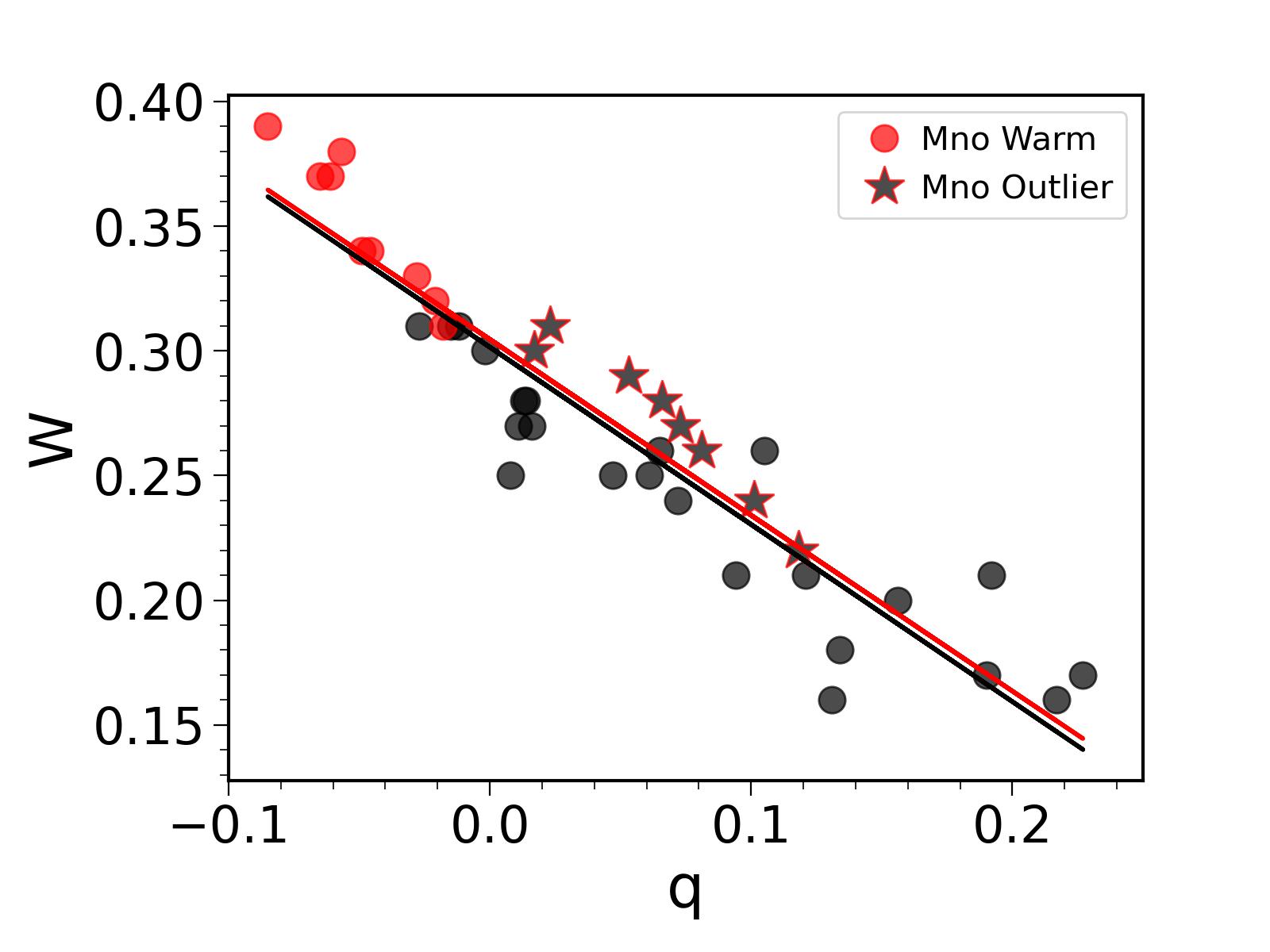}
  \includegraphics[width=4.9cm,trim={0cm 0.5cm 2cm 1.3cm},clip]{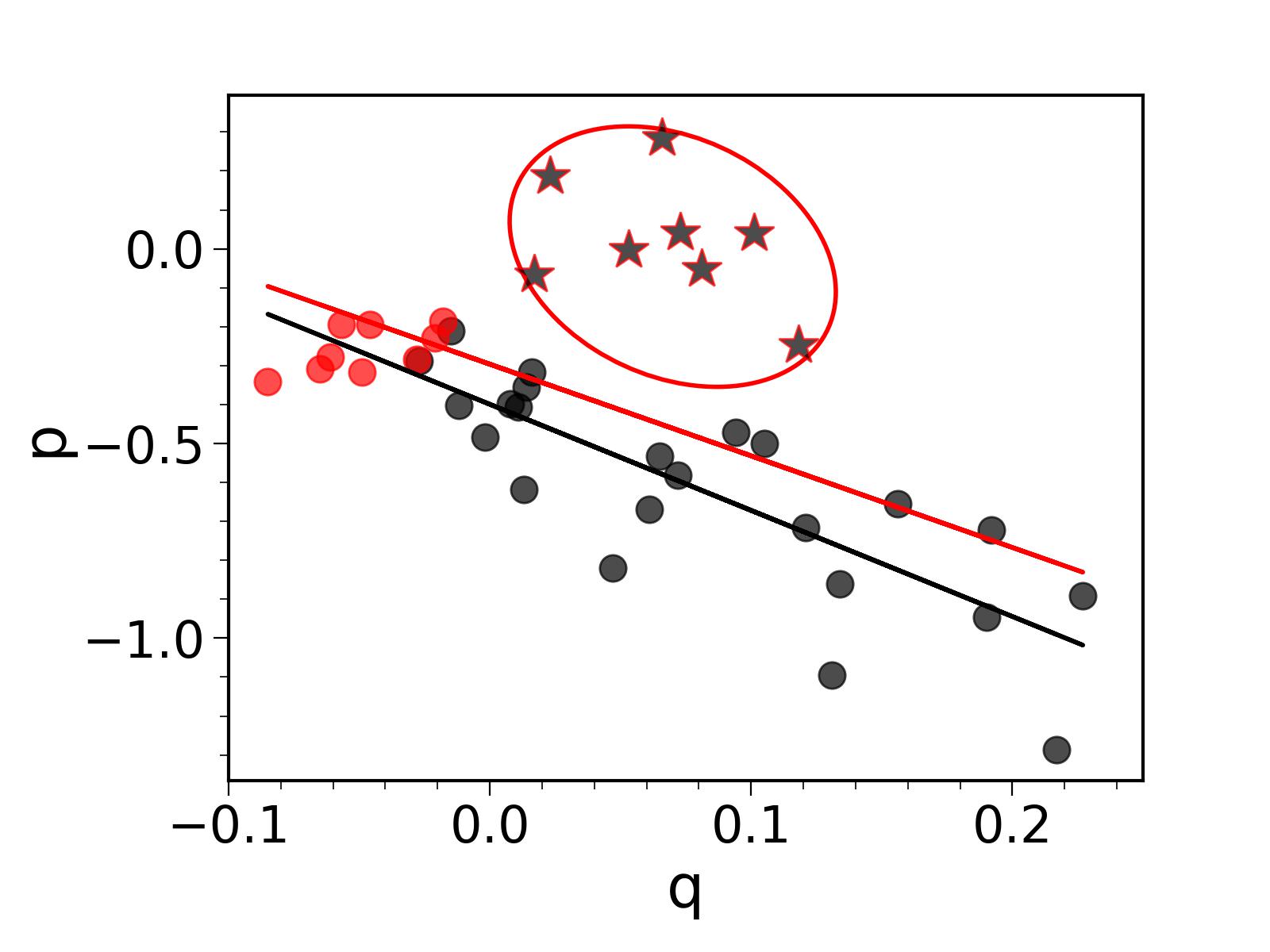}
  \includegraphics[width=4.9cm,trim={0cm 0.5cm 2cm 1.3cm},clip]{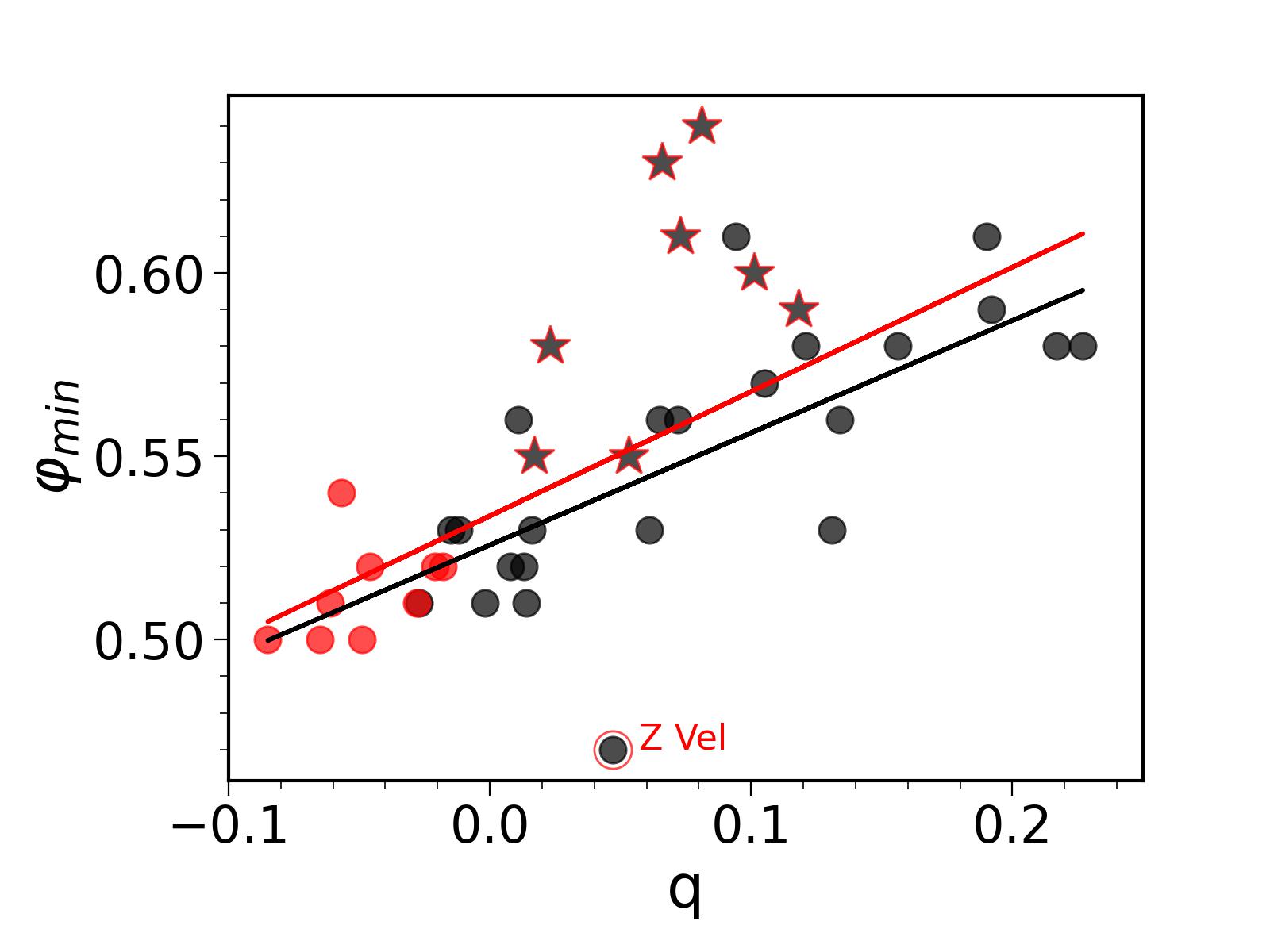}\\
  \includegraphics[width=4.9cm,trim={0cm 0.5cm 2cm 1.5cm},clip]{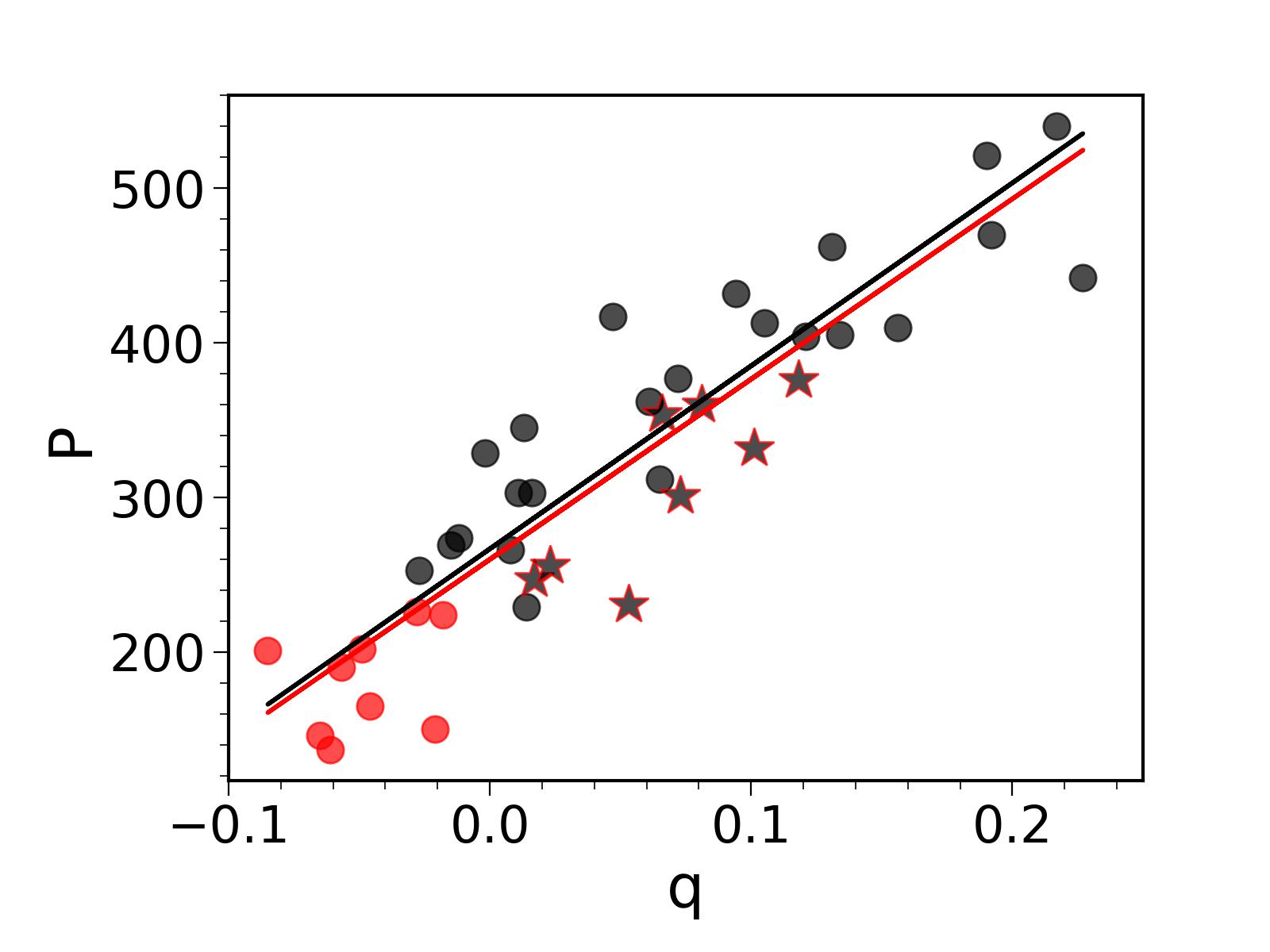}
  \includegraphics[width=4.9cm,trim={0cm 0.5cm 2cm 1.5cm},clip]{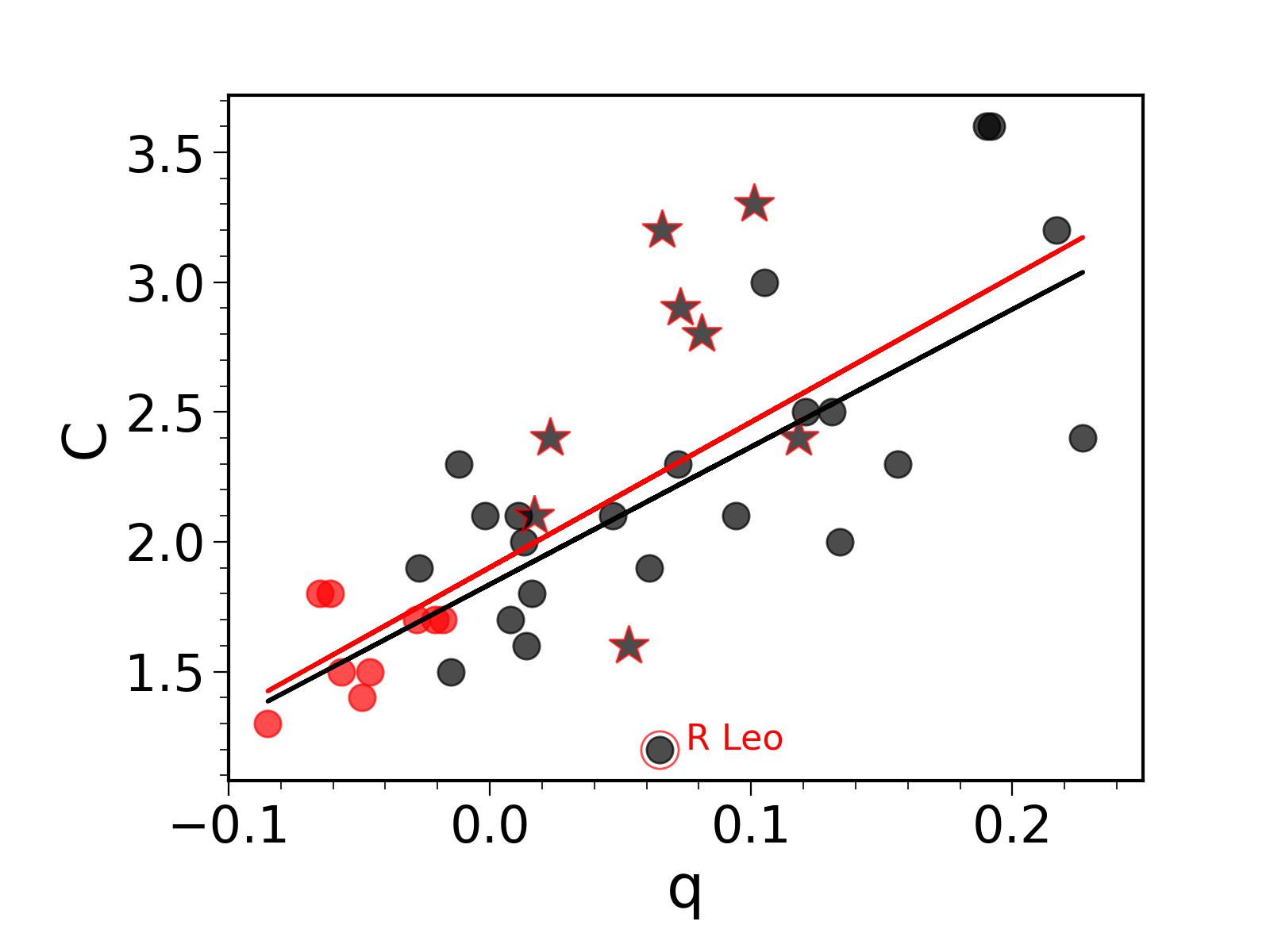}
  \includegraphics[width=4.9cm,trim={0cm 0.5cm 2cm 1.5cm},clip]{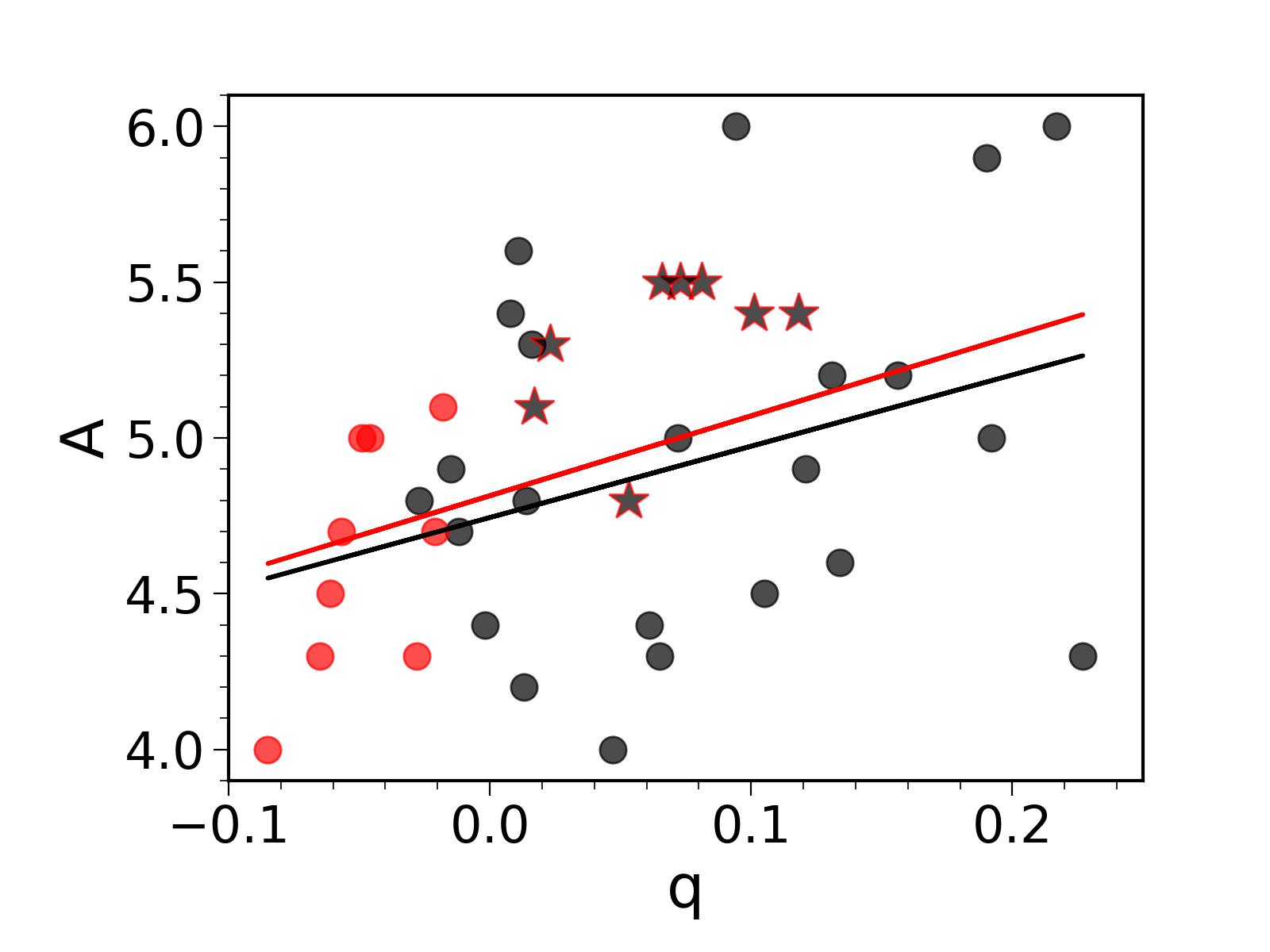}\\
  \includegraphics[width=4.9cm,trim={0cm 0.5cm 2cm 1.5cm},clip]{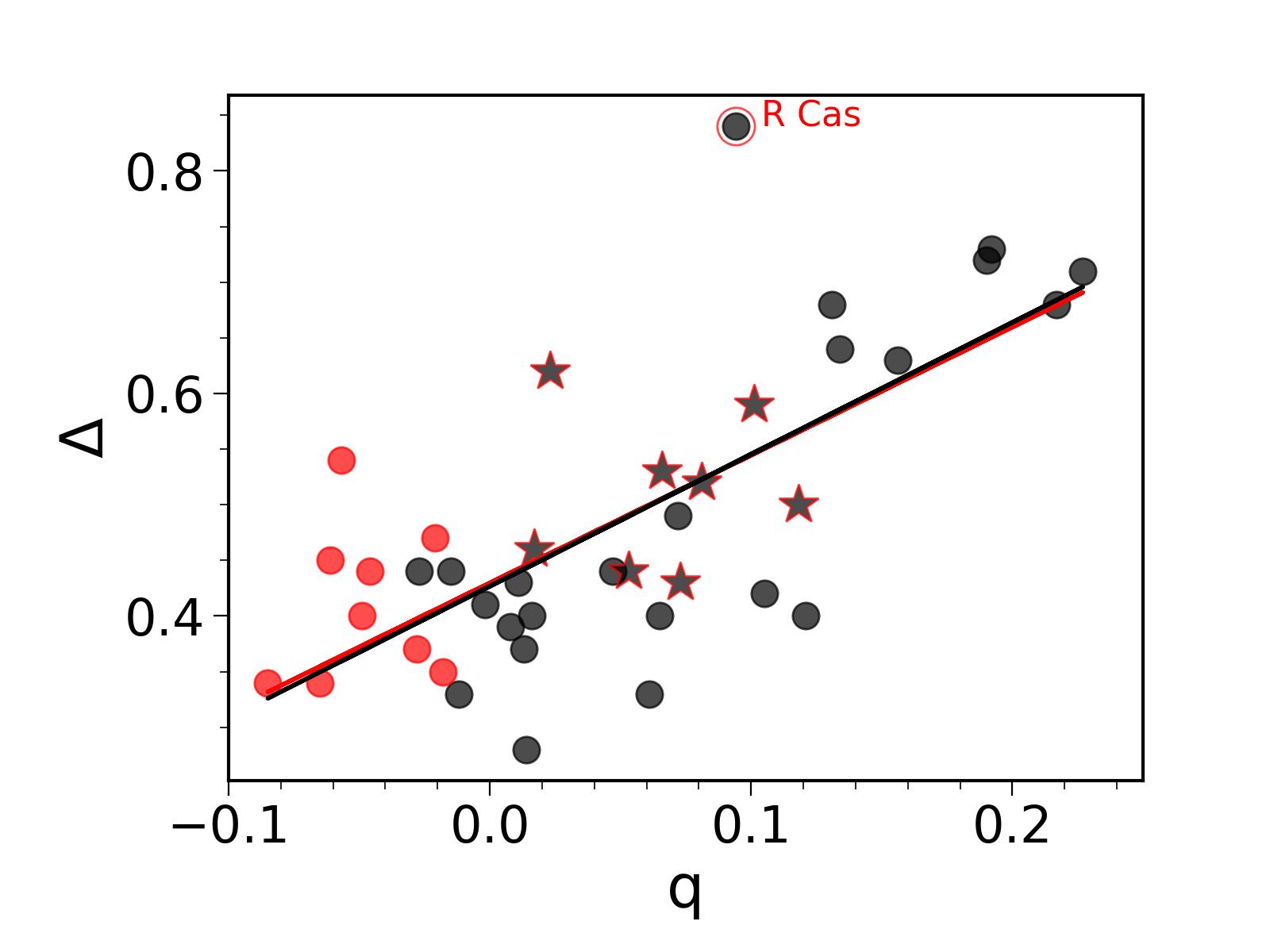}
  \includegraphics[width=4.9cm,trim={0cm 0.5cm 2cm 1.5cm},clip]{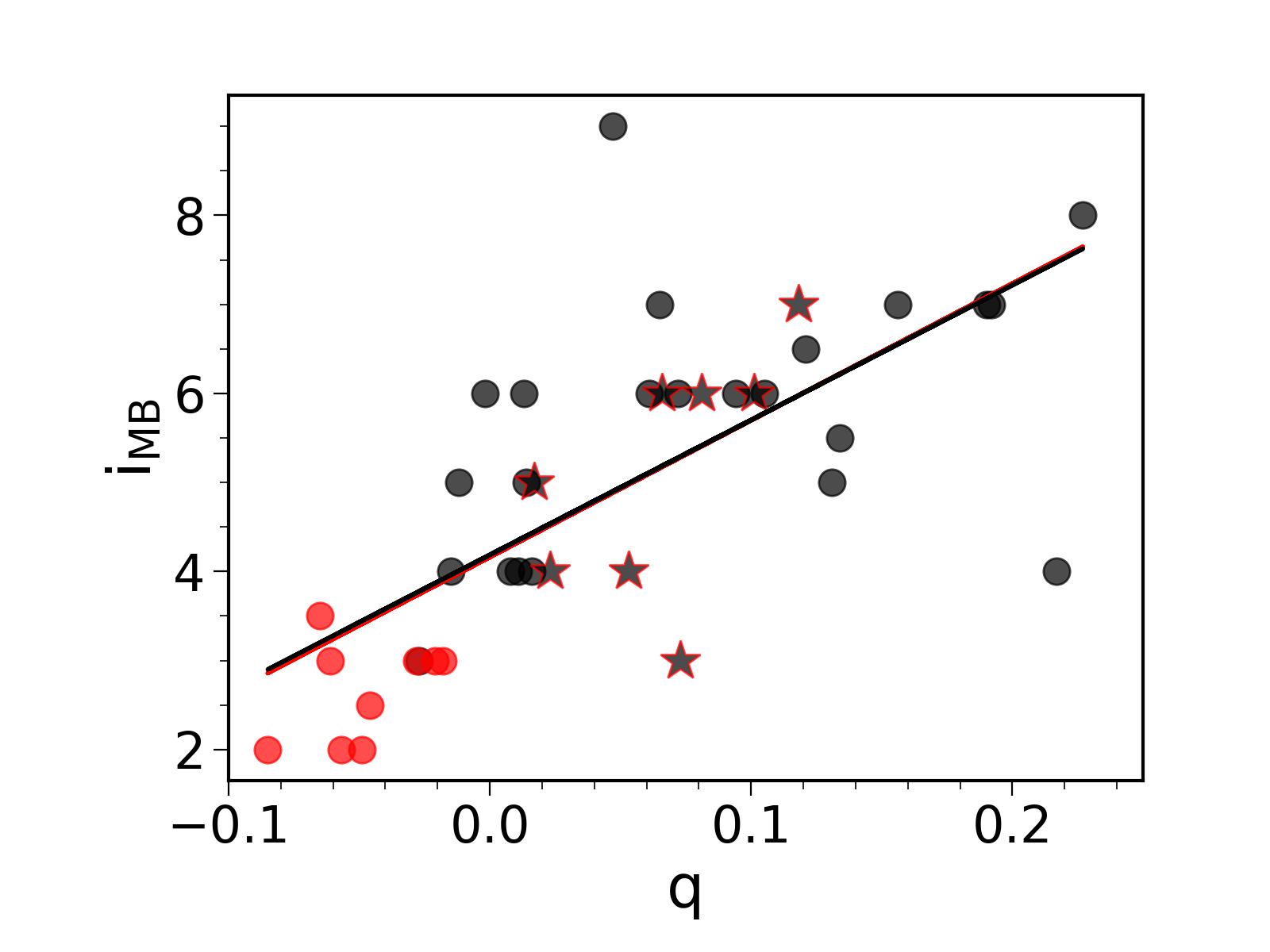}
  \includegraphics[width=4.9cm,trim={0cm 0.5cm 2cm 1.5cm},clip]{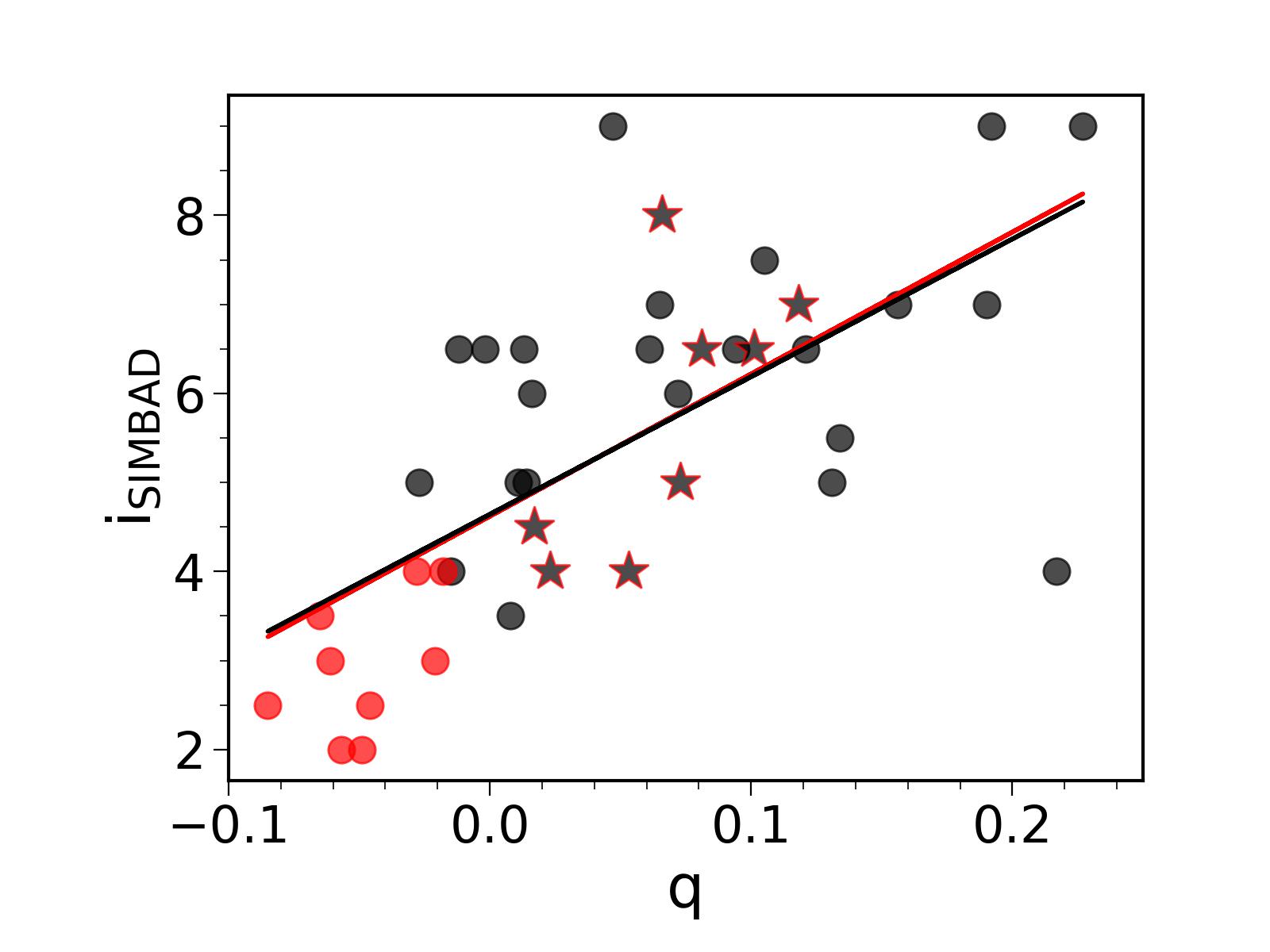}
  \caption{Distribution, in planes containing the $q$ axis, of star and light curve parameters of the Mno members of the sample. The outliers defined in the text are shown as stars. The warmer Mno stars are shown as red circles. The lines show the best linear fit results to the distributions, either excluding (black) or including (red) the outliers.}
  \label{fig10}
\end{figure*}

Figure \ref{fig10} displays, for the whole Mno sample, the dependence on parameter $q$ of all other star and light curve parameters. Clear correlations are observed in all cases, particularly strong for the period, $P$, the index of the M spectral type, $i$, the width parameter $W$, and the other ascending branch parameter, $p$.  While correlations with the latter two can be expected, both $W$ and $p$ depending on the shape of the pulse profile as $q$ does, correlations with $P$ and $i$ are more basic, describing a relation between the shape of the light curve and the state of the star on the AGB. The fact that both $P$ and $i$ increase with $q$ gives strong evidence for $q$ to increase as the star evolves. In spite of significant differences between the spectral type assignments given by the SIMBAD data base and by \citet{MerchanBenitez2023}, both display very similar positive correlations between $i$ and $q$. The location of the warmer stars at the low $q$ ends of the distributions gives support to the idea that these are precursors of the other curves. Namely, when a star is in a state of Mno spectral type, its period increases, its temperature decreases, the width of the pulse decreases and its ascending branch becomes steeper. Qualitatively, such an evolution had already been described in Papers I and II.

\begin{figure*}
  \centering
  \includegraphics[width=5.2cm,trim={0.5cm 0.5cm 1.4cm 1.2cm},clip]{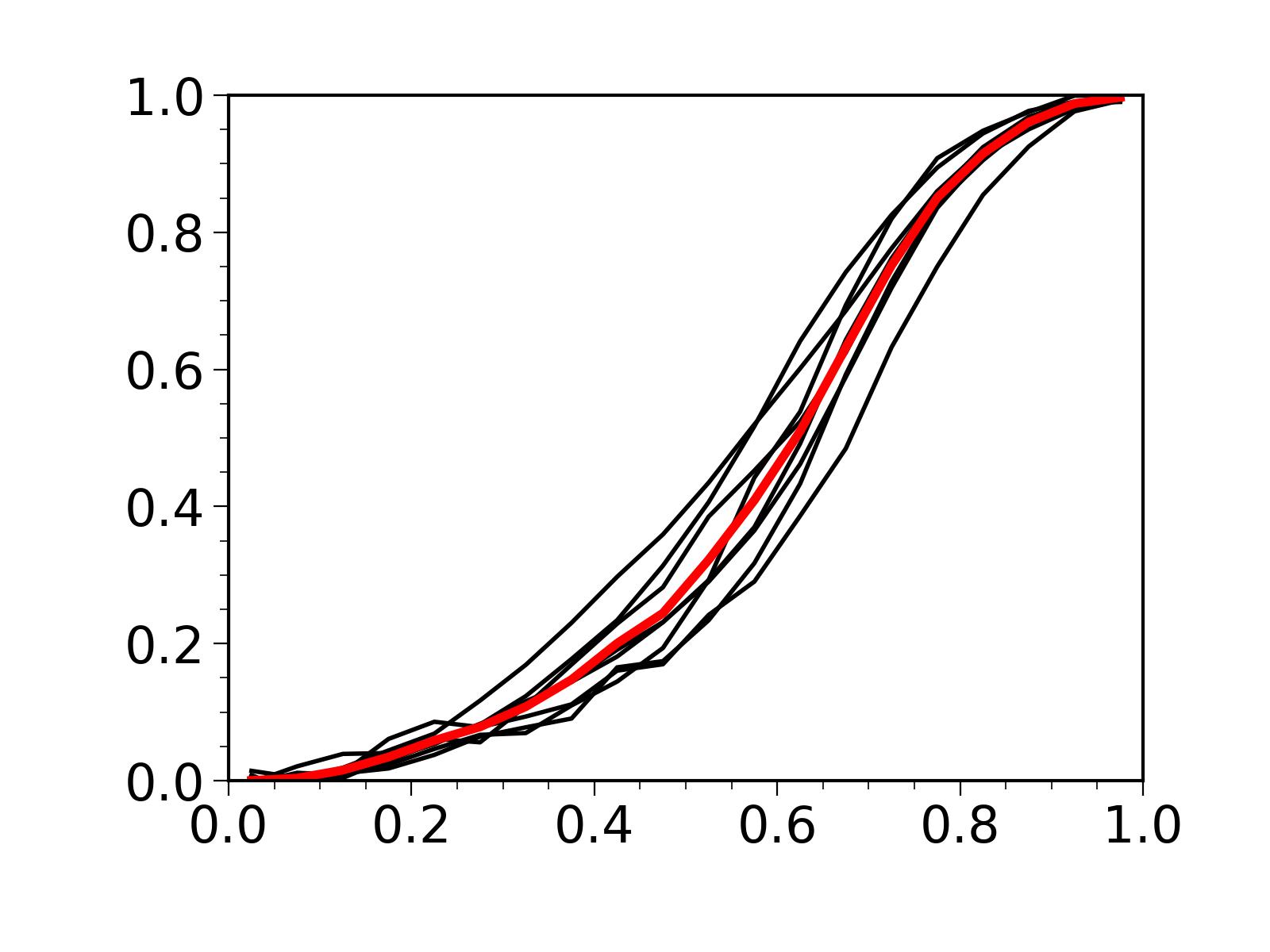}
  \includegraphics[width=5.2cm,trim={0.5cm 0.5cm 1.4cm 1.2cm},clip]{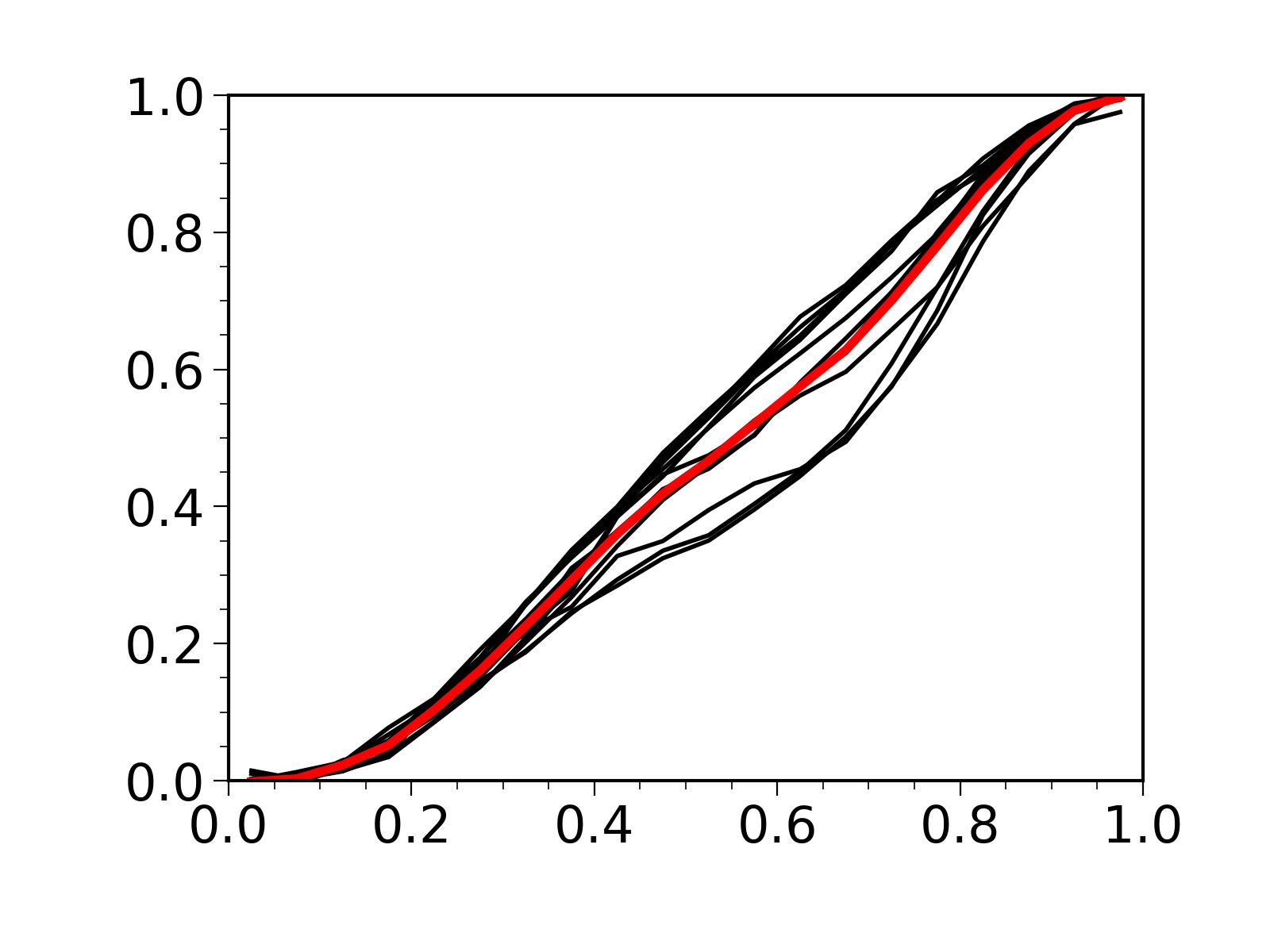}

  \caption{Mno outliers. Superimposed normalized profiles of the ascending branches of the eight outliers (left) and of the twelve other Mno curves having similar periods (right). The red lines show the mean normalized profiles.} 
  \label{fig11}
\end{figure*}

\begin{figure*}
  \centering
  \includegraphics[width=4.5cm,trim={1.2cm 1cm 1.4cm 1.5cm},clip]{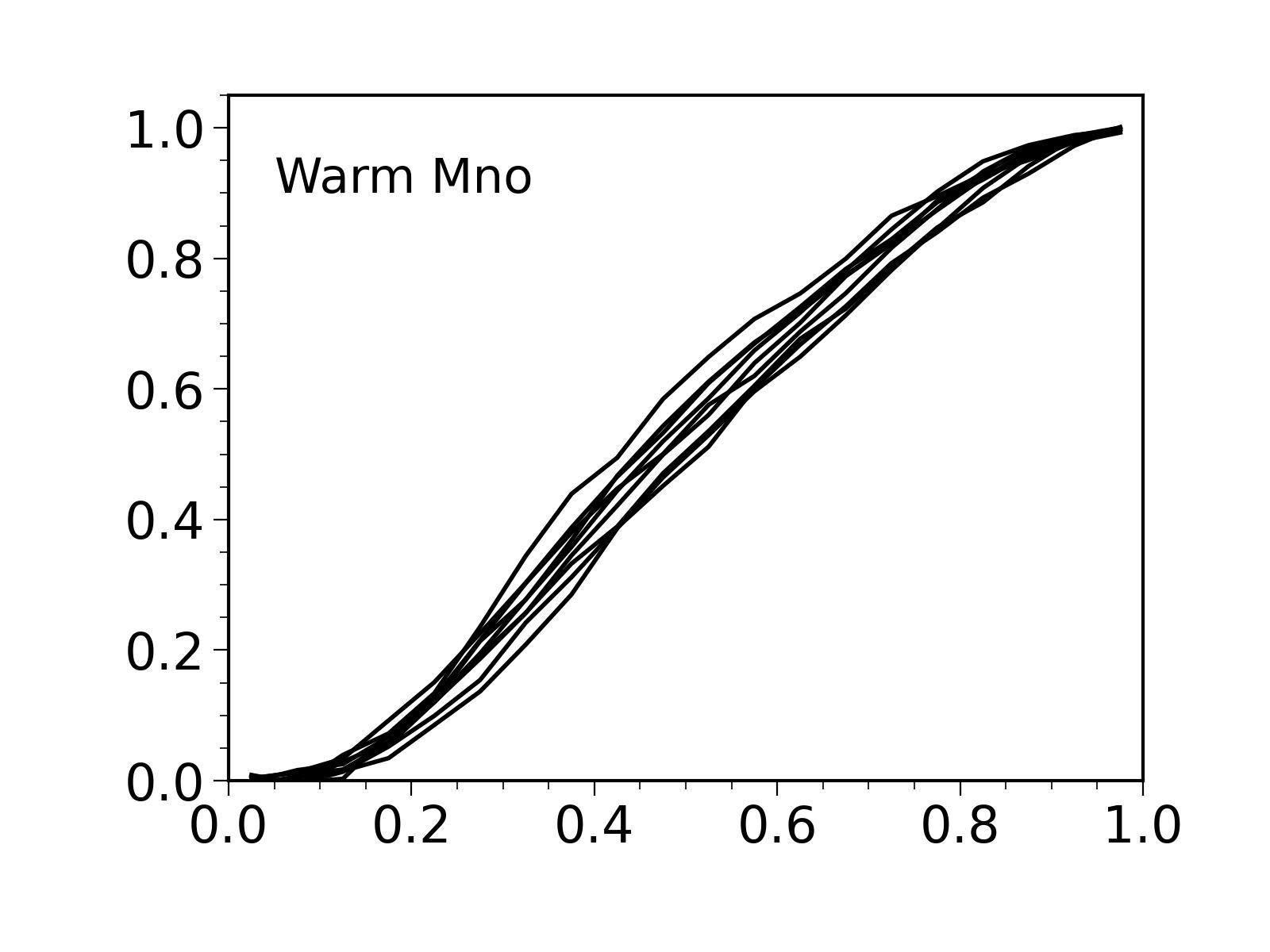}
  \includegraphics[width=4.5cm,trim={1.2cm 1cm 1.4cm 1.5cm},clip]{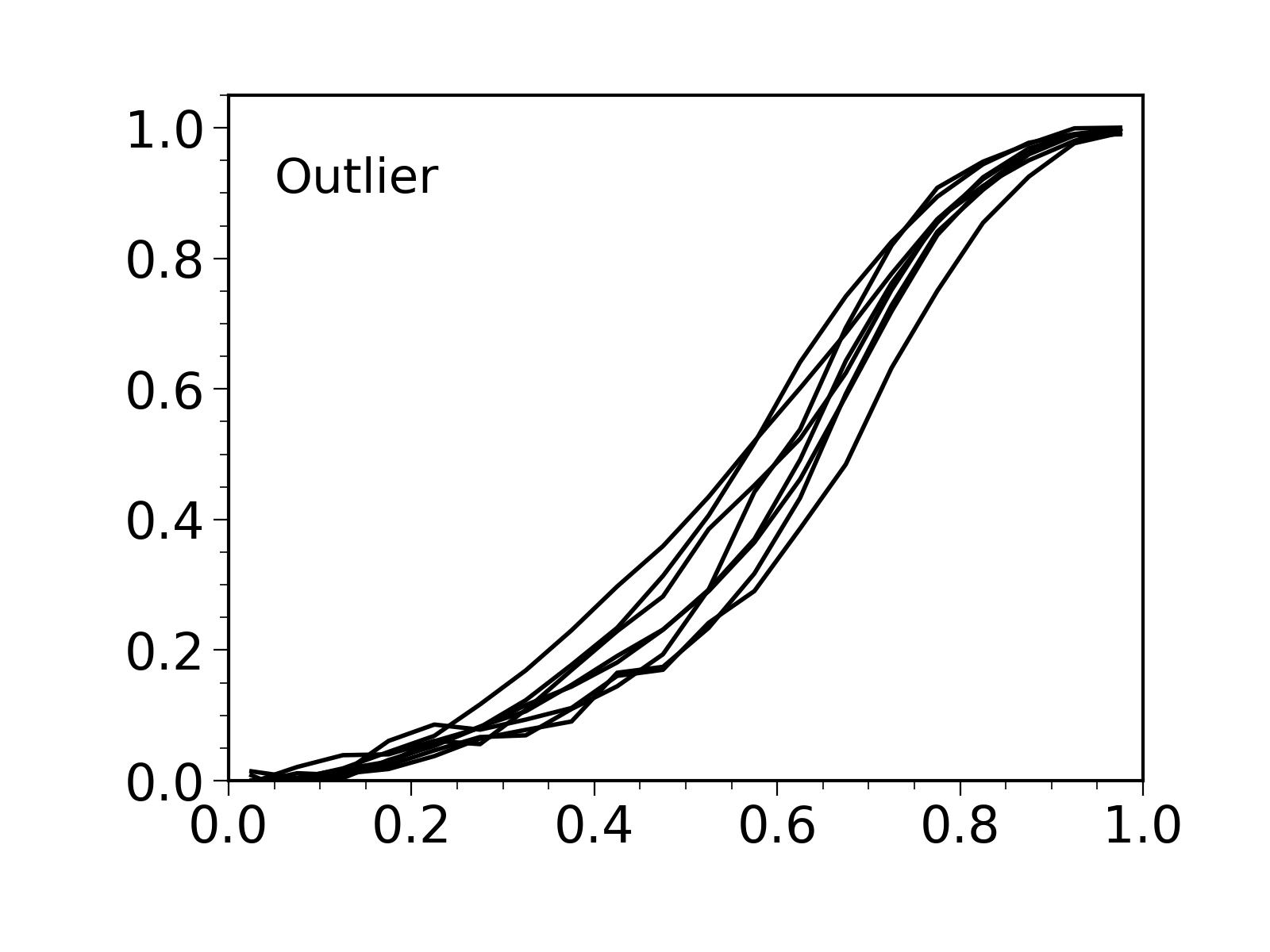}
  \includegraphics[width=4.5cm,trim={1.2cm 1cm 1.4cm 1.5cm},clip]{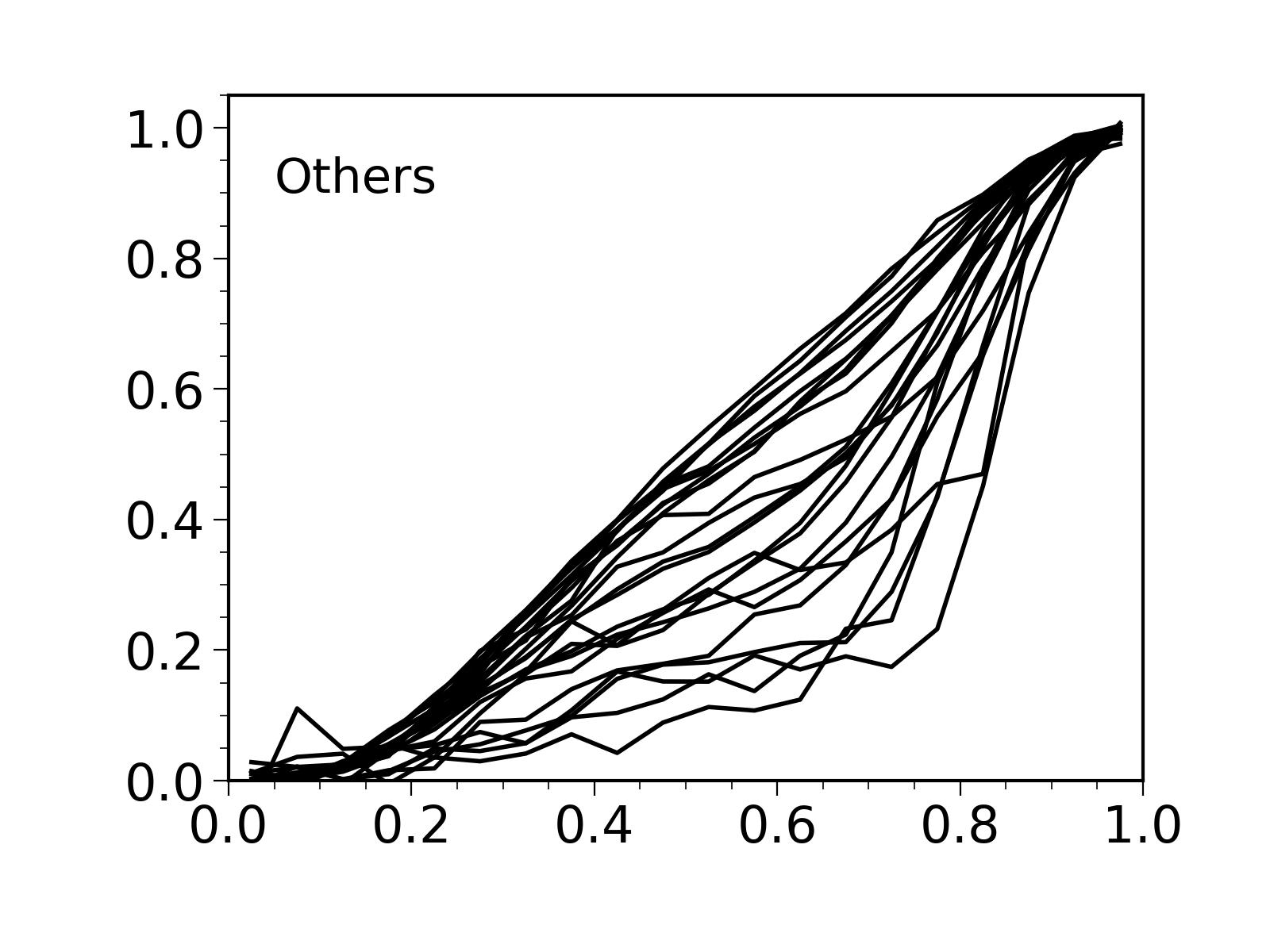}
      
  \caption{Superimposed profiles of the ascending branches of the curves of Mno type: warmer Mno types listed in Table \ref{tab2} (left), outliers (centre) and all other curves (right).}
  \label{fig12}
\end{figure*}

What is new, however, is the presence in the $p$ vs $q$ distribution of a group of 8 clear outliers, shown as red circles in Figure \ref{fig10}: RR UMa, T UMa, RU Hya, RS Vir, W Eri, S CrB, R Dra and R UMa. They do not stand out clearly in the other parameter distributions, which prevented their identification in Papers I and II. They mostly differ from the other Mnos by the shape of their ascending branches, as illustrated in Figure \ref{fig11}, and we failed to find evidence for another significant difference, which might have helped with providing a plausible interpretation. As we lack a clear understanding of the physics behind the distinction between outliers and other Mno curves, we show in Figure \ref{fig10} the best fit results of linear fits, both excluding and including the outliers. Their coefficients are listed in Table \ref{tab3}. The profiles of the ascending branches of the Mno sample are illustrated in Figure \ref{fig12}.

Four curves are seen to stand out from the general trend in at least one of their parameters: Z Vel, X Cep, R Leo and R Cas.

Z Vel and X Cep have respectively high and low values of their spectral type indices, both with the SIMBAD data base and the Merch\'{a}n Benitez list. They are of unknown technetium content and were assigned an Mno spectral type from the values of the other star and curve parameters. Z Vel has a low value of the phase at minimum light. However, its low amplitude of oscillation would make an Myes assignment even less acceptable than an Mno assignment.

R Leo, of spectral type M7, has the lowest value of $C$ among stars of Mno spectral type, $C$=1.2 mag. It is quoted in Paper I as having K-[22]=1.3 mag, [3.4]-[22]=1.0 mag and a $^{12}$C/$^{13}$C isotopic ratio of 10. The other parameters are consistent with the general trend. \citet{Ramstedt2014} quote a mass-loss rate of $10^{-7}$ M$_\odot$ yr$^{-1}$. \citet{Hoai2023} discuss the complexity of the morpho-kinematics of the atmosphere, which may explain the very low values of its colour indices.

R Cas, of spectral type M6.5, was seen in Paper I to have very large values of the irregularity parameters $\Delta{M_{\rm{max}}}$ and $\Delta'{M_{\rm{max}}}$, respectively 0.75 and 0.93 mag, giving the mean value $\Delta$=0.84 mag used in Paper II. $\Delta{M_{\rm{max}}}$ is the rms deviation from its mean of the magnitude at maximum light, $\Delta'{M_{\rm{max}}}$ is the mean value of the difference of magnitude between two successive light maxima. Such large values are unusual for a star of Mno spectral type, being instead more common for stars of S spectral type. But R Cas is clearly of Mno spectral type and has a $^{12}$C/$^{13}$C isotopic ratio of only 12 compared with values larger than 30 for the S stars. The light curve has been measured over more than a century with observations of excellent quality and high density and its irregularity remained all that time at the same level. However, as shown in the right column of Figure \ref{fig11}, the longer period Mno stars, such as X Cep and RU Aur, are also quite irregular. Compared with them, R Cas has a lower value of the colour index and a different profile of the ascending branch (Figure \ref{fig6}).  Its large amplitude of oscillation may suggest that it is close to evolve to Myes spectral type, if it ever will.

\begin{deluxetable*}{ccc ccc c}
\tablenum{3}
\tablecaption{Results of the linear fits of the form $a$+$bq$ to the parameters of Mno curves.  \label{tab3}}
\tablewidth{0pt}
\tablehead{
 \colhead{Parameter}&\multicolumn{3}{c}{Excluding outliers} &\multicolumn{3}{c}{Including outliers}\\
  \colhead{}&\colhead{$a$}&\colhead{$b$}&\colhead{rms}&\colhead{$a$}&\colhead{$b$}&\colhead{rms}
}
\startdata
$p$ &	$-$0.39 &$-$2.82 &	0.14 &	$-$0.28 &$-$2.49 &0.28\\
$P$ &	265&	1203&	43&	258&	1189 &	44\\
$C$ &	1.84&	5.25&	0.37&	1.91&	5.53 &	0.43\\
$A$ &	4.78&	1.97&	0.49&	4.85&	2.23 &	0.48\\
$W$ &	0.3&	$-$0.69&	0.022&	0.3&	$-$0.69&0.021\\
$\Delta$&	0.43&	1.19&	0.099&	0.43&	1.16&	0.094\\
$\varphi_{\rm{min}}$&	0.53&	0.31&	0.022&	0.53&	0.34&	0.03\\
$i_{\rm{MB}}$&	4.24&	14.6&	1.67&	4.21&	14.9&	1.56\\
$i_{\rm{SIMBAD}}$&	4.69&	15&	2.13&	4.66&	15.5&	1.95\\
\enddata
\end{deluxetable*}

\section{Stars of Myes and S spectral type}
\subsection{Overview}
The distribution of Myes, S and C stars in the curve parameter space is complex and more difficult to interpret than that of the Mno stars: it needs to be scrutinized in detail. In order to ease the discussion, we find it useful to group together curves sharing similar parameters and use such groups as references along the evolution of the stars and of their light curves; we define five such groups:

- a group My1 of Myes stars close to the Mno band including RW And, RU Her, R Hor, S Pic, R Ser and S Vir;

- a group S1 of S stars close to the Mno band including X And, U Cas, R And, W And, W Aql, S Cas, chi Cyg and R Cyg;

- a group My2 of Myes stars close to the warmer Mno's, including T Cep, S CMi, S Her, R Hya, Z Peg and U UMi;

- a group S2 of S stars close to the warmer Mno's including V Cnc, Z Del, R Gem and R Lyn;

- a group S3 of S stars of low $q$ values and low colour indices including R Cam, T Cam, T Gem and S UMa.

\begin{figure*}
  \centering
  \includegraphics[height=4.2cm,trim={0.cm 0cm 1.4cm 1.2cm},clip]{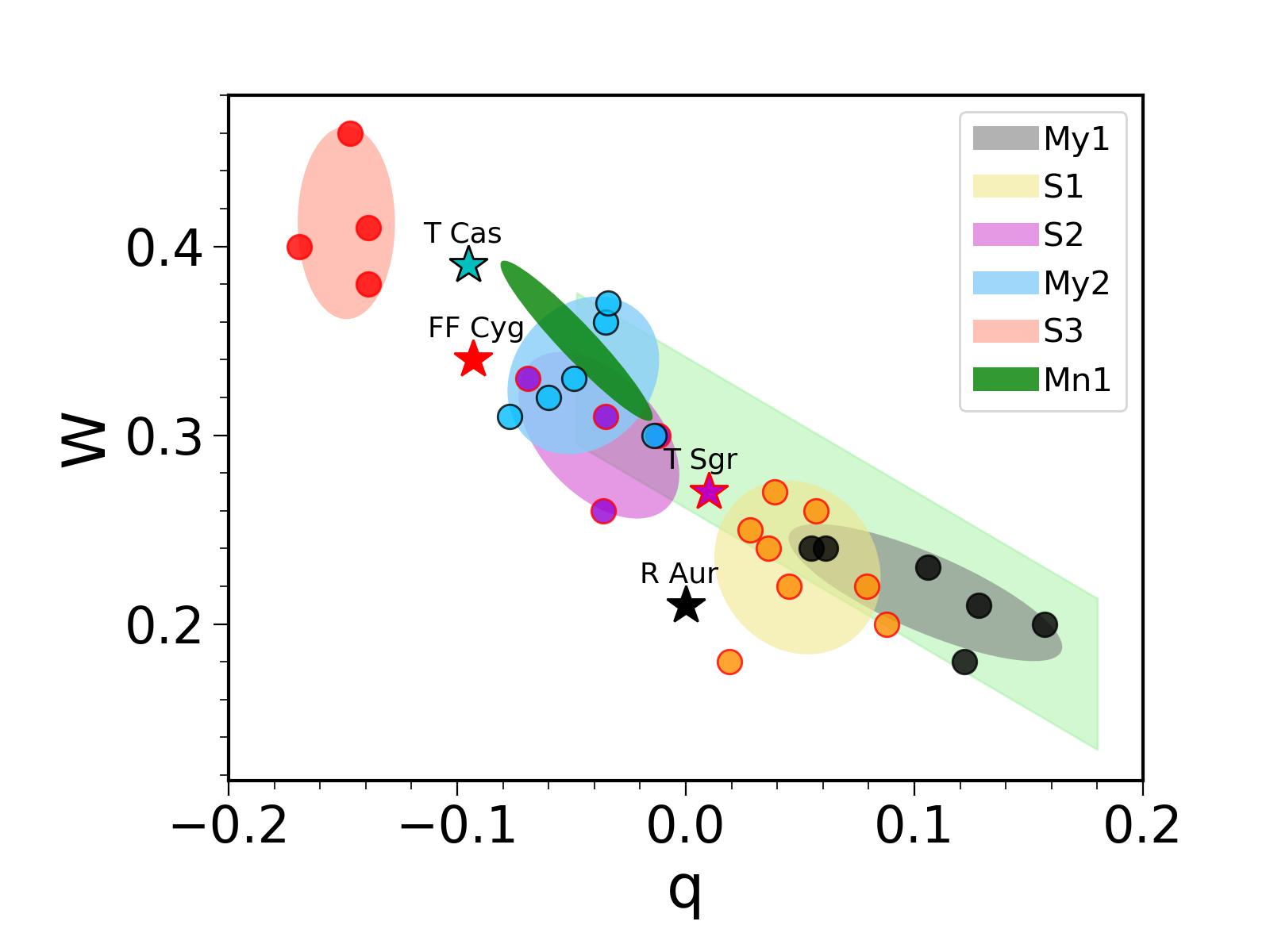}\\
  \includegraphics[height=4.2cm,trim={0.cm 0cm 1.4cm 1.2cm},clip]{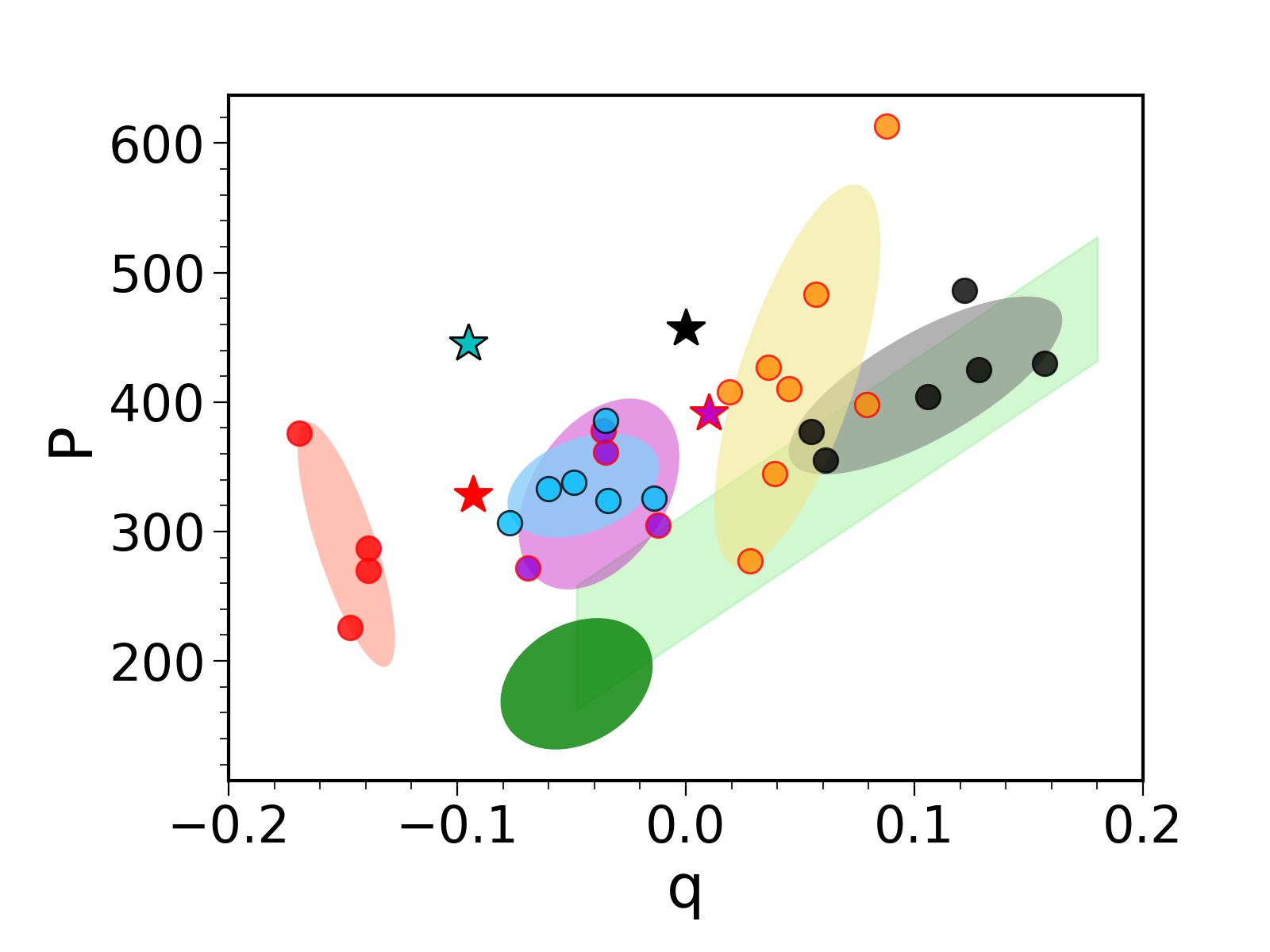}
  \includegraphics[height=4.2cm,trim={0.cm 0cm 1.4cm 1.2cm},clip]{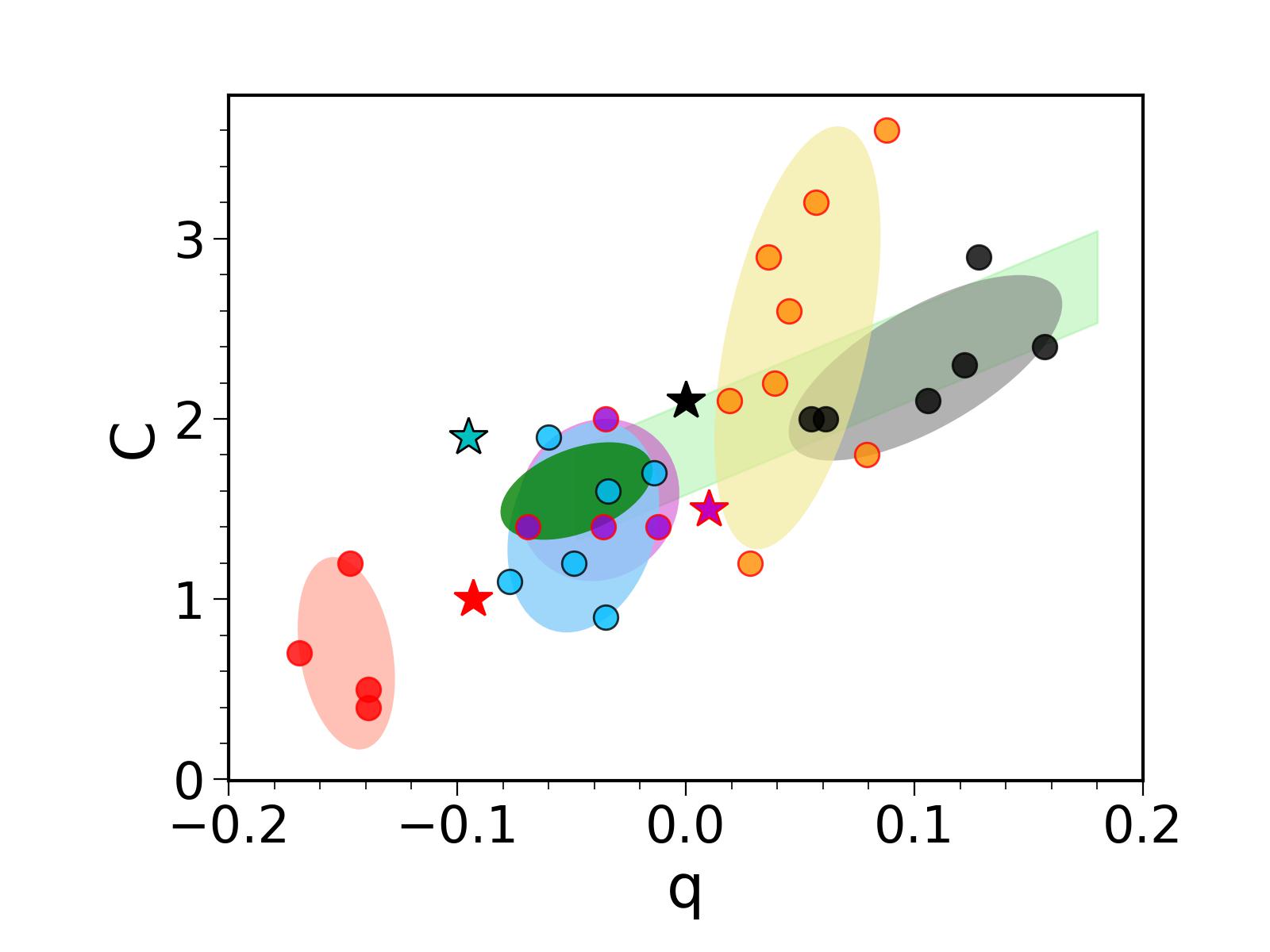}
  \includegraphics[height=4.2cm,trim={0.cm 0cm 1.4cm 1.2cm},clip]{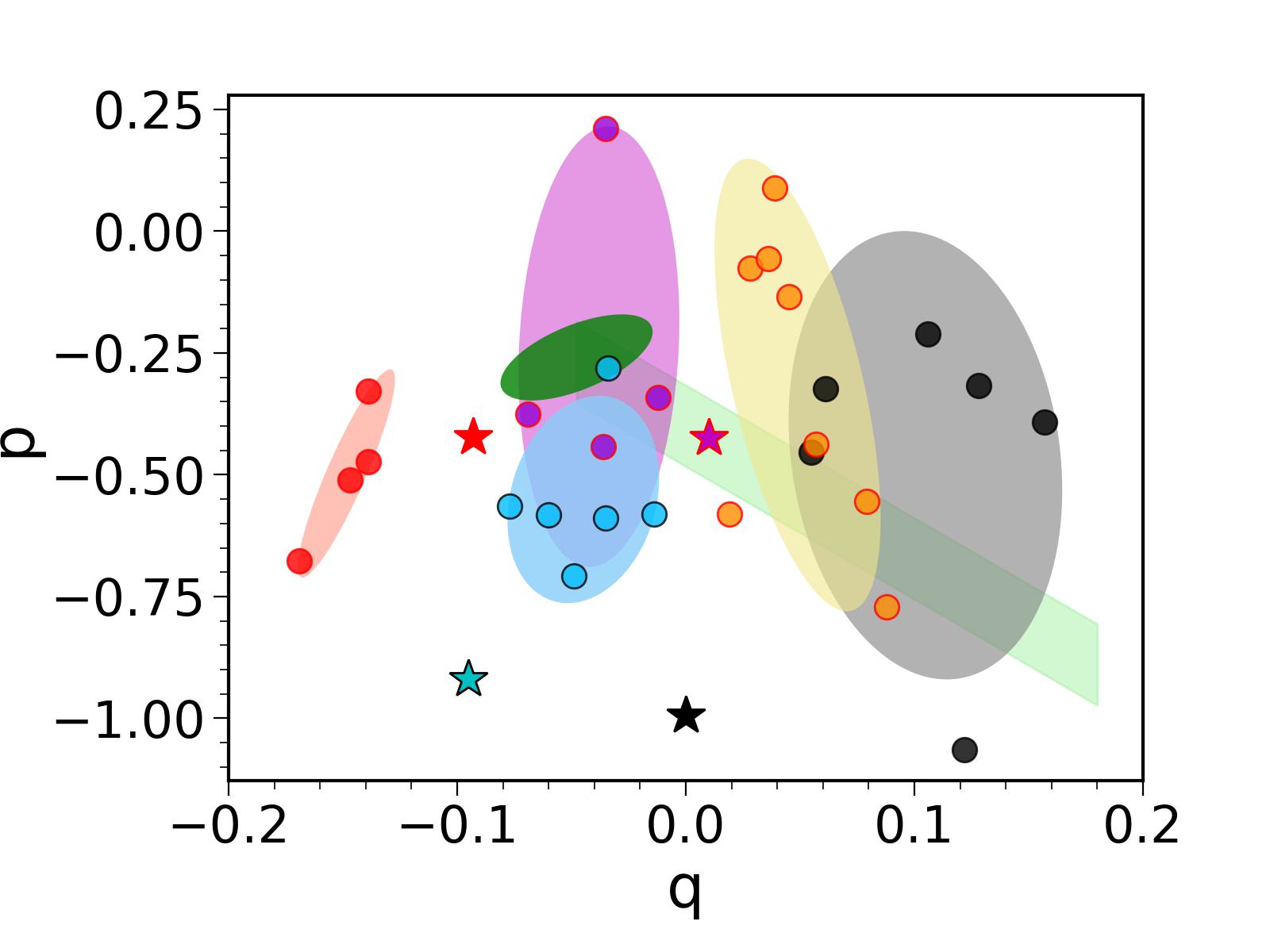}\\
  \includegraphics[height=4.2cm,trim={0.cm 0cm 1.4cm 1.2cm},clip]{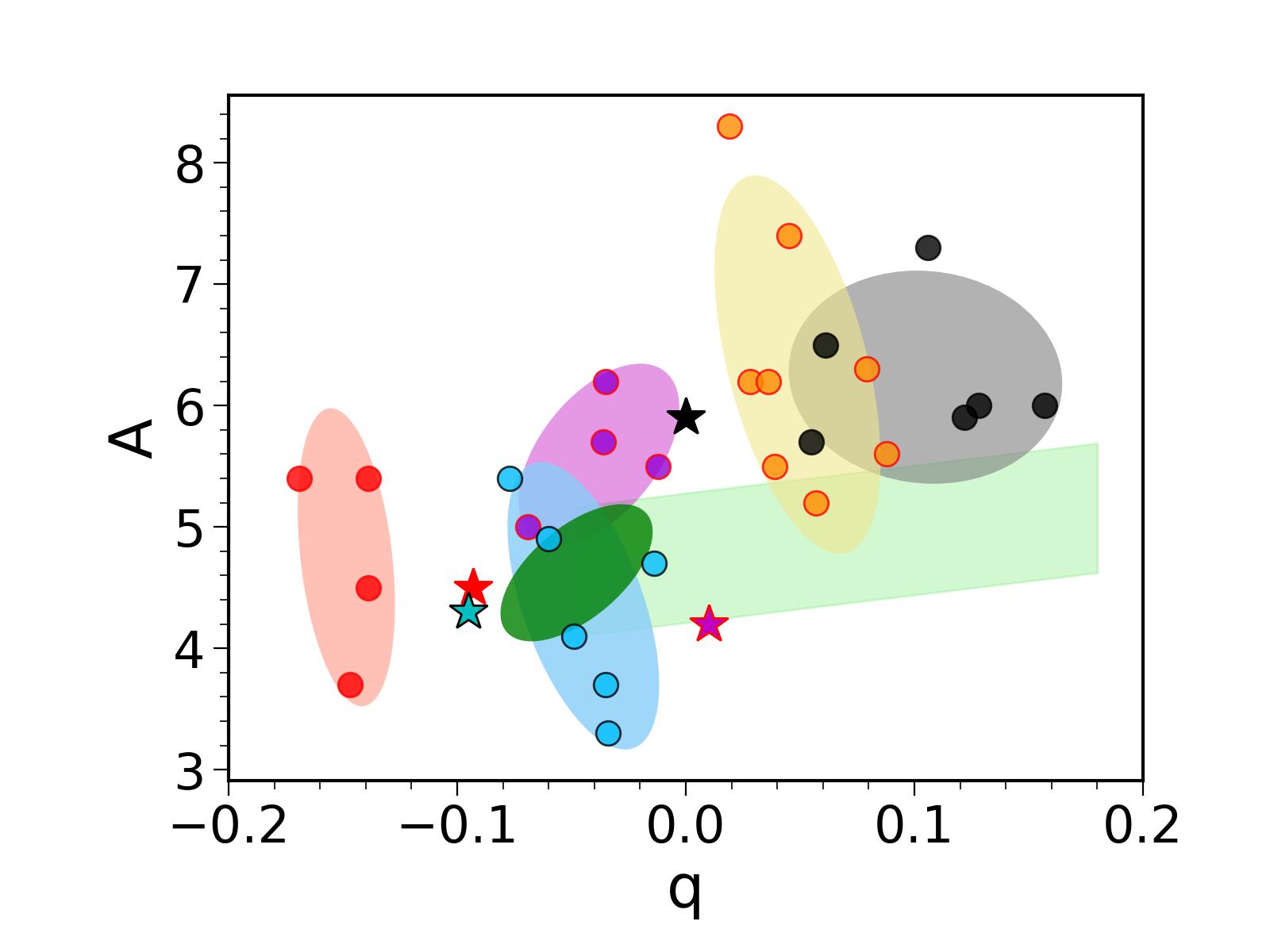}
  \includegraphics[height=4.2cm,trim={0.cm 0cm 1.4cm 1.2cm},clip]{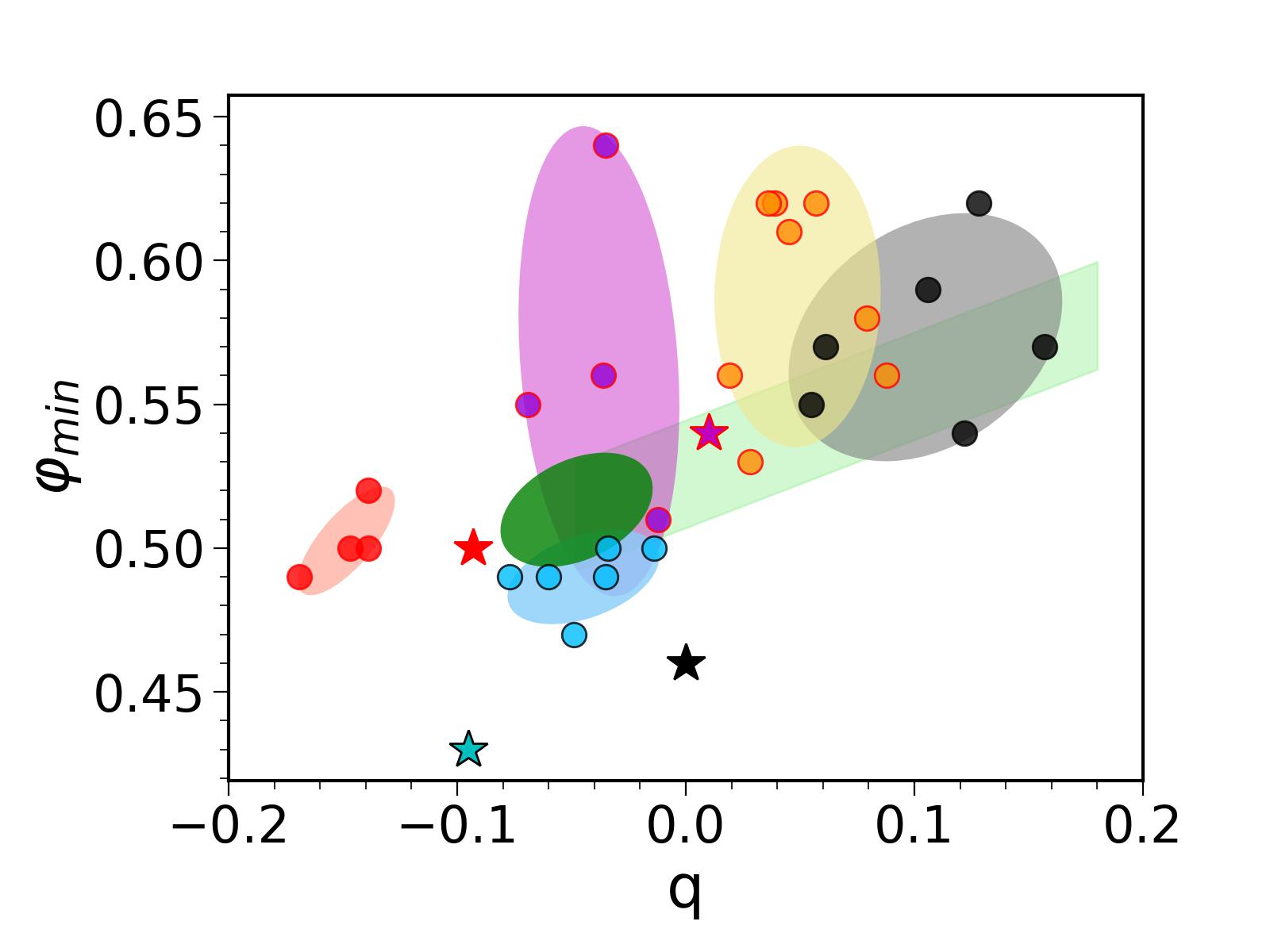}
  \includegraphics[height=4.2cm,trim={0.cm 0cm 1.4cm 1.2cm},clip]{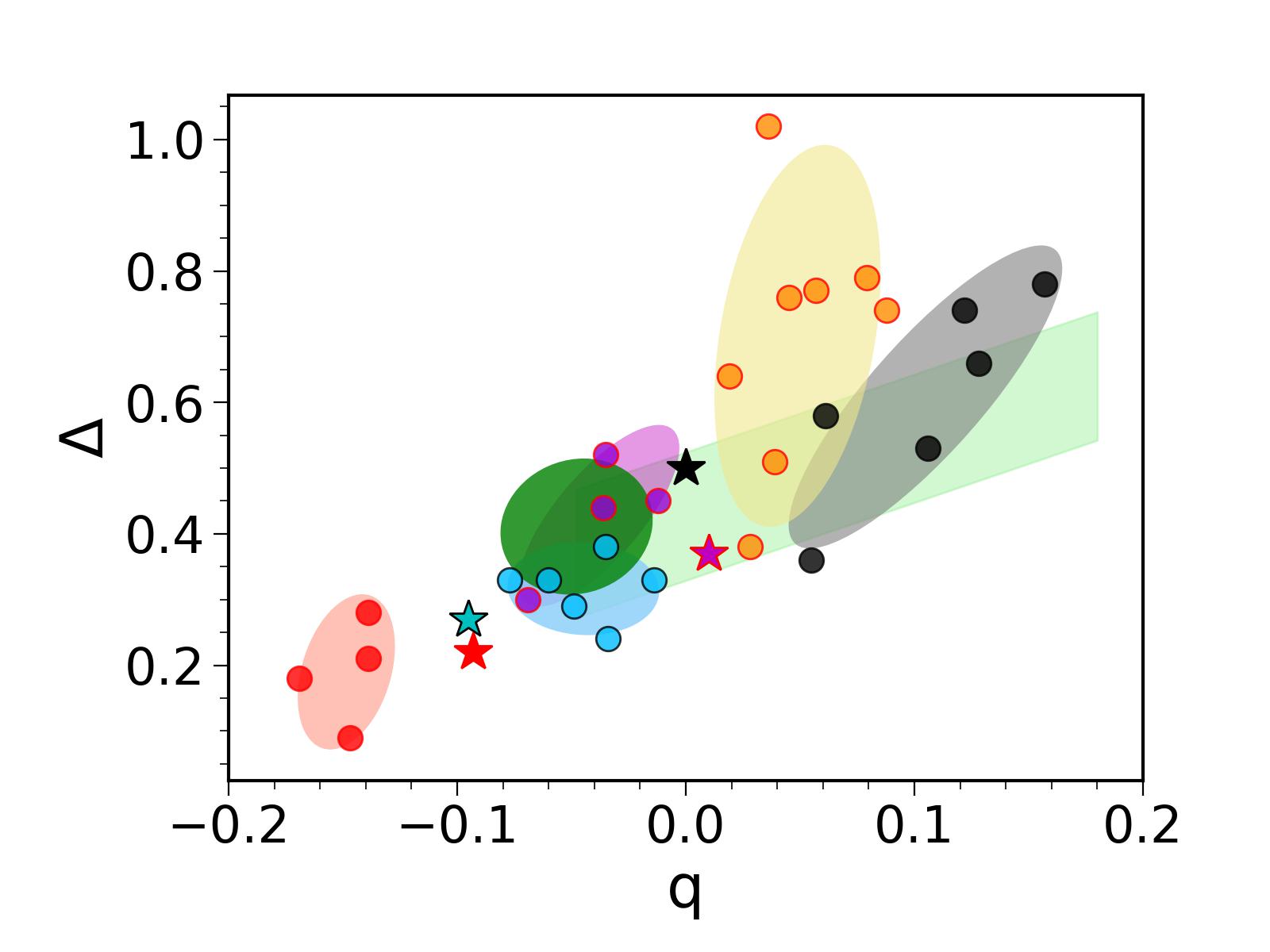}

  \caption{Curves of Myes and S spectral types: dependence on $q$ of all other star and light curve parameters. Different colours are used for the groups defined in Table \ref{tab4}, as indicated in the insert of the upper-left panel. Coloured ellipses having major and minor half-axes equal to 1.5 times the rms deviation from the mean are shown for each group and for the Mno curves (Table \ref{tab3}). The four stars excluded from the groups of Table \ref{tab4} are: FF Cyg (red star), T Cas (cyan star), R Aur (black star) and T Sgr (purple star). The light-green stripe shows the approximate location of the Mno curves.}
  \label{fig13}
\end{figure*}

\begin{figure*}
  \centering
  \includegraphics[width=4.2cm,trim={0.5cm 0 1.4cm 1.2cm},clip]{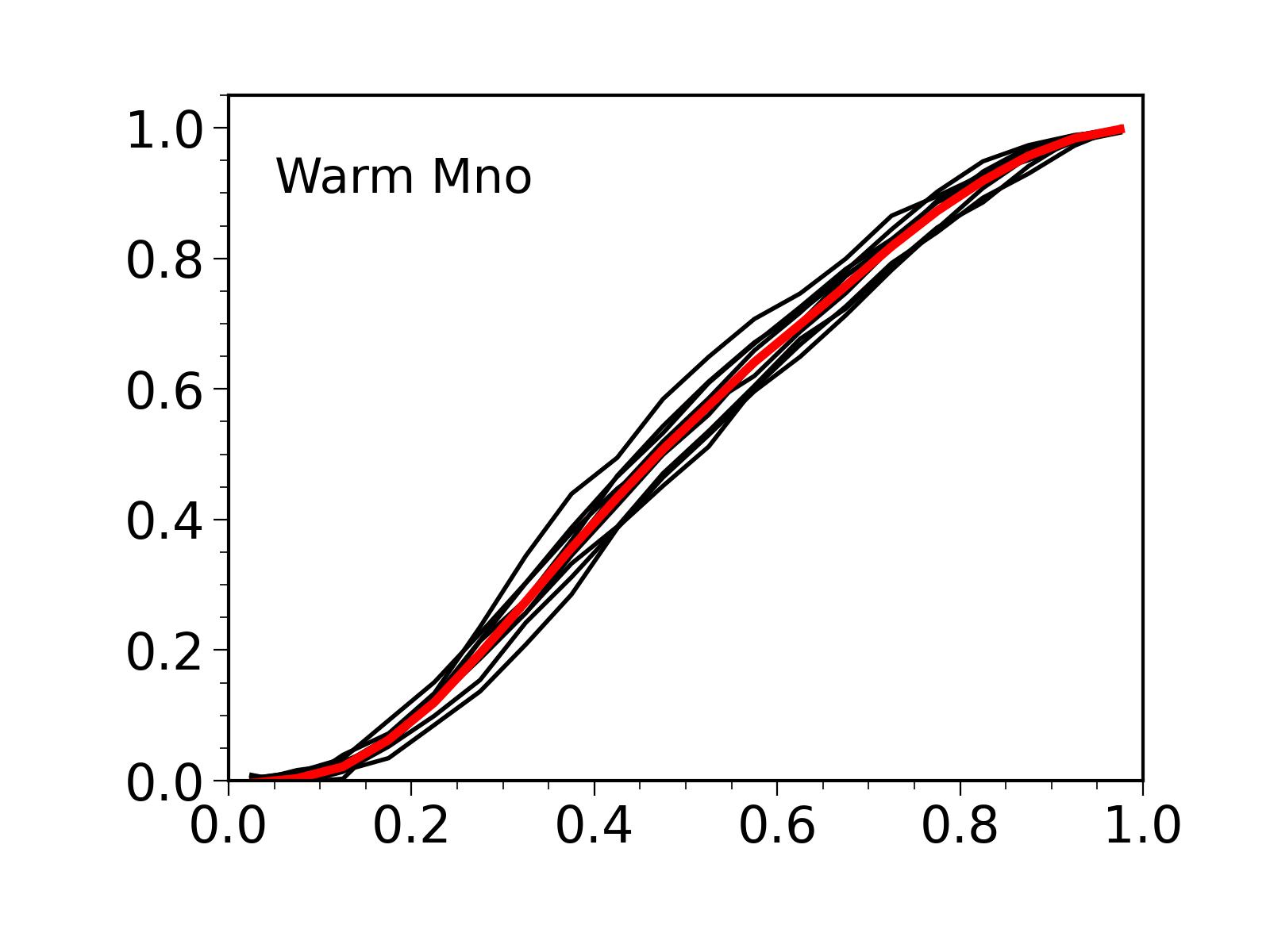}
  \includegraphics[width=4.2cm,trim={0.5cm 0 1.4cm 1.2cm},clip]{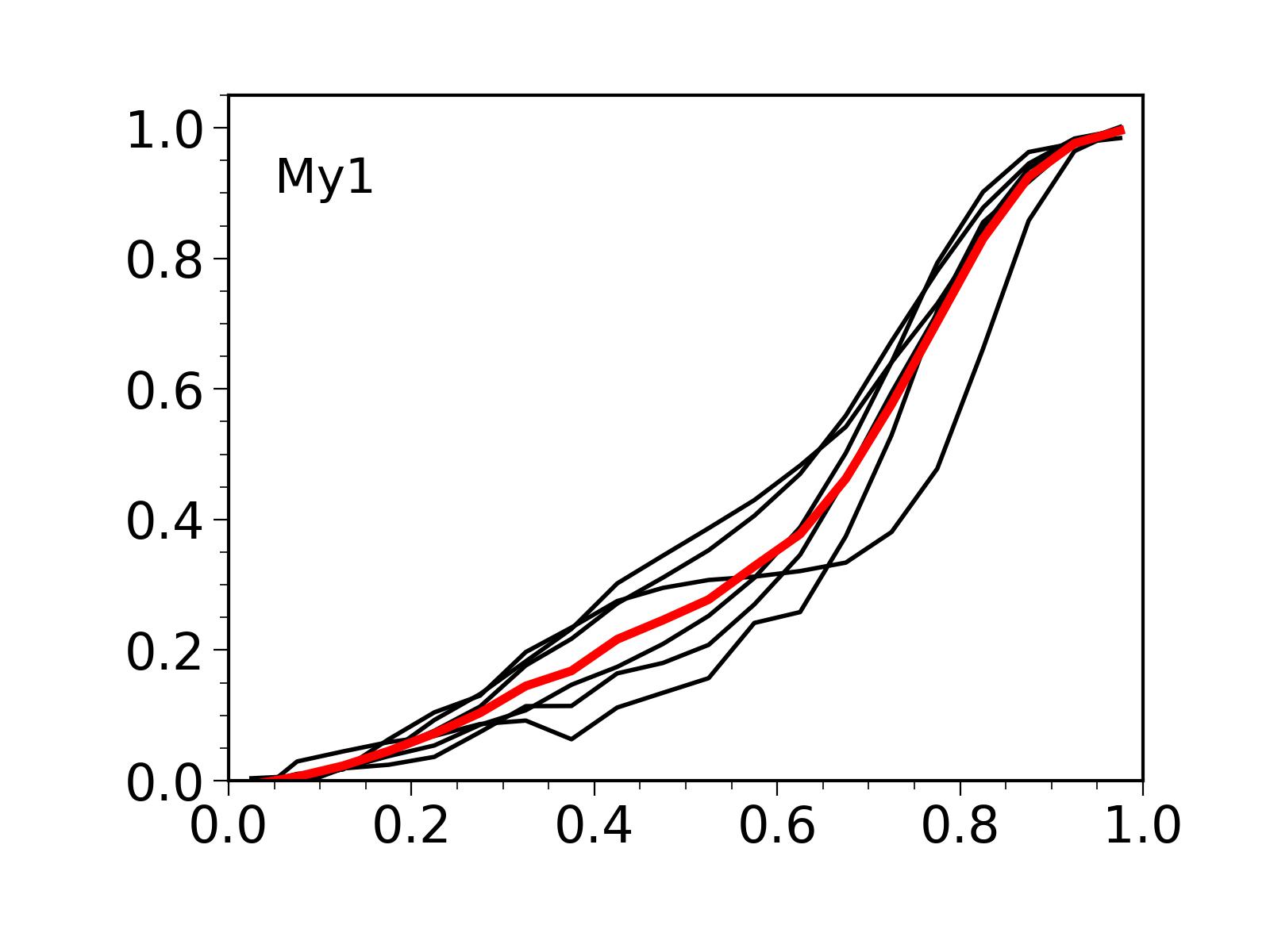}
  \includegraphics[width=4.2cm,trim={0.5cm 0 1.4cm 1.2cm},clip]{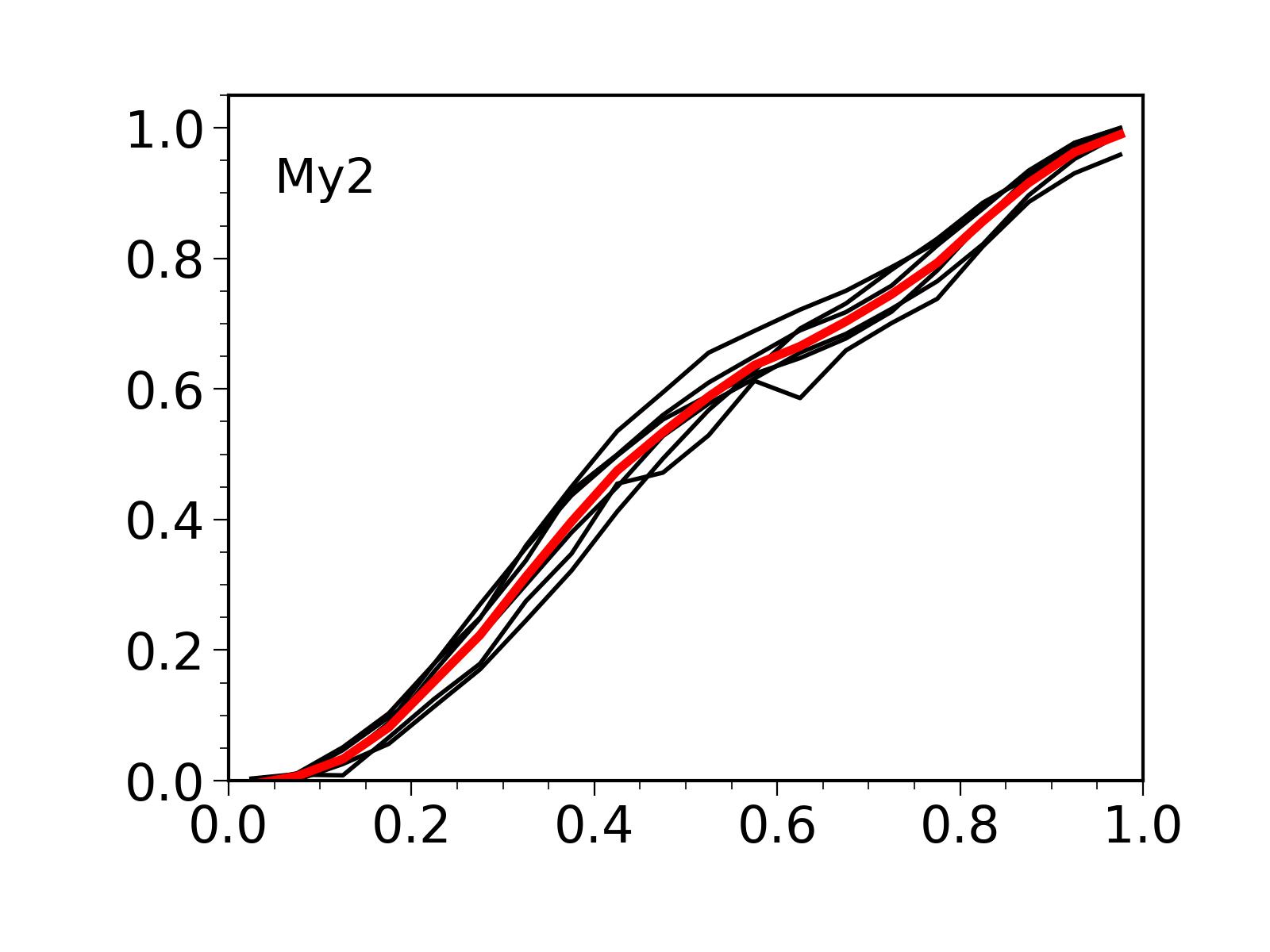}\\
  \includegraphics[width=4.2cm,trim={0.5cm 0 1.4cm 1.2cm},clip]{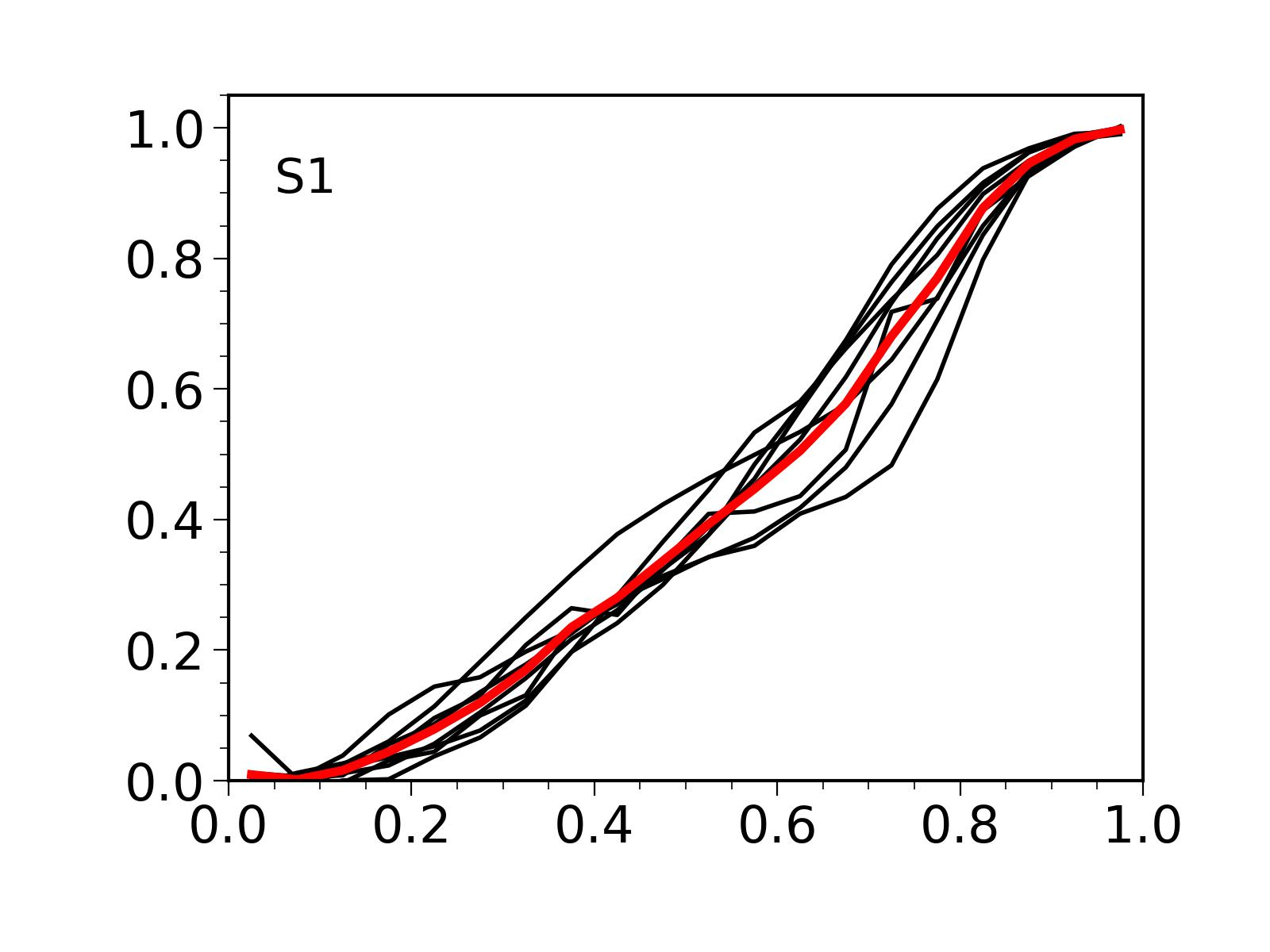}
  \includegraphics[width=4.2cm,trim={0.5cm 0 1.4cm 1.2cm},clip]{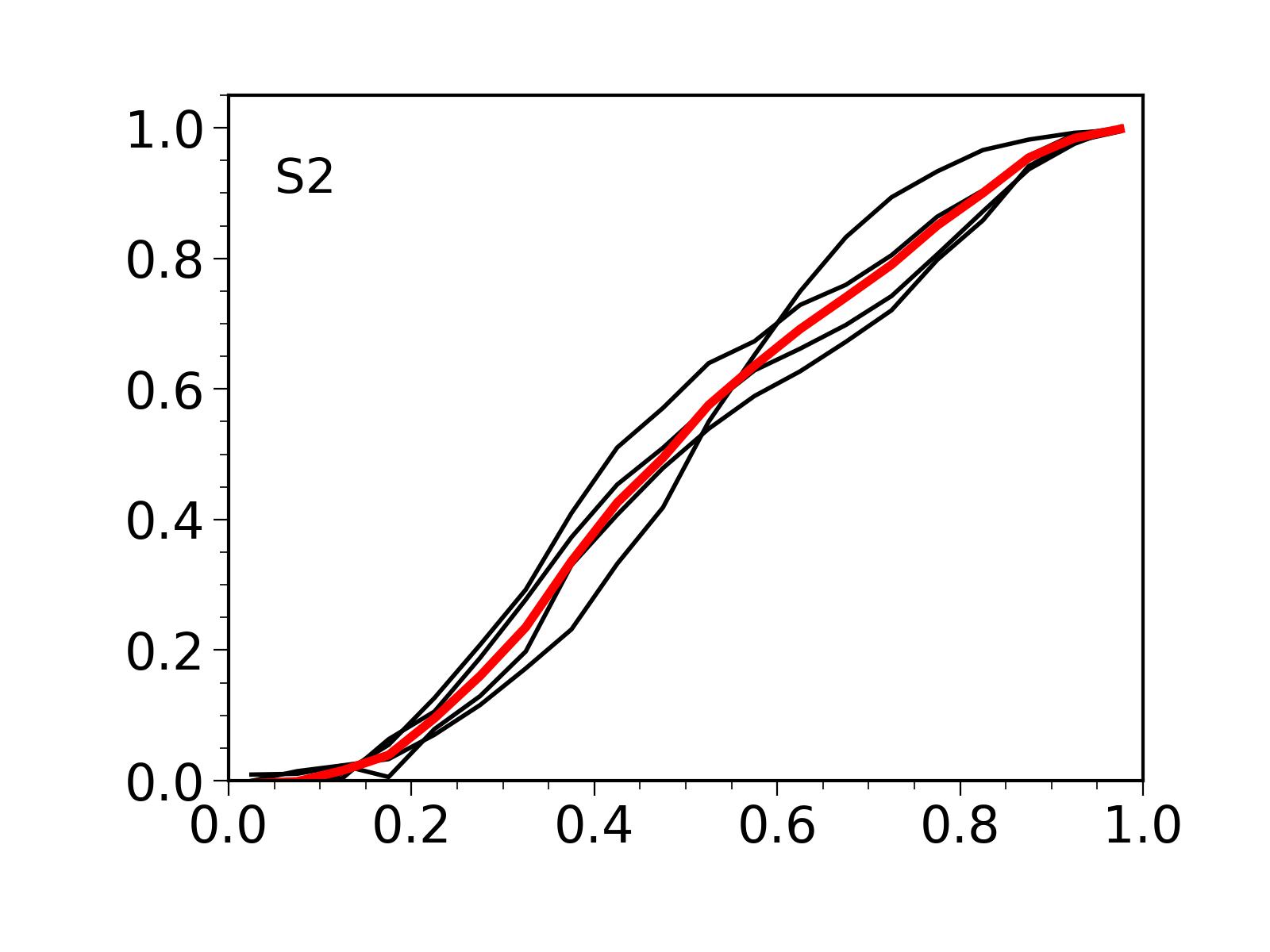}
  \includegraphics[width=4.2cm,trim={0.5cm 0 1.4cm 1.2cm},clip]{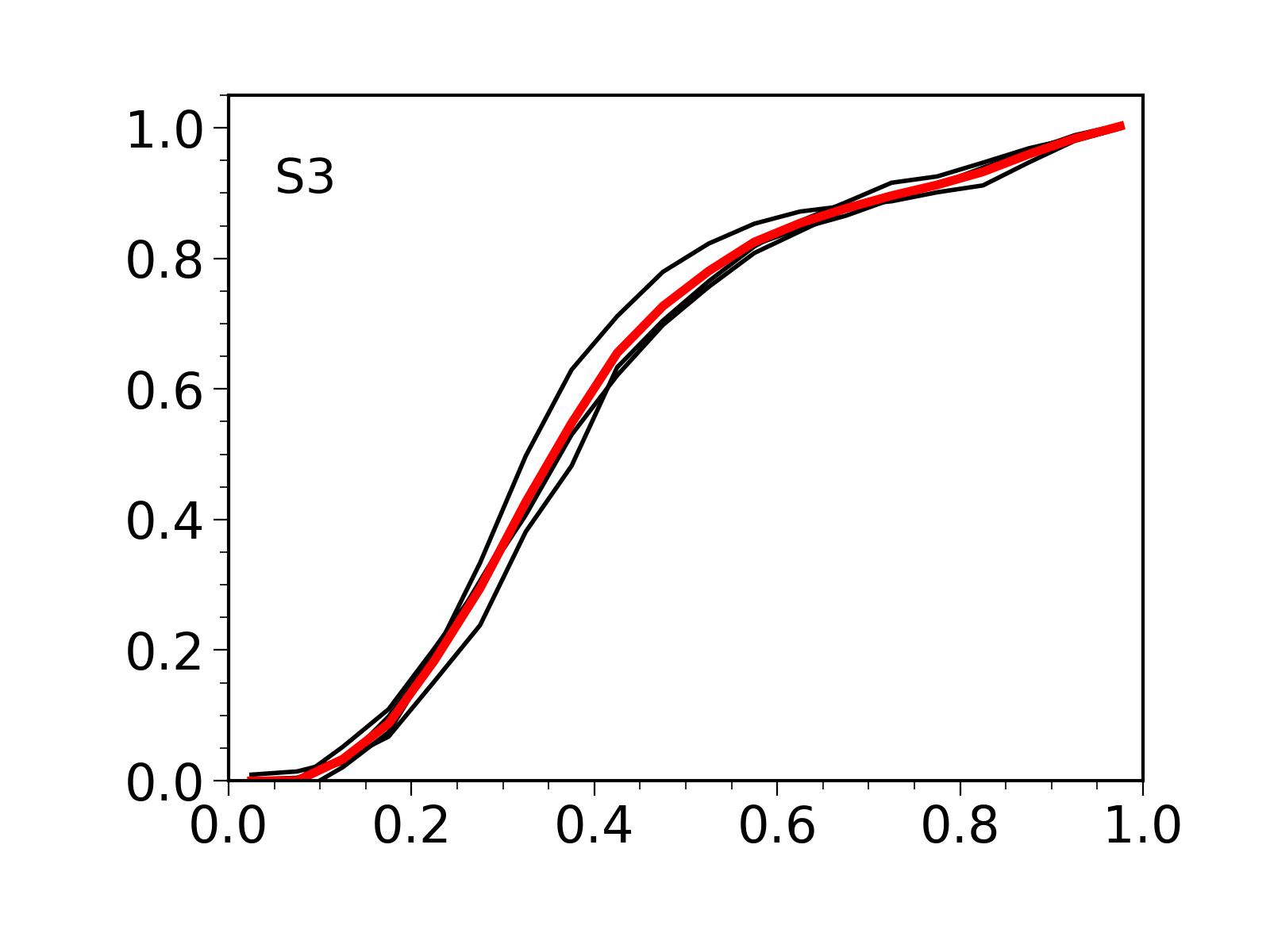}
 
    \caption{Superimposed profiles of the curves of each of the five groups of Myes and S spectral types. The red lines show the mean profiles. Also shown for reference are the superimposed profiles of the curves of the warmer Mno types (upper left panel).}
  \label{fig14}
\end{figure*}

Figure \ref{fig13} displays the dependence on $q$ of all other star and light curve parameters with different colours identifying the five groups separately. Table \ref{tab4} lists the mean values and rms deviations from the mean of the five groups and Figure \ref{fig14} displays the profiles of their ascending branches. The aim of the grouping is only to organize the discussion; one must refrain, at this stage, from giving it a deeper meaning. In particular, one should not think of these groups as being separated from each other but rather as milestones on a continuous evolution

The proposed grouping excludes four stars: an S star, FF Cyg and an Myes star, T Cas, both between groups My2 and S3, suggesting that they are in the process of transiting from one to the other; they are discussed in Subsections 5.3 and 5.4. Another S star, T Sgr, with a companion likely to distort the light curve, is discussed in Subsection 5.2. The fourth star, R Aur, of Myes spectral type, is discussed in Subsection 5.4.

\begin{deluxetable*}{ccc ccc ccc c}
\tablenum{4}
\tablecaption{S and Myes curves: mean locations and rms deviations from the mean in the parameter space. Also shown are the same quantities for the group of warm Mno stars listed in Table \ref{tab2} (Mn1) and the C1 group of cool C stars defined in Subsection 6.1. \label{tab4}}
\tablewidth{0pt}
\tablehead{
 \colhead{Group}&\colhead{Type}&\colhead{$P$}&\colhead{$C$}&\colhead{$A$}&\colhead{$W$}&\colhead{$\Delta$}&\colhead{$\varphi_{\rm{min}}$}&
 \colhead{$p$}&\colhead{$q$}\\
  \colhead{}&\colhead{}&\colhead{(day)}&\colhead{(mag)}&\colhead{(mag)}&\colhead{}&\colhead{(mag)}& \colhead{}&\colhead{$\times$10$^3$}&\colhead{$\times$10$^3$}
}
\startdata
My1&	Myes&	413/42&	2.3/0.3&	6.2/0.5&	0.22/0.02&	0.61/0.14&	0.57/0.03&	$-$460/280&	105/36\\
S1&	S&	420/92&	2.5/0.7&	6.3/1.0&	0.23/0.03&	0.68/0.18&	0.59/0.03&	$-$316/290&	49/23\\
My2&	Myes&	336/24&	1.4/0.4&	4.4/0.7&	0.33/0.03&	0.32/0.04&	0.49/0.01&	$-$551/130&	$-$45/20\\
S2&	S&	329/43&	1.6/0.3&	5.6/0.4&	0.30/0.03&	0.43/0.08&	0.57/0.05&	$-$237/261&	$-$38/20\\
S3&	S&	290/55&	0.7/0.3&	4.8/0.7&	0.41/0.03&	0.19/0.07&	0.50/0.01&	$-$497/123&	$-$149/12\\
Mn1&	Mno&	180/33&	1.6/0.1&	4.7/0.3&	0.34/0.02&	0.42/0.06&	0.52/0.01&	$-$249/50&	$-$43/17\\
C1&	C&	417/64&	2.7/0.6&	3.7/0.2&	0.37/0.04&	0.49/0.13&	0.55/0.03&	$-$249/201&	41/19\\
\enddata
\end{deluxetable*}

\begin{figure*}
  \centering
  \includegraphics[width=4.5cm,trim={0.5cm 0 1.9cm 1.5cm},clip]{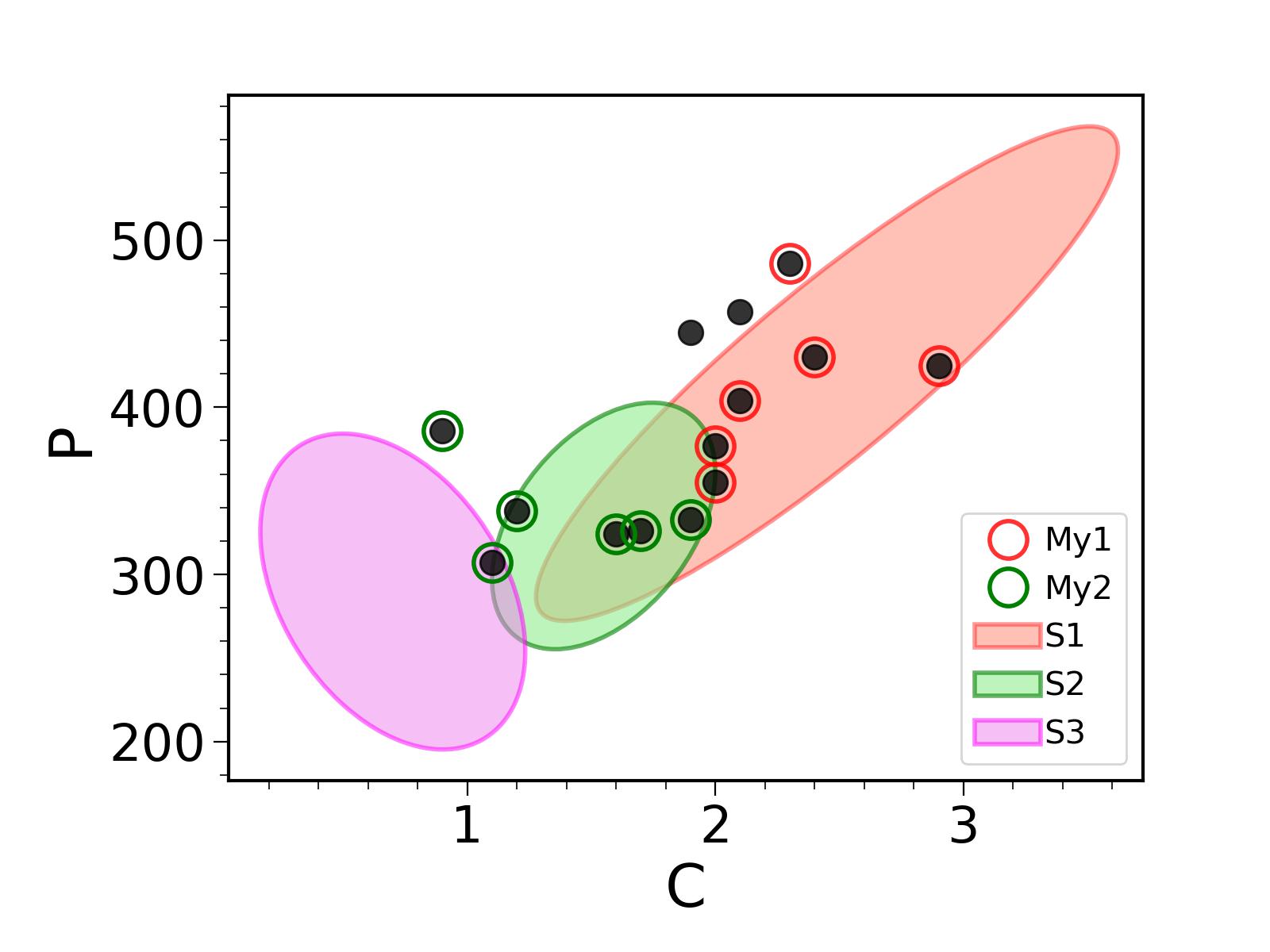}
  \includegraphics[width=4.5cm,trim={0.5cm 0 1.9cm 1.5cm},clip]{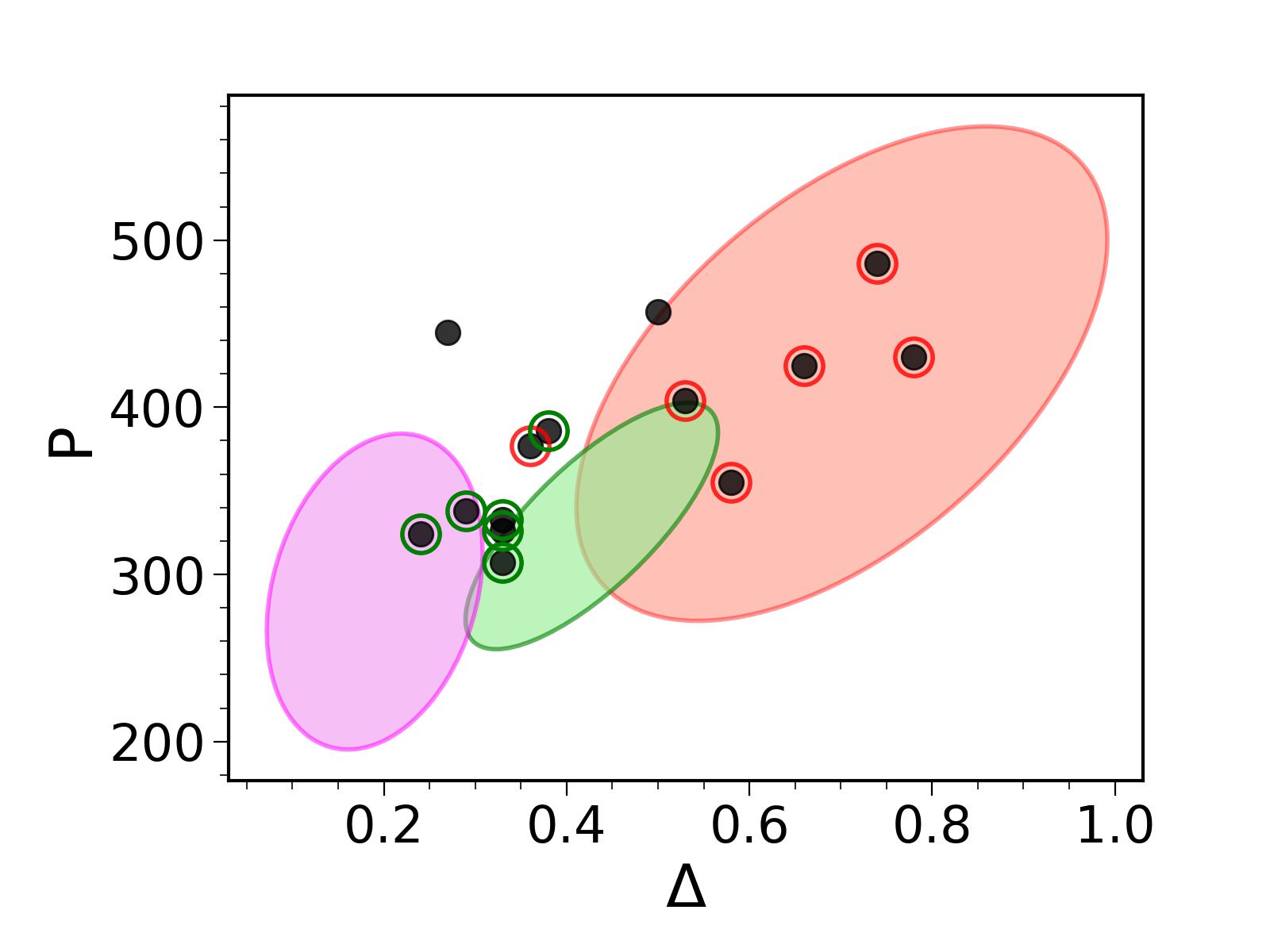}
  \includegraphics[width=4.5cm,trim={0.5cm 0 1.9cm 1.5cm},clip]{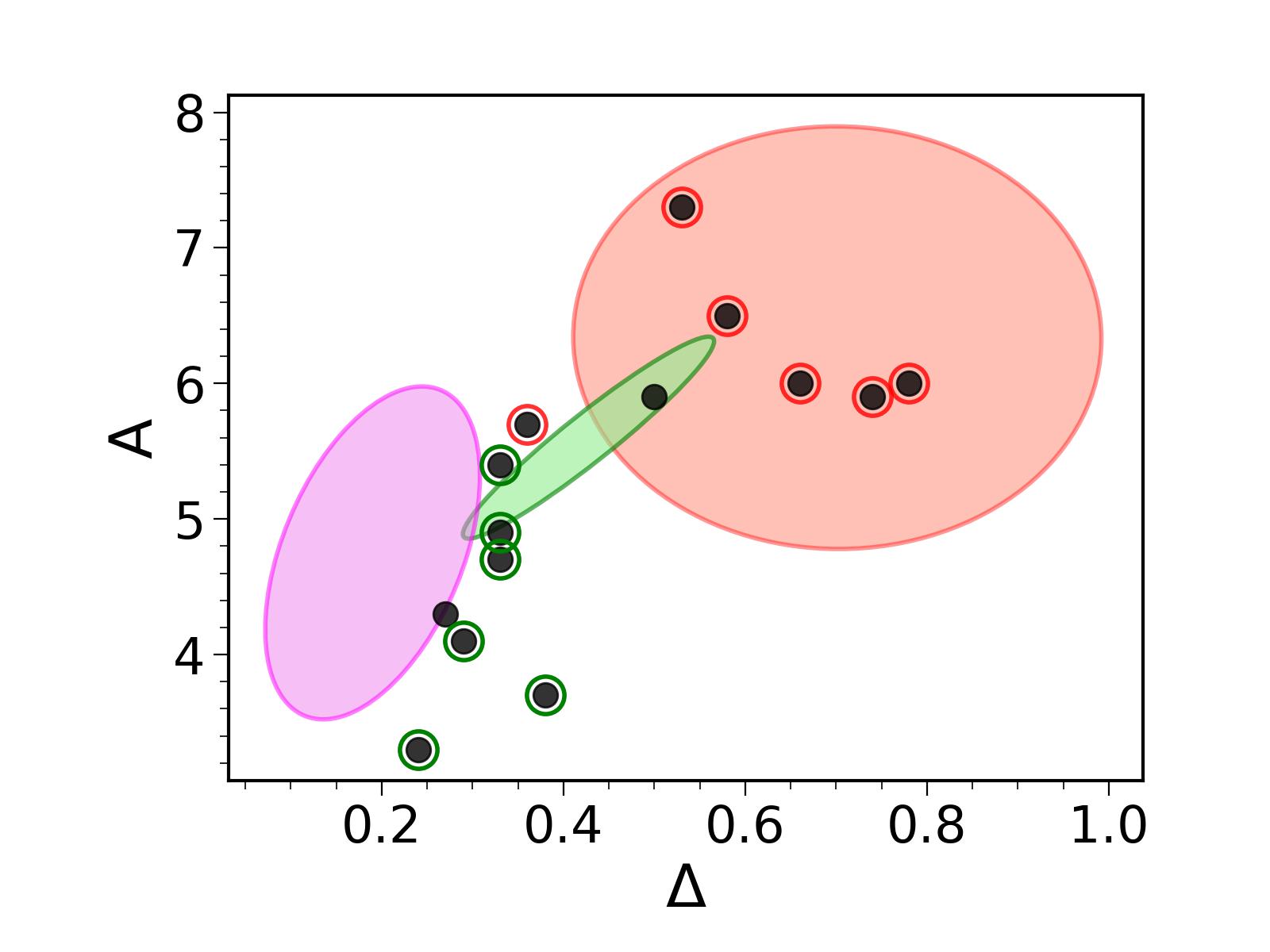}\\
  \includegraphics[width=4.5cm,trim={0.5cm 0 1.9cm 1.5cm},clip]{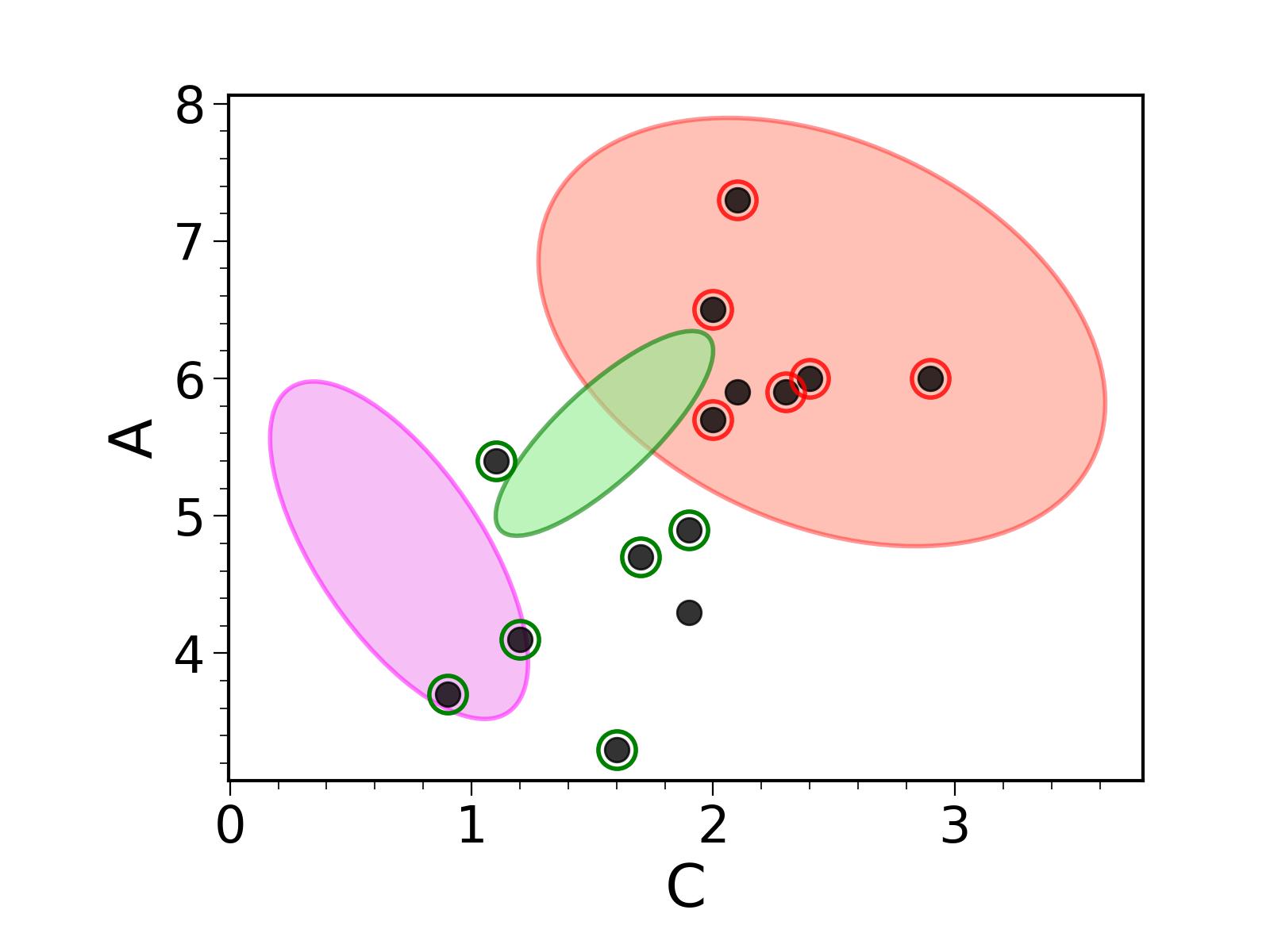}
  \includegraphics[width=4.5cm,trim={0.5cm 0 1.9cm 1.5cm},clip]{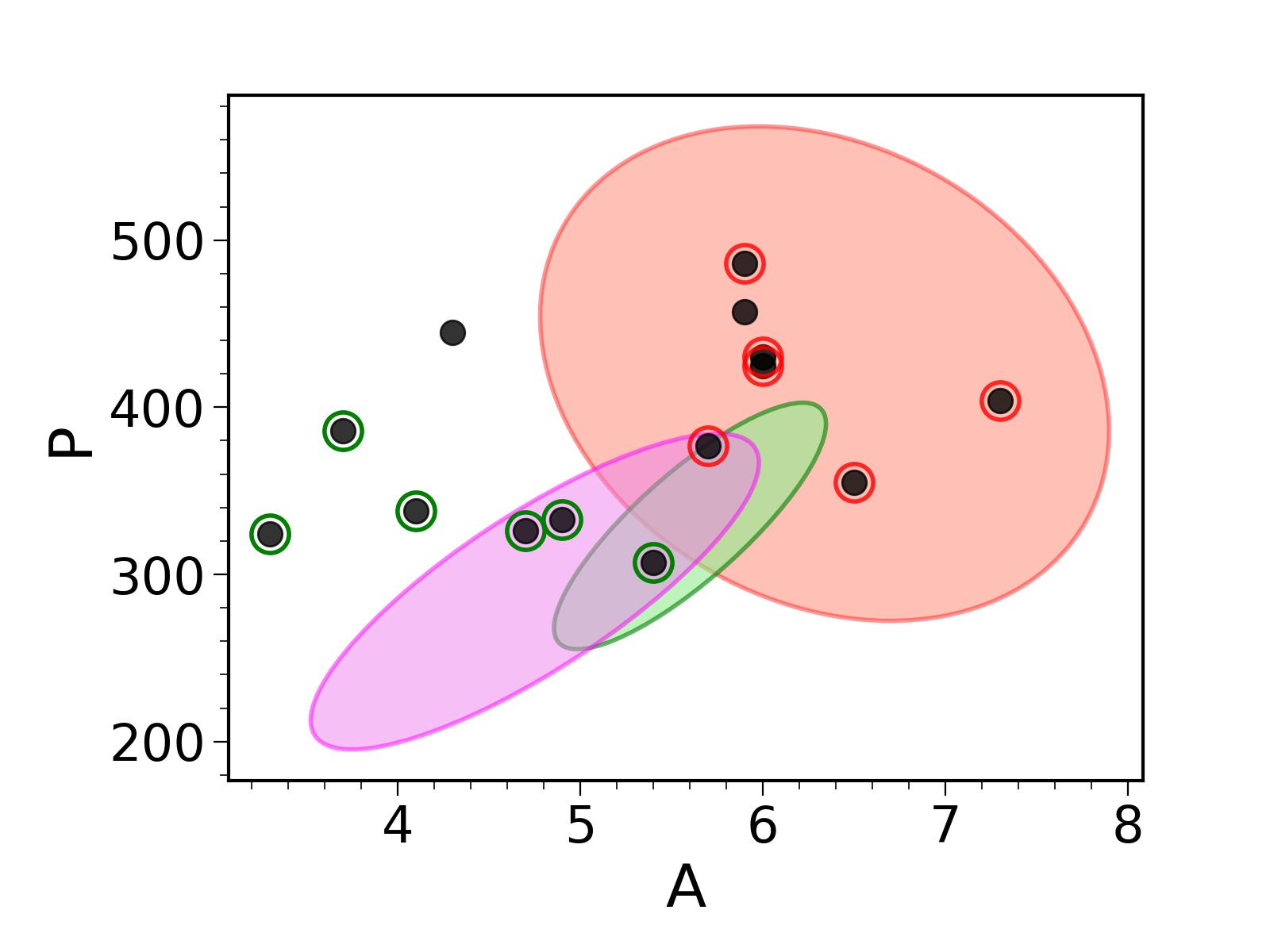}
  \includegraphics[width=4.5cm,trim={0.5cm 0 1.9cm 1.5cm},clip]{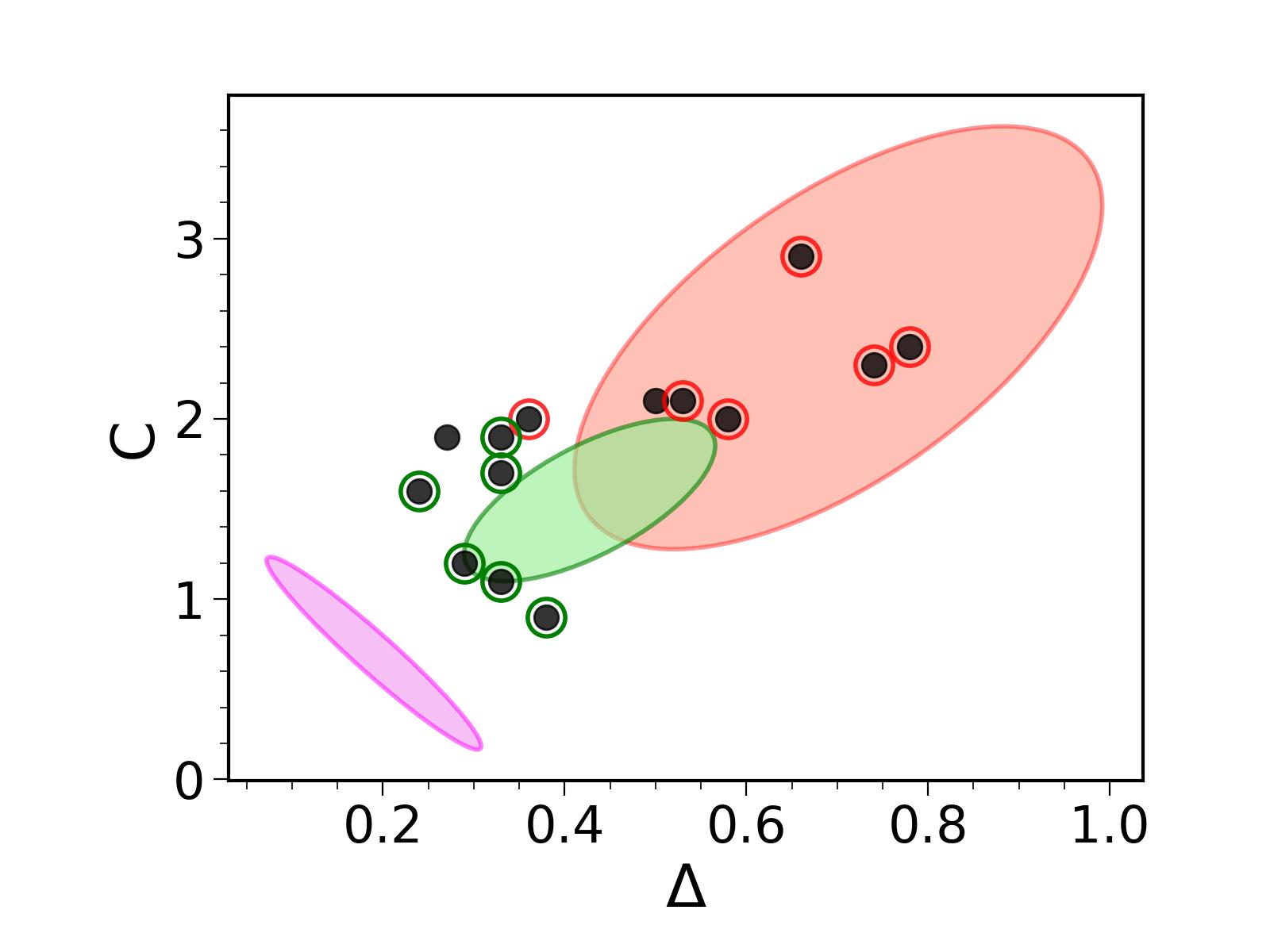}

  \caption{Projections of the parameters of Myes spectral types on planes spanned by shape-insensitive parameters: full dots circled in red for group My1 and circled in green for group My2. Ellipses show the locations of the groups S1, S2 and S3.}
  \label{fig15}
\end{figure*}

\begin{figure*}
   \centering
   \includegraphics[width=4.5cm,trim={0.5cm 0 1.9cm 1.5cm},clip]{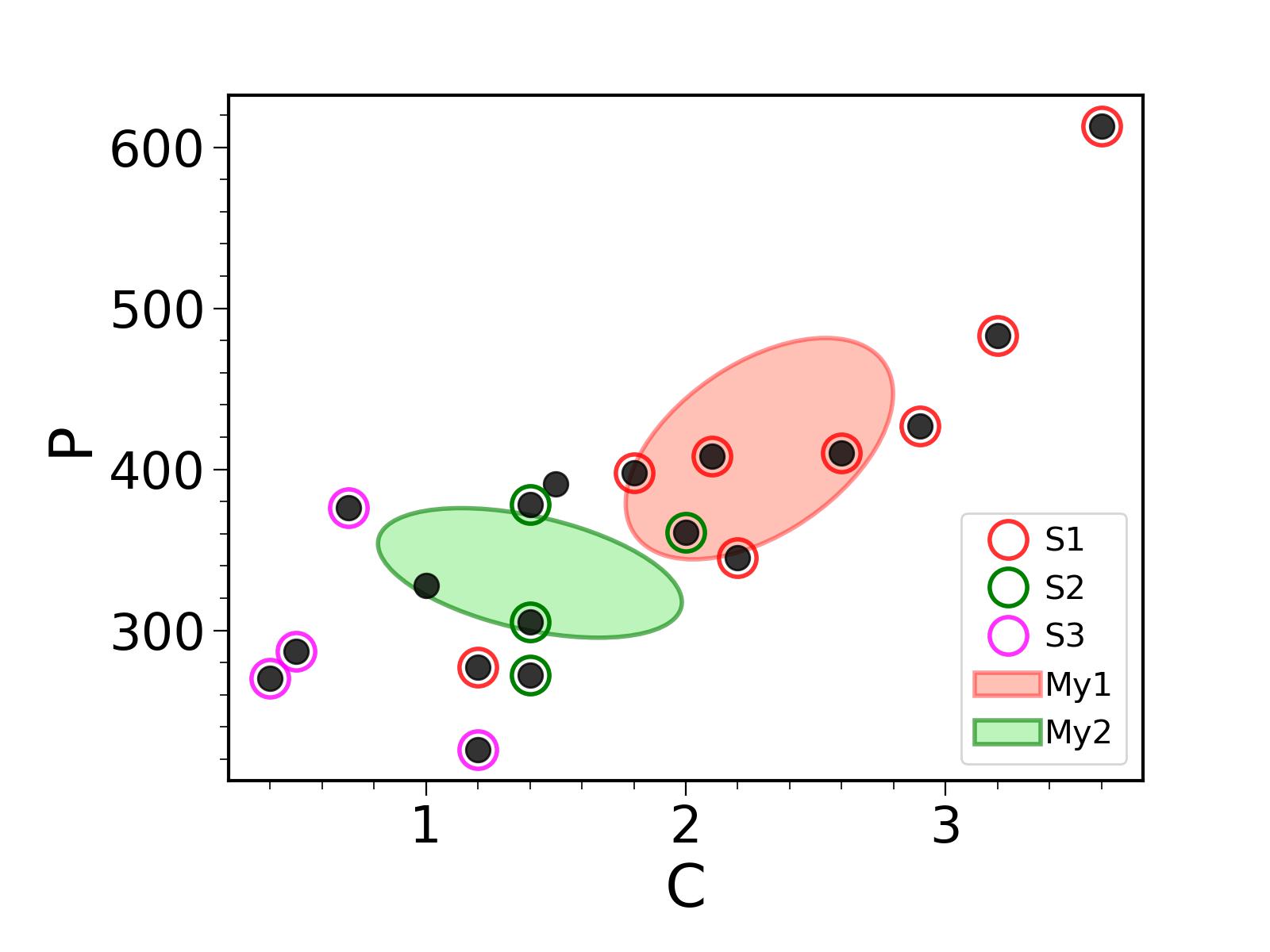}
  \includegraphics[width=4.5cm,trim={0.5cm 0 1.9cm 1.5cm},clip]{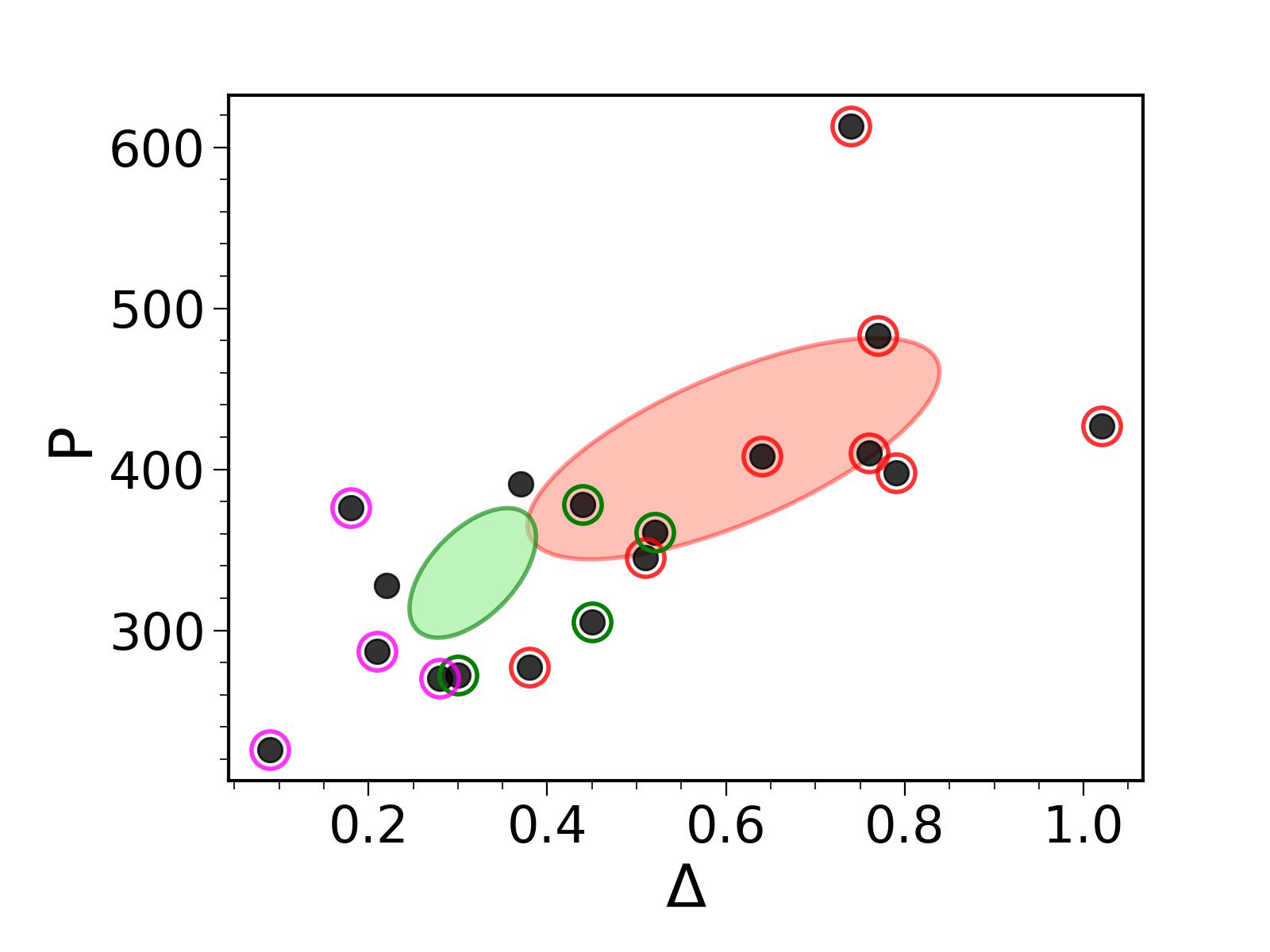}
  \includegraphics[width=4.5cm,trim={0.5cm 0 1.9cm 1.5cm},clip]{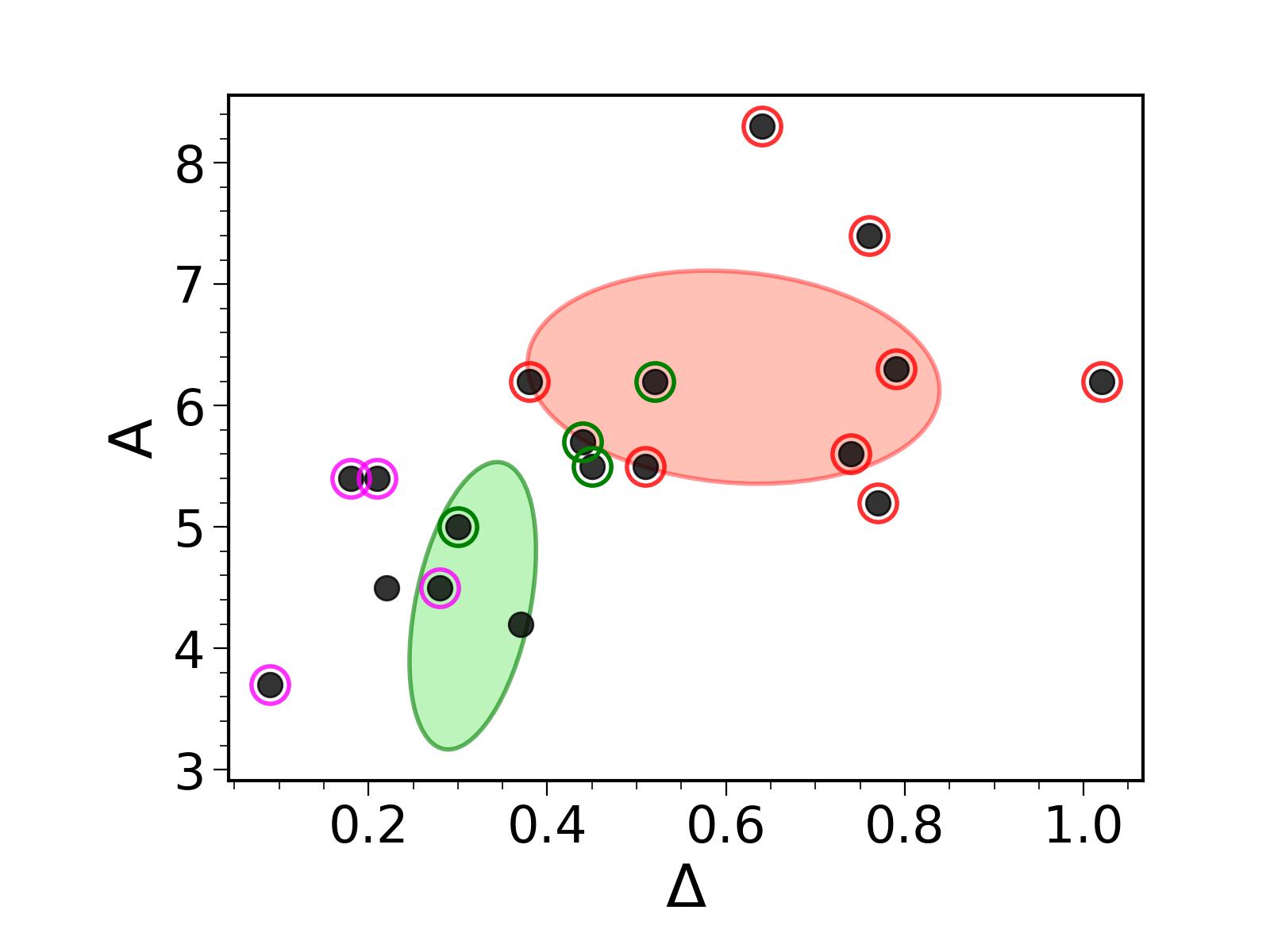}\\
  \includegraphics[width=4.5cm,trim={0.5cm 0 1.9cm 1.5cm},clip]{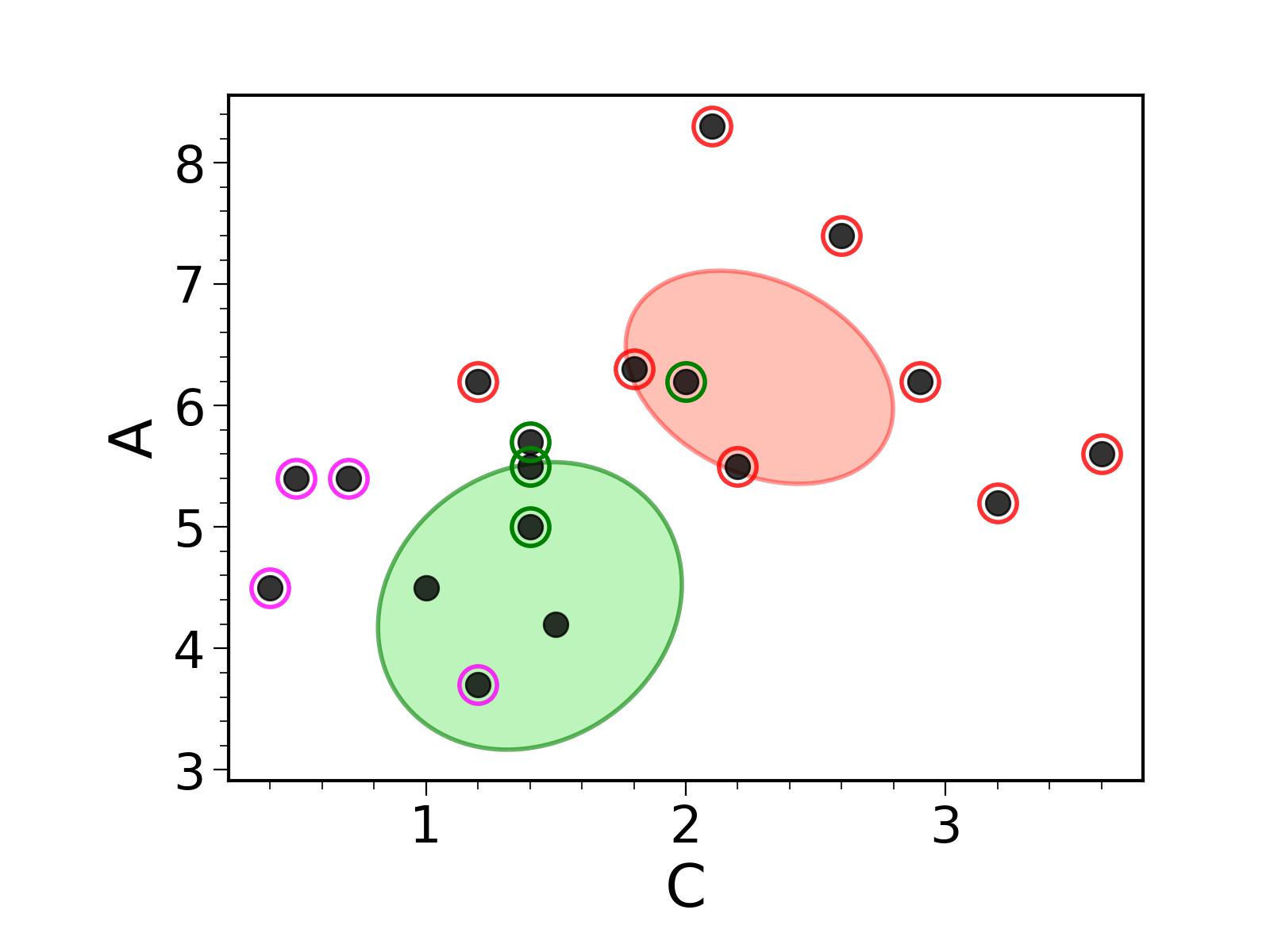}
  \includegraphics[width=4.5cm,trim={0.5cm 0 1.9cm 1.5cm},clip]{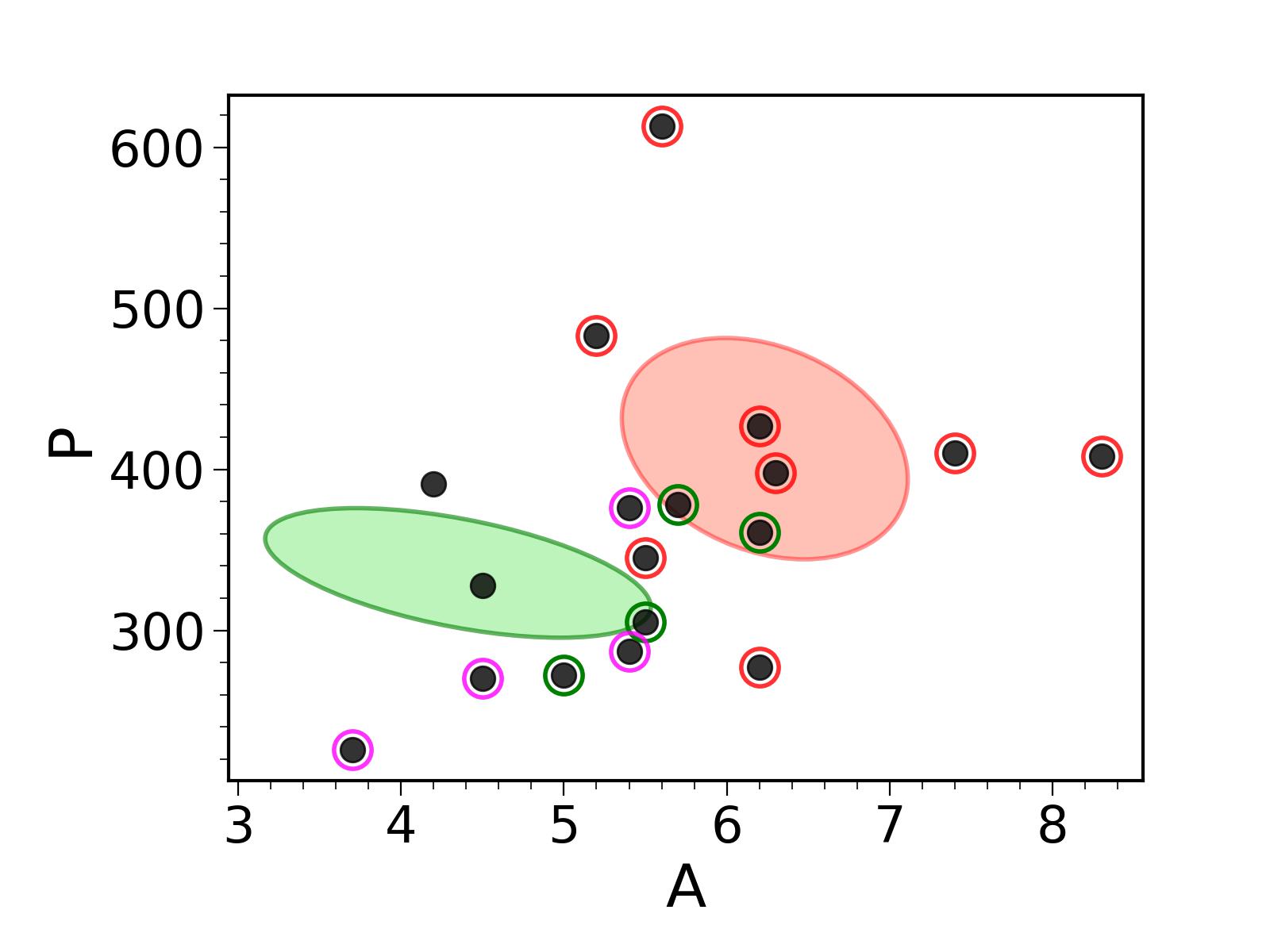}
  \includegraphics[width=4.5cm,trim={0.5cm 0 1.9cm 1.5cm},clip]{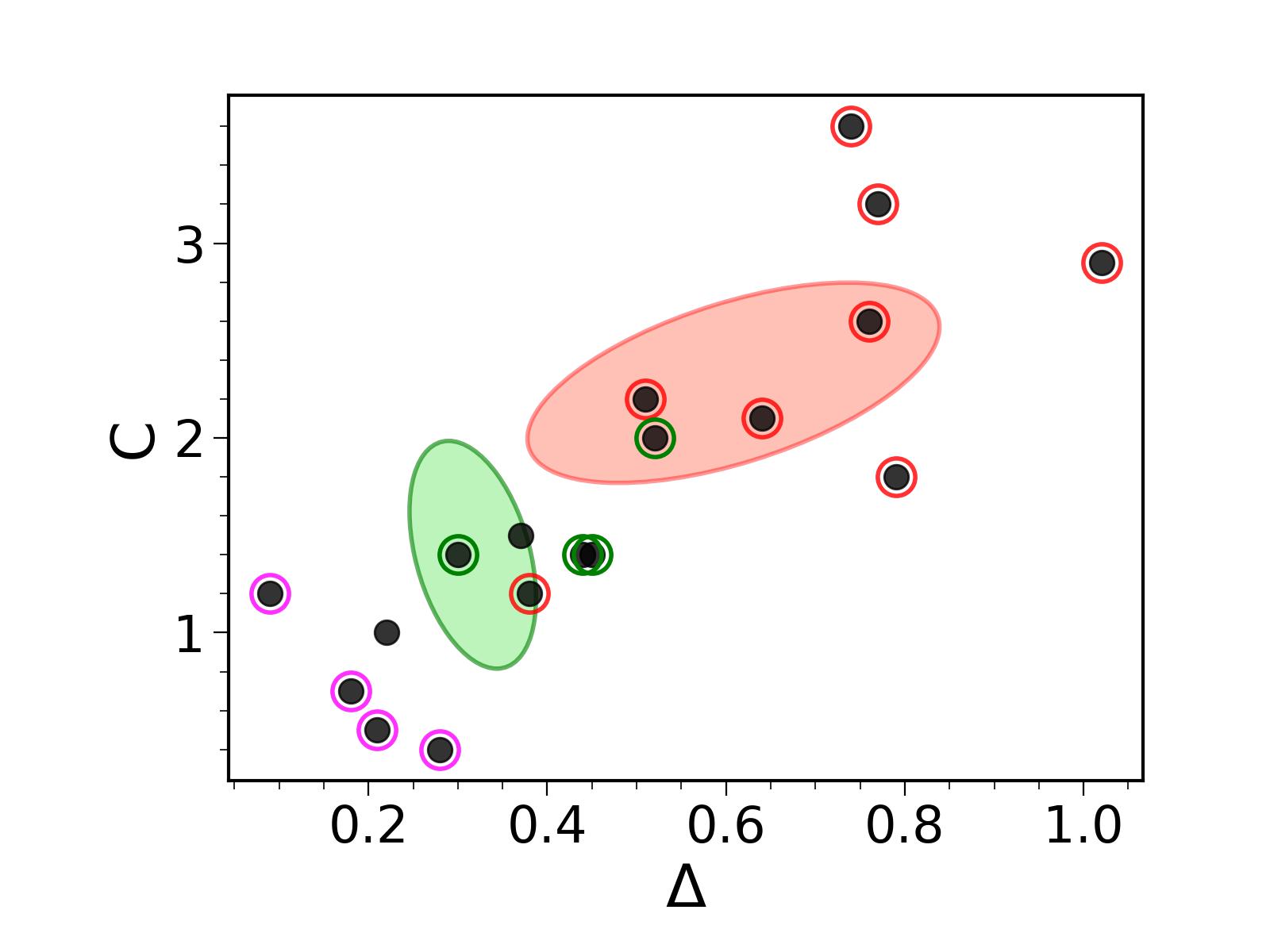}
 
  \caption{ Projections of the parameters of S spectral types on planes spanned by shape-insensitive parameters: full dots circled in red are for S1, in green for S2 and in purple for S3. Ellipses show the locations of groups My1 and My2. }
  \label{fig16}
\end{figure*}

Figures \ref{fig15} and \ref{fig16} display maps, in planes spanned by shape-insensitive parameters, of the parameters of Myes and S spectral types, respectively. They show that the groupings in My1-My2 and S1-S2-S3, although having been made as a function of the evolution of the curves with respect to $q$, retain clear identities: My1 (S1) members have larger values of $P$, $C$, $A$ and $\Delta$ than My2 (S3) members and S2 members are located between the S1 and S3 members. This result gives therefore evidence in favour of a correlation between stellar and light curve parameters for the Myes and S stars, as was found in Section 4 for the Mno stars.

Papers I and II have shown that the transition from Mno to Myes spectral types occur at different values of the shape parameter $W$, some very close to the warm Mnos, some farther away, the former possibly evolving toward larger values of $W$, meaning smaller values of $q$, the latter to smaller values of $W$, meaning larger values of $q$. This is clearly seen to be the case on the maps shown in Figure \ref{fig13}: the locations of the My1 and S1 groups suggest that the associated stars left the Mno spectral type at different states of evolution toward larger $q$ values, while the location of the My2 group, nearly coincident with that of the warmer Mno stars, suggests an evolution toward the S2 and S3 groups, the latter reaching very low $q$ values. In all cases, such a scenario is consistent with all stars being successors of the warmer Mno group. From Mno to My2 and S2, the curves evolve without significant variation of the parameters, the only exceptions being the spectral type and the period, both proxies of the evolution of the star along the AGB.

In all planes, the evolution in the curve parameter space from Mno to My1 and S1 shows an Mno-S-Myes rather than Mno-Myes-S succession as one would expect; namely the extreme $q$ values are reached in the Myes rather than S spectral type. Figures \ref{fig15} and \ref{fig16}, which display the five groups in projections on planes spanned by coordinates insensitive to the shape of the light curve ($P$, $A$, $C$ and $\Delta$) show that in these planes, the Myes and S groups nearly overlap: the unexpected succession is caused by the evolution of the shape of the light curves, with shape-sensitive parameters $p$, $q$, $\varphi_{\rm{min}}$ and $W$ being affected; as Figure \ref{fig13} uses $q$ as axis of abscissa, S curves are seen to precede Myes curves in all panels.

\subsection{Groups My1 and S1}
As shown by Figure \ref{fig13}, the curves of the My1 and S1 groups are close in parameter space to those of Mno spectral type, giving support for their being successors of those that have a large enough initial mass to experience strong TDU events. In contrast with the other parameters, which display modest variations, the oscillation amplitude is observed to increase by over one magnitude when the star transits from the Mno to the Myes spectral type. There is one exception, T Sgr, which has an oscillation amplitude of only 4.2 mag, even smaller than the mean value for Mno stars at the same $q$ value, 4.8 mag. The reason is probably the presence of an unresolved F companion \citep[][and references therein]{Jorissen2019}. But, apart from T Sgr, all other stars have oscillation amplitudes in excess of 5.2 mag, reaching up to 8.3 mag in the case of chi Cyg. There is no clear difference between the amplitudes of Myes and S curves, 6.3 mag on average in both cases (excluding T Sgr). The values of the other curve and star parameters are in most cases close to those of the Mno spectral type, with larger values of $P$, $\varphi_{\rm{min}}$ and $\Delta$, and smaller values of $W$. This suggests that strong TDU events essentially result, for a same interval of $q$, in variations of the parameters that are slightly larger than what they would have been if the star had stayed of Mno spectral type. Accordingly, successive TDU events cause a deviation from the Mno group that keeps increasing.  

The groups My1 and S1 are compact in most panels of Figure \ref{fig13}; only RU Her, of spectral type M6, stands out as having a long period and a clearly different profile of its ascending branch; accordingly, it has the lowest values of $p$, $W$ and $\varphi_{\rm{min}}$ in group My1. Its case is further discussed in Subsection 5.4.
In group My1, apart from their oscillation amplitude, R Ser and S Vir, of spectral types M5 and M6, respectively, are located in the same region of the parameter space as Mno spectral types. The $^{12}$C/$^{13}$C isotopic ratio has been measured for only R Ser, with a low value of 14, typical of Mno stars. This suggests that these two stars have only recently started to experience strong enough TDU events, the increased oscillation amplitude and the presence of technetium in the spectrum being the only detectable consequences.

In summary, we can assume with reasonable confidence that:

- Stars having curves of the My1 and S1 groups have evolved from the warmer Mno stars and have followed the general trend described in Section 4 before leaving the Mno spectral type. The transition from Mno spectral types to group My1 goes together with a major and immediate increase of the oscillation amplitude.

- In general, as a function of $q$, the evolution from Mno to My1 retains approximately the same direction in parameter space as followed by Mno spectral types but is accelerated, meaning that parameters that are decreasing (respectively increasing) decrease (respectively increase) faster and become therefore smaller (respectively larger) than what they would have been if the star had stayed of Mno spectral type. This is particularly clear for the parameters $P$, $\varphi_{\rm{min}}$, $\Delta$ and $W$. Successive TDU events cause therefore an increasing deviation from the evolution of Mno curves as a function of $q$.

- When projected on planes excluding shape-sensitive parameters, groups My1 and S1 are nearly indistinct. However, when projected on planes including shape-sensitive parameters, group My1 is farther away from the warmer Mno stars than group S1 is, showing that the shape of the ascending branch reaches maximal distortion before transiting to S spectral type, at which time the shape returns to a form closer to that of curves of less evolved stars.

- Some cases of large irregularity are observed in group S1 (R And, W And, W Aql and R Cyg) and My1 (RW And). This is a nearly exclusive feature of these groups: in the whole sample, the only other case of a $\Delta$ parameter in excess of 0.75 is that of Mno star R Cas, with $\Delta$=0.84, which was discussed in Section 4. Members of the S1 group reach several of the most extreme values of their parameters in the whole sample: R And and chi Cyg with respective oscillation amplitudes of 7.4 and 8.3 mag, S Cas with a period of 613 days and a colour index of 3.6 mag and R Cyg with an irregularity parameter of 1.02 mag. The four stars having $C$$>$2.5 mag, $q$$>$0.04 and $\Delta$$>$0.73 mag are also those having a mass loss rate larger than 6$\times$10$^{-7}$ M$_\odot$ yr$^{-1}$ \citep{Ramstedt2009}.

- There is a big $q$ gap of 0.07 between My1 and My2. But the scarcity of Myes curves prevents making its presence a robust result. It may be instead that there is continuity in the distribution of the values of $q$ over which a curve moves away from the warmer Mno curves, before experiencing strong TDUs and switching to Myes. In such a case larger statistics would reveal the presence of Myes curves filling the gap. With the present data sample, continuity can neither be excluded nor ascertained.

\subsection{Groups My2, S2 and S3}
Figure \ref{fig13} shows that, apart from a longer period, by some 150 days, both My2 and S2 are very close to, and often overlap, the region of the parameter space covered by the warmer Mno stars. This suggests that they evolved from this region, with their periods increasing and all other parameters displaying only small variations. However, there is no strong evidence for the validity of this statement; it might be that the My2 stars are successors of shorter period stars having a pulsation regime not regular enough for them to be accepted as Mira variables.

The only exception, as clearly shown by Figure \ref{fig13}, is the curve of R Gem, which differs significantly from the curves of the three other members of group S2 by having larger values of $\varphi_{\rm{min}}$ (0.64 instead of 0.54) and $p$ (+0.21 instead of $-$0.39). However, its diameter and temperature are the same as those of R Lyn, one of the other members of group S2 \citep{Ramstedt2009}. We failed to find, in the published literature, arguments that might explain the different shape of the ascending branch compared with stars having similar stellar parameters.

Group S3, in strong contrast with group S2, is clearly distinct from group My2. The presence of FF Cyg and T Cas at intermediate $q$ values may suggest an evolution from the latter to the former, but the case for this is weakened by the long period of T Cas and the significantly different shape of its light curve (lower values of $p$ and $\varphi_{\rm{min}}$, see Subsection 5.4). The specificity of group S3 had been already noted and discussed in Paper II. In particular, it was remarked that these stars being successors of group My2 would imply them moving away from the carbon stars of the sample: their colour index and irregularity parameters decrease and the amplitude of their oscillations increase. Figure \ref{fig13} displays very clearly such an evolution as a function of $q$. The evolution toward low colour indices was interpreted as being caused by a low dust abundance resulting from the lack of excess of C over O, or of O over C, causing the mass-loss rate to nearly cancel \citep{Hashimoto1998, Marigo2020}. Compared to group S2, the stars are much warmer, typically 3500 K instead of 2500 K \citep{Otto2011}, and have shorter periods, showing that they are younger on the AGB branch. 

\subsection{Four curves having unusual values of parameter $p$}
The $p$ vs $q$ map displayed in Figure \ref{fig13} gives evidence for four curves having unusual values of parameter $p$: T Cas, R Aur and RU Her, low by about 0.5 unit and R Gem, high by about 0.5 unit. The latter is an S star that would fit in group S2 if it were not for such a high $p$ value.

T Cas, R Aur and RU Her are three stars of Myes spectral type spanning a broad range of $q$ values. T Cas stands out as displaying very irregular features that we discuss in Subsection 5.5. But apart from this, the three stars share many properties: similar periods, $\sim$465 days, within $\pm$20 days, similar spectral types, $\sim$M6.5, within $\pm$0.5, similar colour indices, $\sim$2.1, within $\pm$0.2, a same abnormally low value of $p$, $\sim$$-$0.99, within $\pm$0.07, similar $^{12}$C/$^{13}$C isotopic ratios of $\sim$30. Their spectra have been shown to contain technetium, with certitude for R Aur and RU Her and probably for T Cas. They differ by the values of $W$, 0.39, 0.21 and 0.18, and therefore of $q$, $-$0.10, 0.00 and 0.12, of $\Delta$, 0.27, 0.50 and 0.74, and of $\varphi_{\rm{min}}$, 0.43, 0.46 and 0.54, respectively. The amplitudes of the oscillations are 4.3, 5.9 and 5.9 mag, respectively. If it were not for the very low values of the $p$ parameter, RU Her would fit well in My1, T Cas could be interpreted as transiting between My2 and S3, and R Aur as being an example of continuity between My1 and My2.

\begin{figure*}
  \centering
  \includegraphics[height=5.3cm,trim={0.cm 0.5cm 1.4cm .8cm},clip]{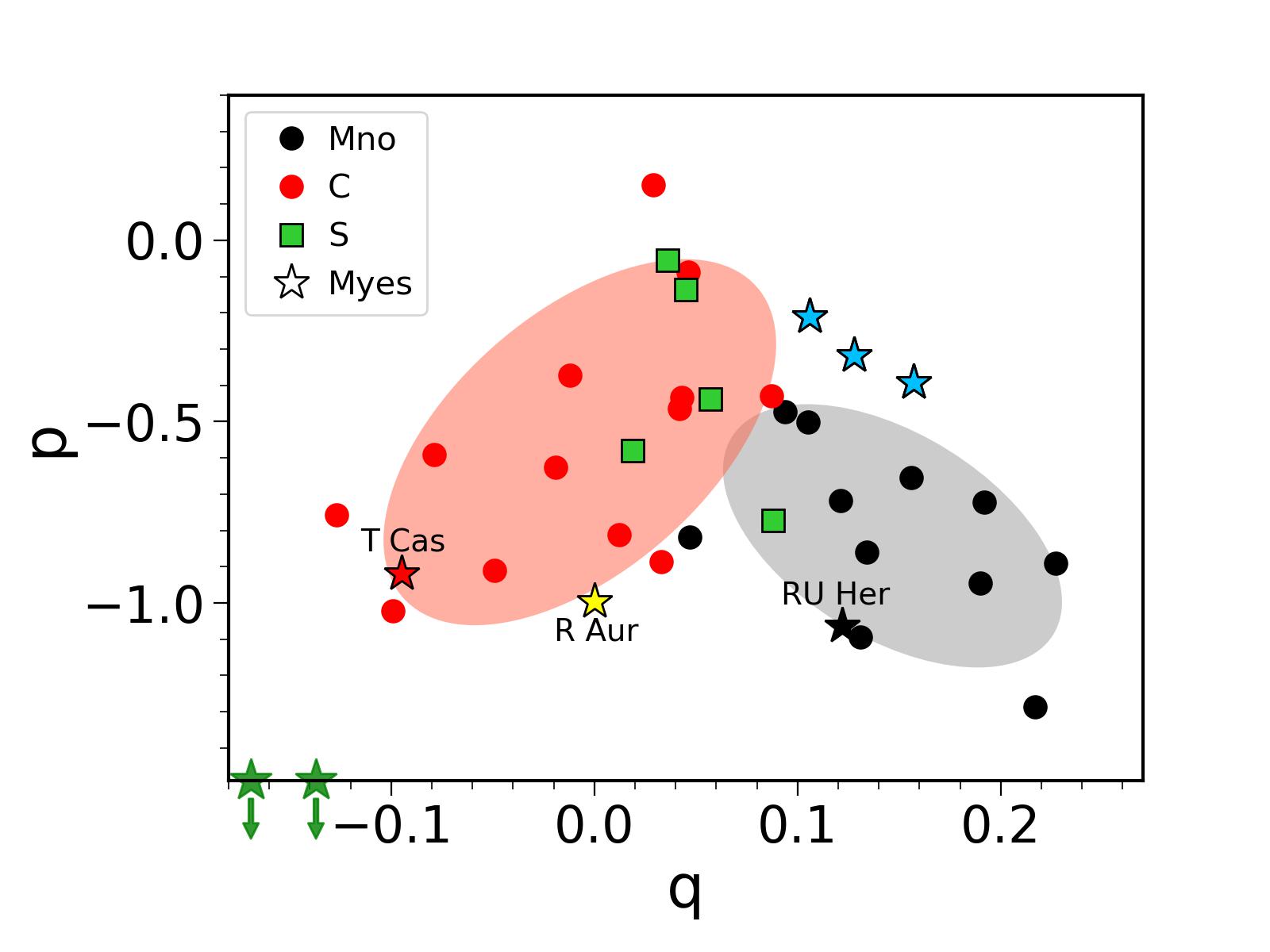}
  \includegraphics[height=5.1cm,trim={0.cm -1.1cm 0cm 0cm},clip]{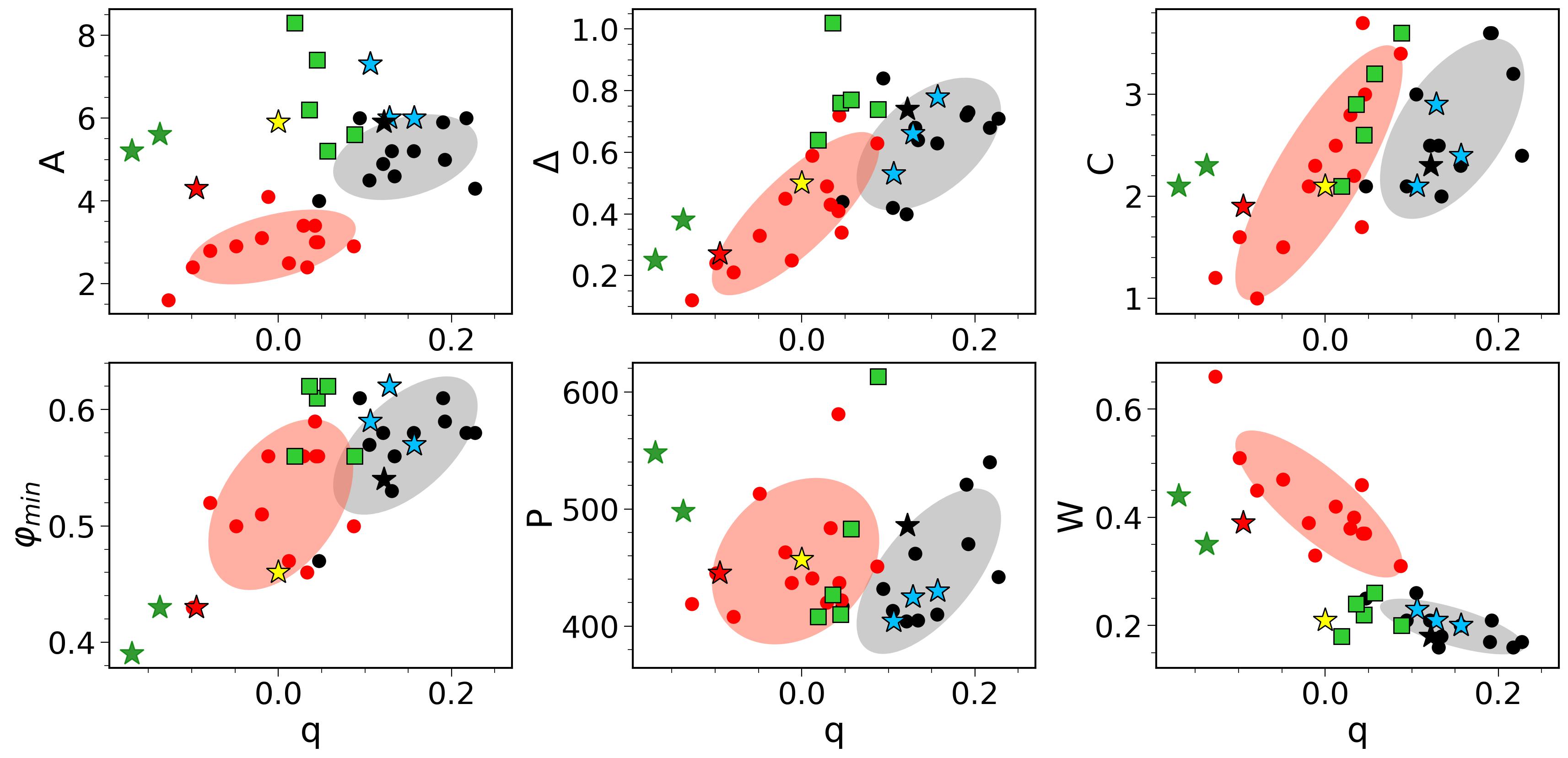}

  \caption{Maps of stars having $P$$>$400 days. Mno: black circles and ellipse. C: red circles and ellipses. Other Myes: blue stars. Other S: green squares. The green stars show the locations of Hot Bottom Burners R Cen and R Nor (see Section 7); in the $p$ vs $q$ map, it is well below the covered area, at $p$$\sim$$-$2. }
  \label{fig17}
\end{figure*}

Another common feature of T Cas, R Aur and RU Her is to have a longer period than the other Myes stars having similar values of $q$: 445 days for T Cas vs 318 days for FF Cyg/S Her, 457 days for R Aur vs 351 days for Z Peg/S Vir and 486 days for RU Her compared with 415 days for R Hor/S Pic. Figure \ref{fig17} displays parameter space maps of all curves having a period in excess of 400 days. The Mno and C curves dominate the sample and occupy well distinct regions of the parameter space. The sample includes only 5 S stars, S Cas, W Aql, R And, chi Cyg and R Cyg and three additional Myes stars, RW And, R Hor and S Pic. The $p$ vs $q$ map illustrates the spectacular difference between the two triplets of Myes stars. RU Her is the only star of the triplet that has been studied in detail. \citet{AlonsoHernandez2024} have found that it stands out in their selected sample of binary candidates as having the largest bolometric luminosity, (20$\pm$7) 10$^3$ solar luminosities. They quote a mass-loss rate of 1.6 10$^{-6}$ M$_\odot$ yr$^{-1}$.

In summary, the four curves having unusual $p$ values would fit in the general picture if it were not the different shape of their ascending branches (Figure \ref{fig6}). 

\subsection{Different forms of irregularity}
\begin{figure*}
  \centering
  \includegraphics[width=3.5cm,trim={0cm 0 0cm 0},clip]{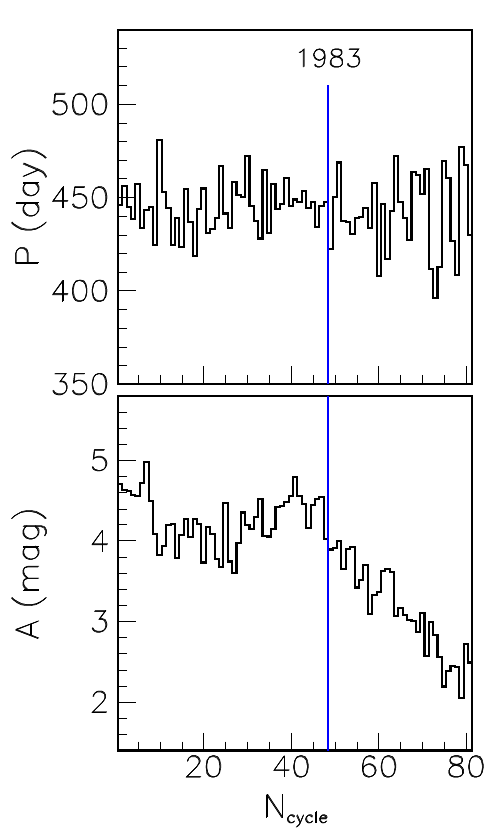}
  \includegraphics[width=3.5cm,trim={0cm 0 0cm 0},clip]{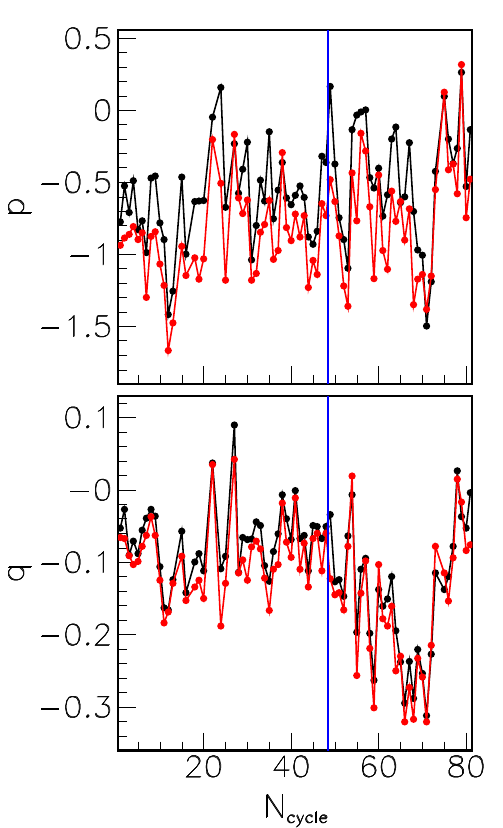}
  \includegraphics[width=6cm,trim={0cm 0 0cm 0},clip]{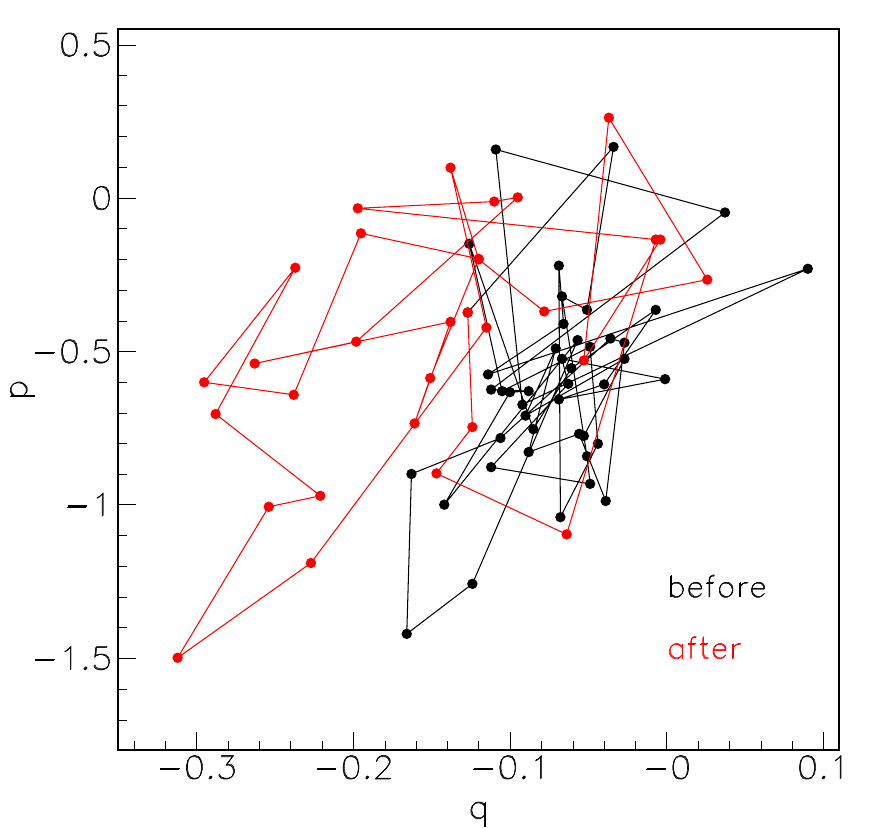}
  \includegraphics[width=4cm,trim={0cm 3.3cm 0cm 0},clip]{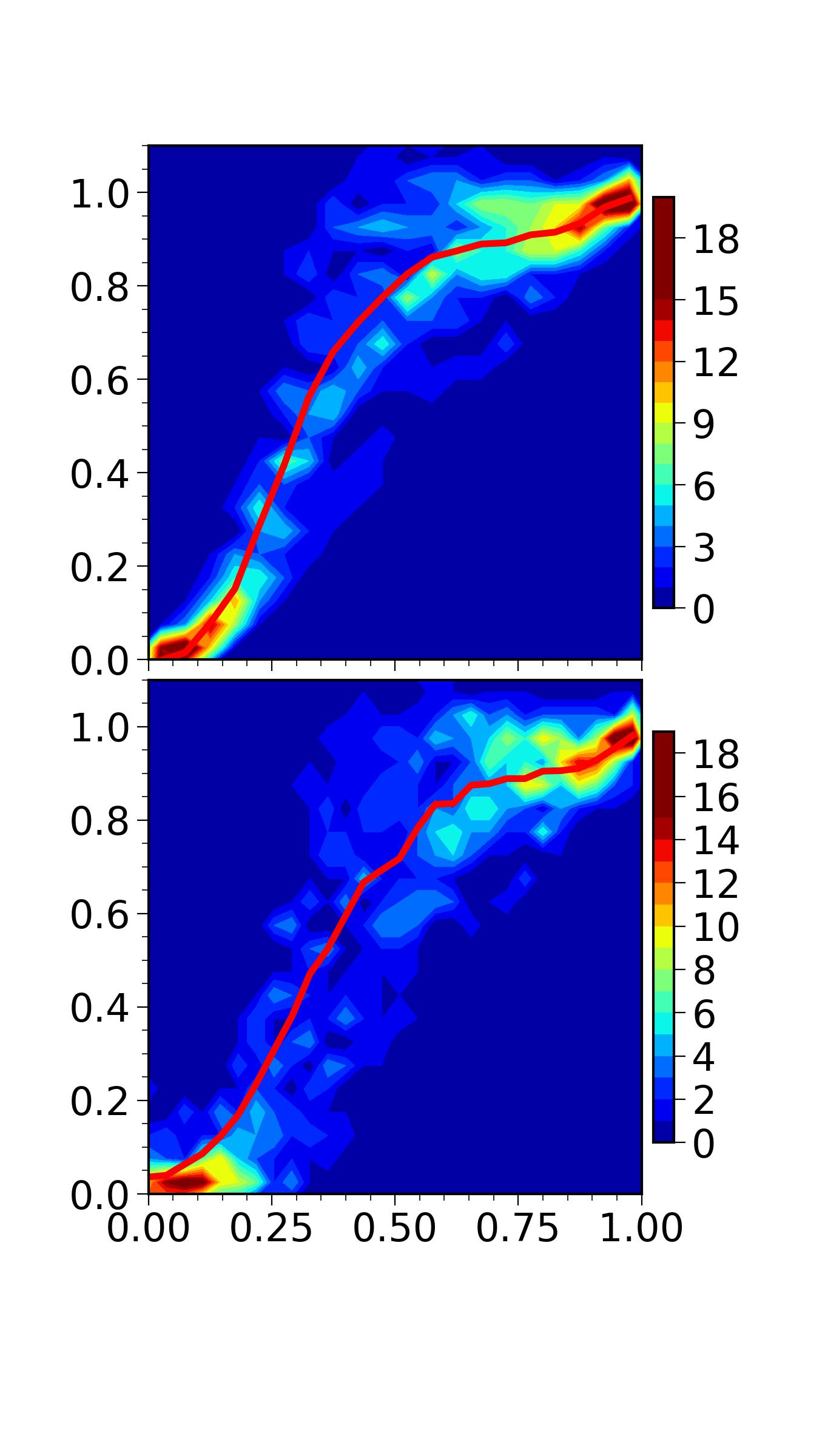}
  
  \caption{The light curves of T Cas in the 1925 to 1983 and 1983 to 2025 epochs (80 cycles). Left: dependence on cycle number of the period $P$ (upper panel) and the amplitude of oscillation $A$ (lower panel). Centre left: dependence on cycle number of the values of $p$ (upper panel) and $q$ (lower panel) calculated with the $\varepsilon$ parameters set to zero (black) or left free (red). Centre right: evolution of the shape parameters in the $p$ vs $q$ plane before (black) and after (red) 1983. Right: superimposed normalized profiles of the ascending branch for the post-1983 epoch (upper panel, giving an rms deviation $\delta$=0.11); the lower panel shows the normalized profiles of the full cycles with the origin of time starting at the mean value of the light minimum, giving a much larger value of $\delta$, 0.16, because of the large variations of $\varphi_{\rm{min}}$. The red lines show the mean profiles.}
  \label{fig18}
\end{figure*}

As mentioned in the preceding subsection, T Cas displays very irregular features illustrated in Figure \ref{fig18} and Table \ref{tab5}: the amplitude of the oscillations of the light curve has dropped by $\sim$2 mag between 1983 and 2025, but the period has stayed approximately the same. \citet{Howarth2001} have analysed the curve before the drop in terms of Fourier components. After the drop, the shape of the ascending branch has experienced very strong variations. Yet the $\Delta M_{\rm{max}}$ and $\Delta'M_{\rm{max}}$ parameters failed to reveal such irregularity and retained reasonably low values. This shows how major changes of shape of the profile can be undetected by the irregularity parameter $\Delta$. We wondered whether other cases of a similar nature were present in the sample of Table \ref{tab1}. To this end, we define a new parameter, $\delta$, as the rms deviation of the normalised profile of the ascending branch with respect to its mean. 

\begin{deluxetable*}{ccc ccc ccc ccc c}
\tablenum{5}
\tablecaption{Parameters of the light curve of T Cas before and after 1983. \label{tab5}}
\tablewidth{0pt}
\tablehead{
 \colhead{}&	\colhead{$P$}&	\colhead{$A$}&	\colhead{$W$}&	\colhead{$M_{\rm {\rm{max}}}$}&\colhead{$M_{\rm{min}}$}&\colhead{$\Delta M_{\rm{\rm{max}}}$}&\colhead{$\Delta'M_{\rm{\rm{max}}}$}&\colhead{$\Delta$}&	\colhead{$\varphi_{\rm{min}}$}&	\colhead{$p$}&	\colhead{$q$}&	\colhead{$\delta$}\\
\colhead{}&\colhead{(day)}&\colhead{(mag)}&\colhead{}&\colhead{(mag)}&\colhead{(mag)}&\colhead{(mag)}&\colhead{(mag)}&\colhead{(mag)}& \colhead{}&\colhead{$\times$10$^3$}&\colhead{$\times$10$^3$}&\colhead{}
}
\startdata
1925 to 1983&	445/13&	4.3/0.3&	0.39&	7.7&	11.9&	0.26&	0.27&	0.27&	0.43/0.03&	$-$919&	$-$95&	0.07\\
1983 to 2025&	441/21&	3.1/0.5&	0.48&	8.3&	11.4&	0.28&	0.17&	0.23&	0.45/0.05&	$-$816&	$-$180&	0.11\\
\enddata
\end{deluxetable*}

\begin{figure*}
  \centering
  \includegraphics[height=3.74cm,trim={0cm -0.2cm 0cm 0},clip]{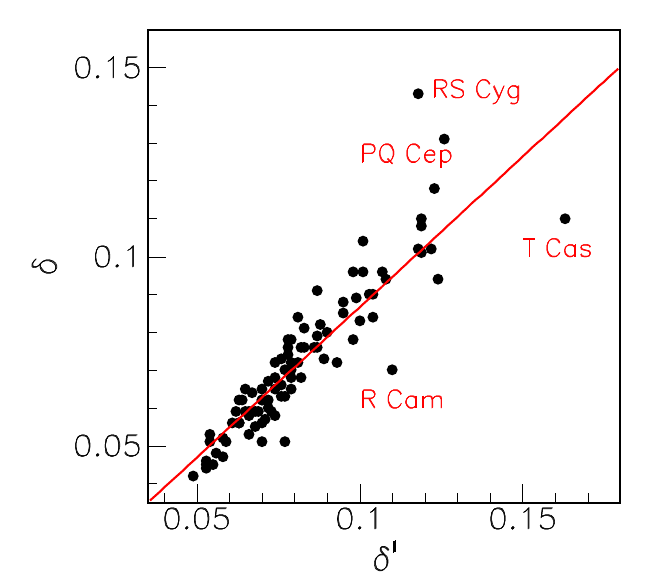}
  \includegraphics[height=3.77cm,trim={0cm 0.3cm 0cm -1cm},clip]{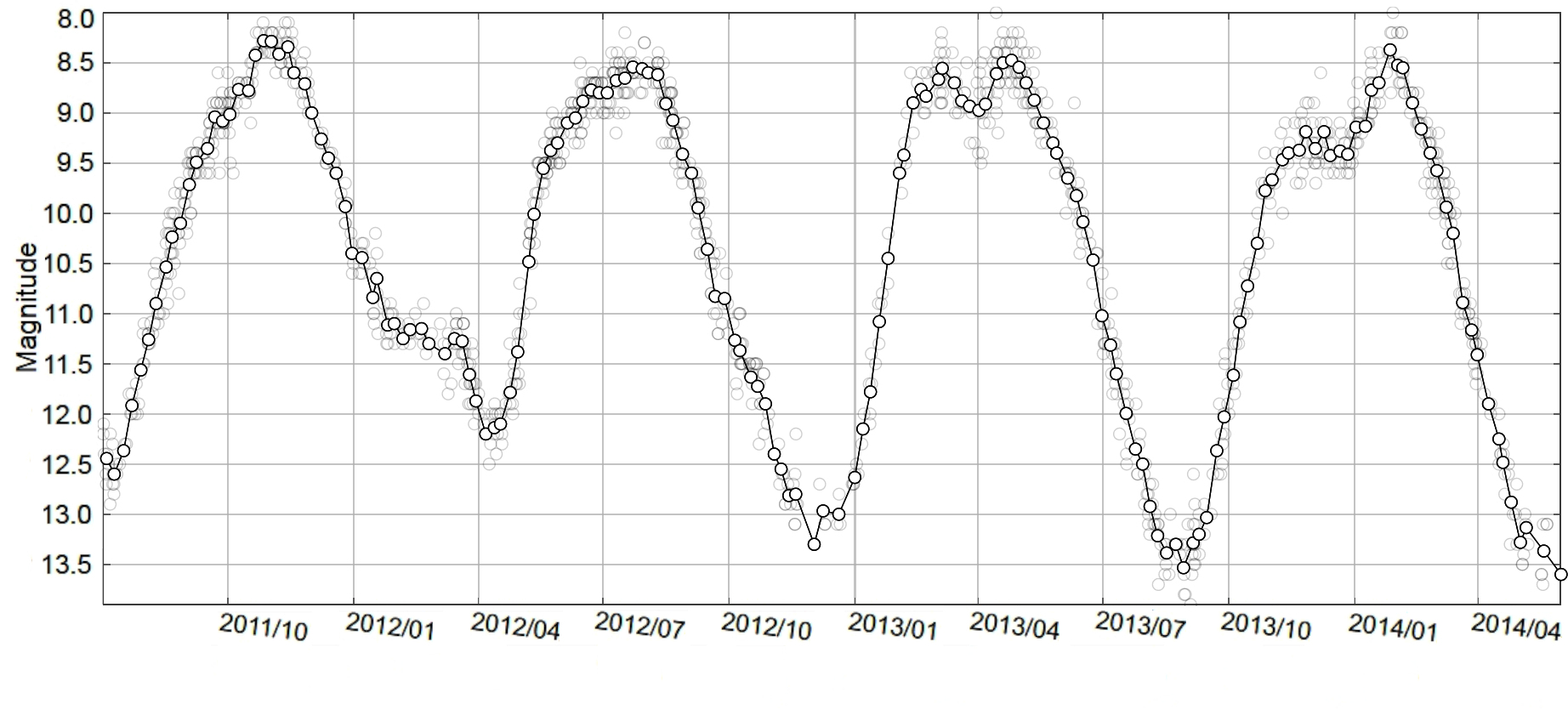}
  \includegraphics[height=3.7cm,trim={0cm 0.7cm 0cm 1cm},clip]{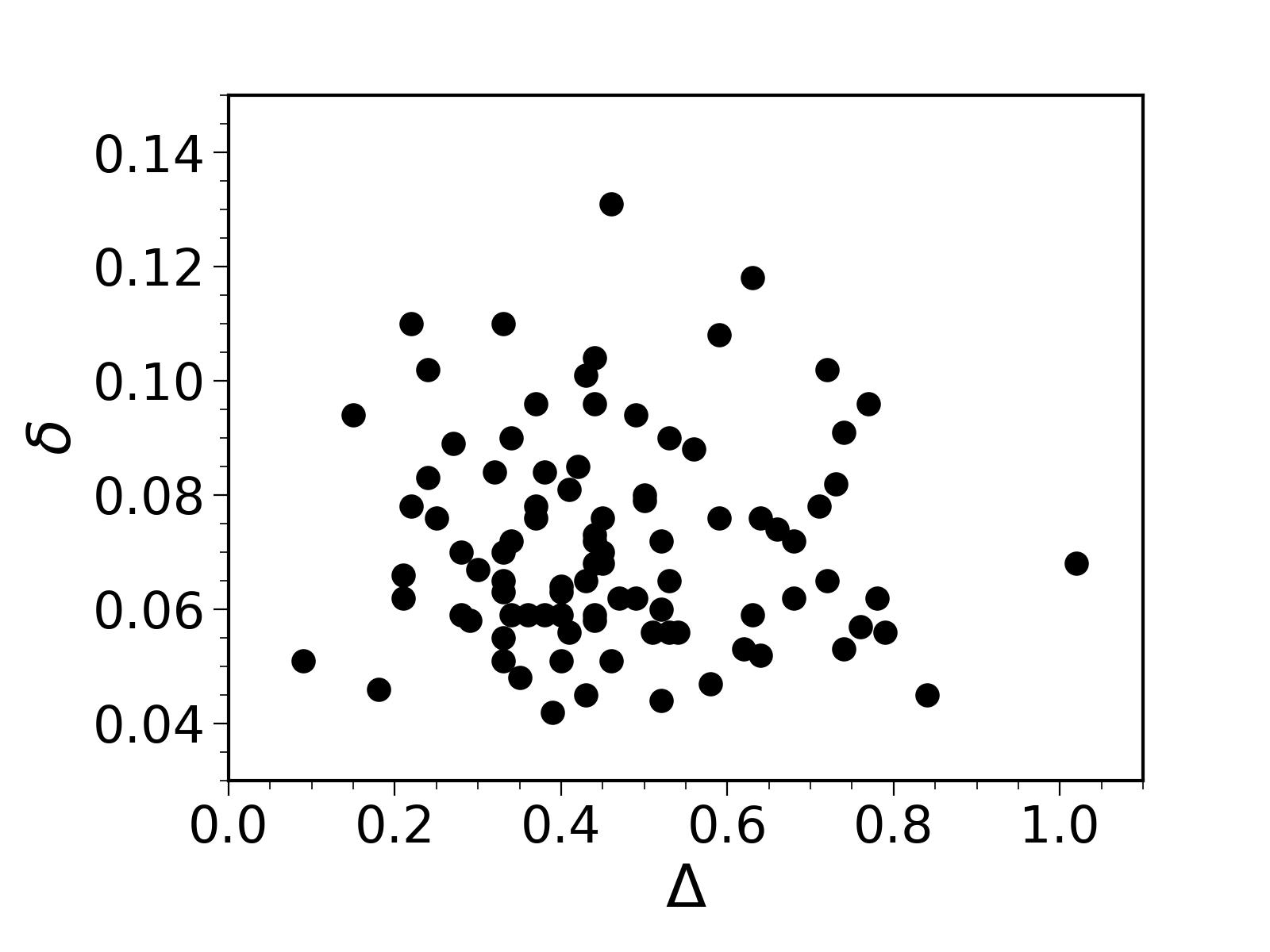}
  \caption{ Left: $\delta$ (ordinate) and $\delta'$ (abscissa) for the whole sample. The linear fit is $\delta$=0.007+0.796 $\delta'$. T Cas is shown only after 1983. Centre: light curve of R Cam in years 2011 to 2014 showing successively a hump on the descending branch, a hump-free single maximum, a double maximum and a hump on the ascending branch. Right: dependence of $\delta$ on $\Delta$ for the whole sample.}
  \label{fig19}
\end{figure*}

As illustrated in the right column of Figure \ref{fig18}, $\delta$ should be averaged over normalized ascending branch profiles. If instead one averages it over the mean normalized profile, variations of $\varphi_{\rm{min}}$ may add important contributions; this is the case for T Cas where the value of $\delta$ calculated in the latter way is $\delta'$=0.16 instead of 0.11.  Figure \ref{fig19} compares the values of $\delta$ and $\delta'$  for the whole sample. On average, $\delta$$\sim$0.8$\delta'$, showing that the mean effect of the fluctuations of $\varphi_{\rm{min}}$ is a 20\% increase of $\delta$. In addition to T Cas, three other stars are seen to stand out: RS Cyg, PQ Cep and R Cam. RS Cyg is a very low amplitude C star with a high temperature of 3100 K and a low mass loss rate of 2$\times$10$^{-7}$ M$_{\odot}$ yr$^{-1}$ \citep{Guandalini2006}, which was commented on in Paper II. Its light curve has been extensively studied by \citet{Cadmus2022, Cadmus2024}. It often displays a hump on the ascending branch, close to maximum light, possibly producing a double-maximum profile preventing a reliable definition of $\varphi_{\rm{min}}$. It was included in Table \ref{tab1} in spite of the irregularity of its light curve and, as can be seen in Section 6, it fits well in the general trend obeyed by warm carbon-rich stars. Its standing out in the $\delta$ vs $\delta'$ plot of Figure \ref{fig19} is not surprising in such a context. PQ Cep is another carbon-rich star with parameters taking less extreme values than RS Cyg, but deviating from the mean in the same direction. Its light curve was not measured in the past century, the density of observations in the present century is less than for RS Cyg and it has not been much studied.

R Cam is one of the four members of group S3, for which the parameters match well the specific features. R Cam was noted in Paper II as occasionally displaying a double maximum and/or a small hump on the descending branch (together with S Cam, T Cas, U Cyg, T Dra and U UMi). The central panel of Figure \ref{fig19} shows four successive cycles of its light curve illustrating the difficulty to reliably define the ascending branch. It is not surprising, in such a case, that one finds different values for $\delta$ and $\delta'$.  Yet, the impact on the definition of parameters $p$ and $q$ is modest: values obtained by fitting different epochs do not vary by more than 0.07 and 0.02, respectively. The right panel of Figure \ref{fig19} displays the dependence of $\delta$ on $\Delta$; the absence of correlation shows clearly the different natures of the irregularity of the shape of the light curve and the irregularity of the magnitude at maximal light.

\section{Carbon-rich stars}
An abundant literature is devoted to the observation and study of carbon-rich stars. Recent articles offering an overview and addressing questions of direct relevence to the present work are \citet{Tanaka2007, Nowotny2011, Rau2017, Abia2020,Straniero2023}. They show that only stars with initial masses between approximately 1.5 and 3.5 solar masses can become carbon-rich; those with lower masses do not experience TDU events, those with larger masses become Hot Bottom Burners. Most are Long Period Variables, but only some of them have large enough oscillation amplitudes to be classified as Miras. Indeed, the transition from S to C spectral type goes together with a major drop of the oscillation amplitude as well as with a significant increase of the period, the latter resulting from the expansion of the star and the decrease of its effective temperature \citep{Lebzelter2007}.

\begin{figure*}
  \centering
  \includegraphics[height=3.4cm,trim={0.cm 0cm 1.4cm 1.2cm},clip]{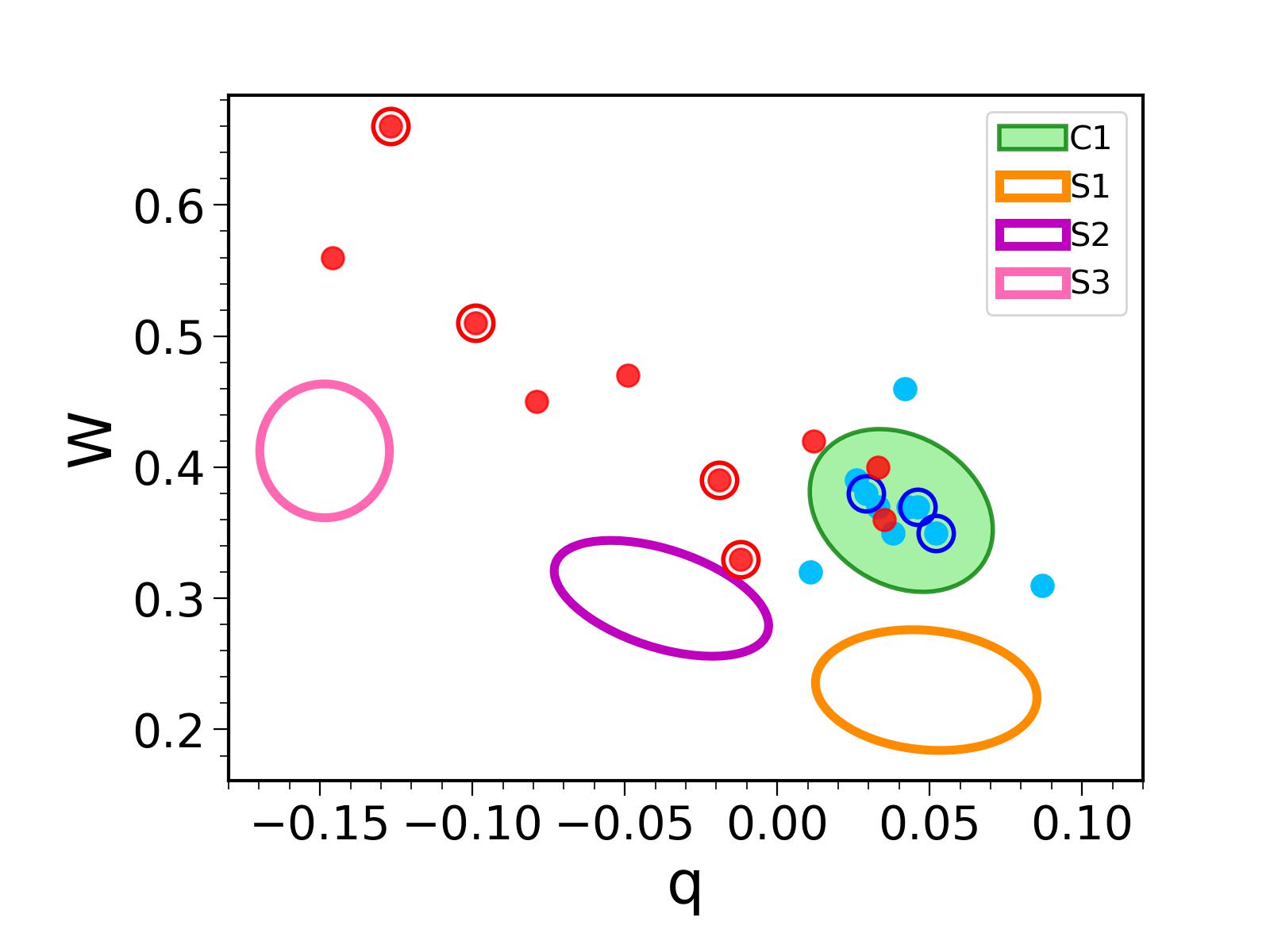}
  \includegraphics[height=3.4cm,trim={0.cm 0cm 1.4cm 1.2cm},clip]{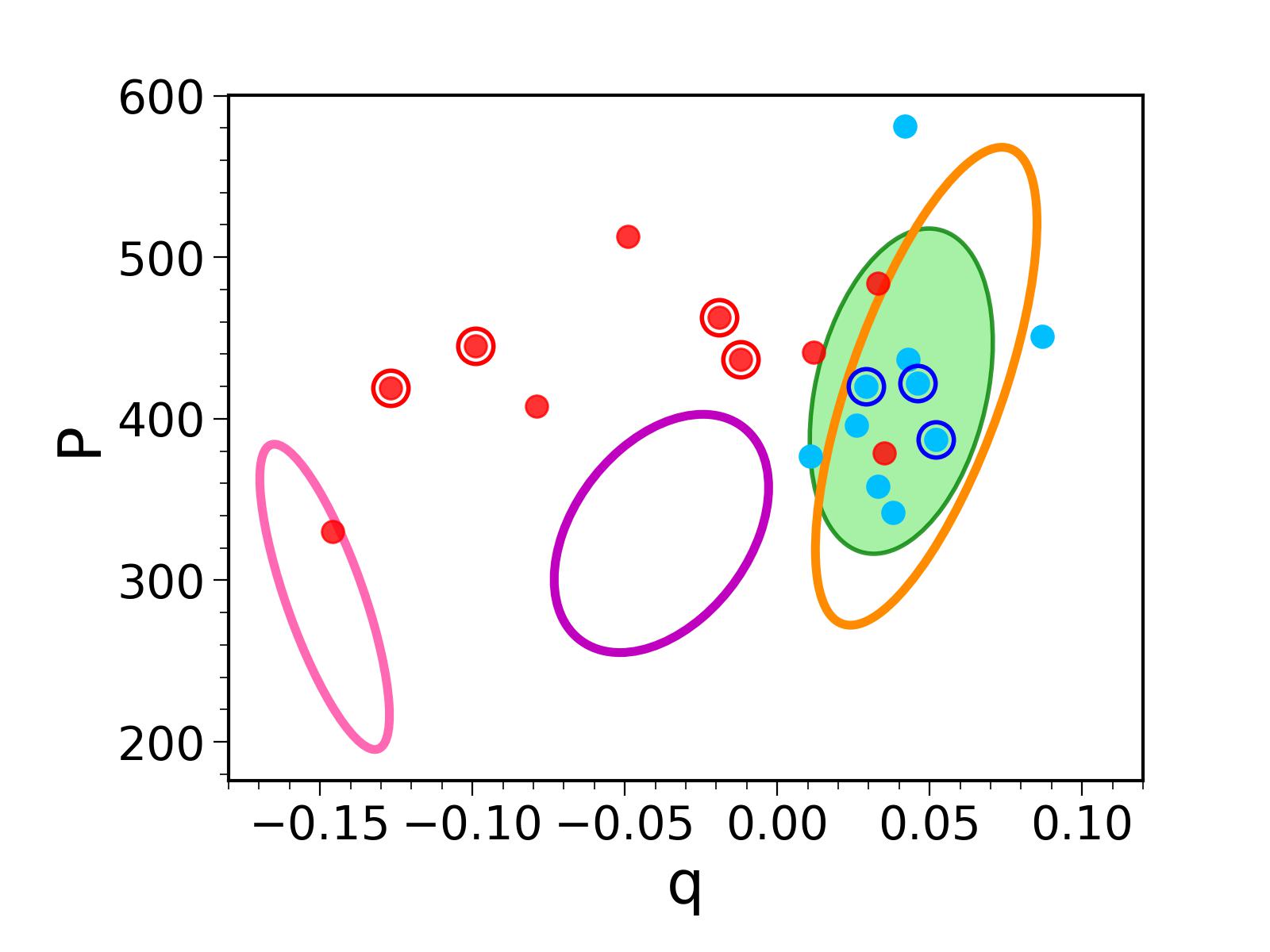}
  \includegraphics[height=3.4cm,trim={0.cm 0cm 1.4cm 1.2cm},clip]{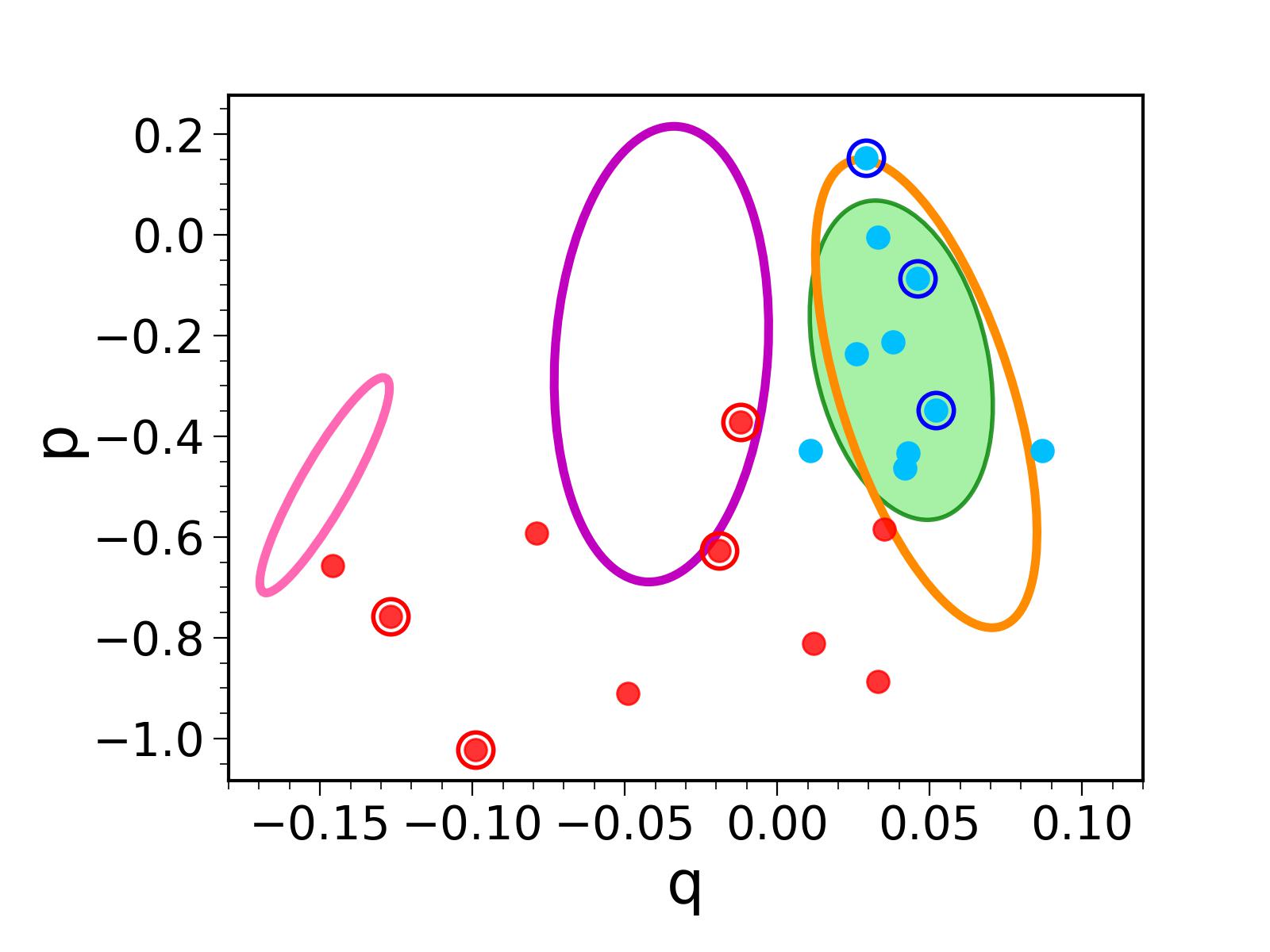}\\
  \includegraphics[height=3.4cm,trim={0.5cm 0cm 1.9cm 1.2cm},clip]{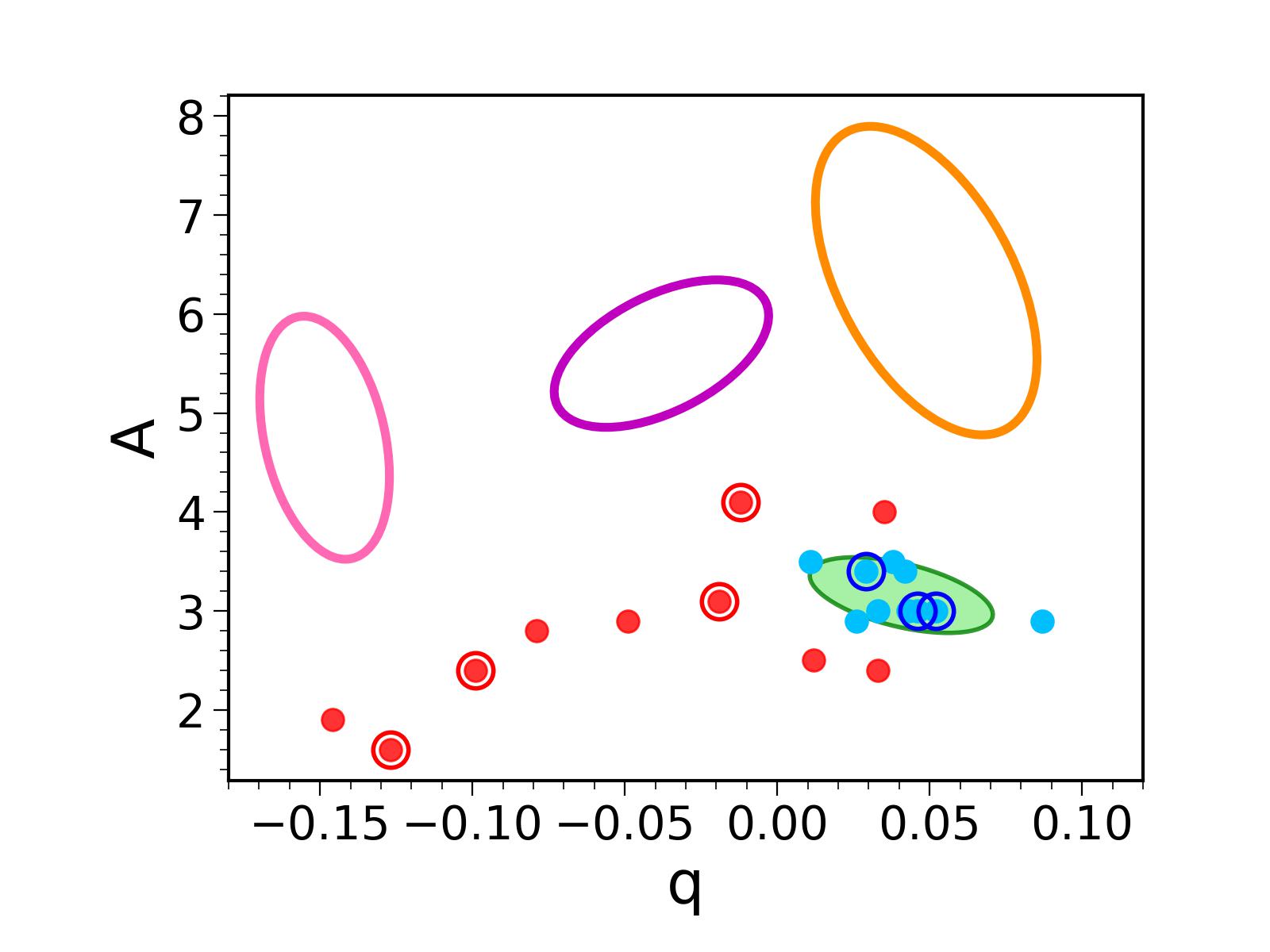}
  \includegraphics[height=3.4cm,trim={0.5cm 0cm 1.9cm 1.2cm},clip]{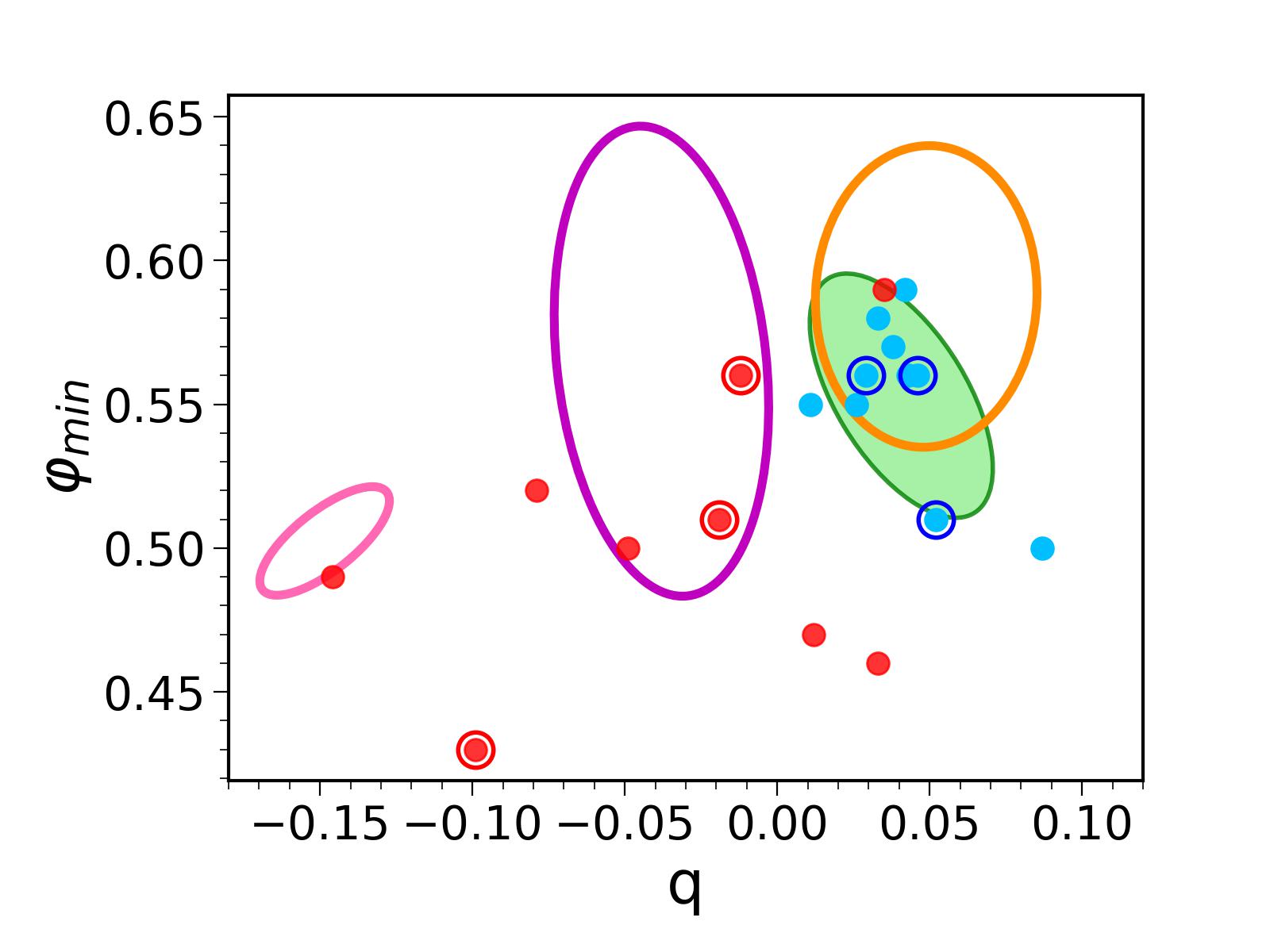}
  \includegraphics[height=3.4cm,trim={0.5cm 0cm 1.9cm 1.2cm},clip]{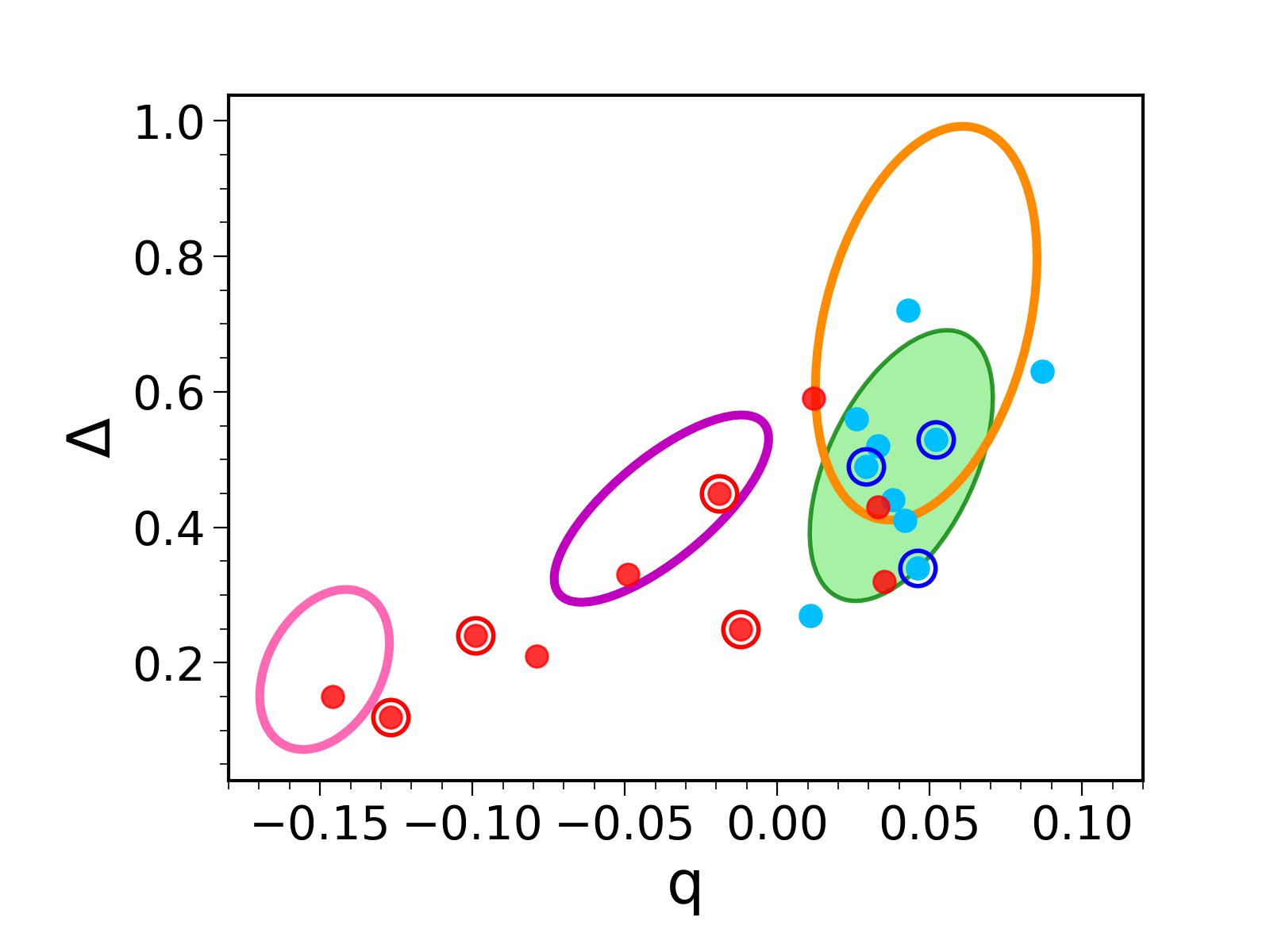}
  \includegraphics[height=3.4cm,trim={0.5cm 0cm 1.9cm 1.2cm},clip]{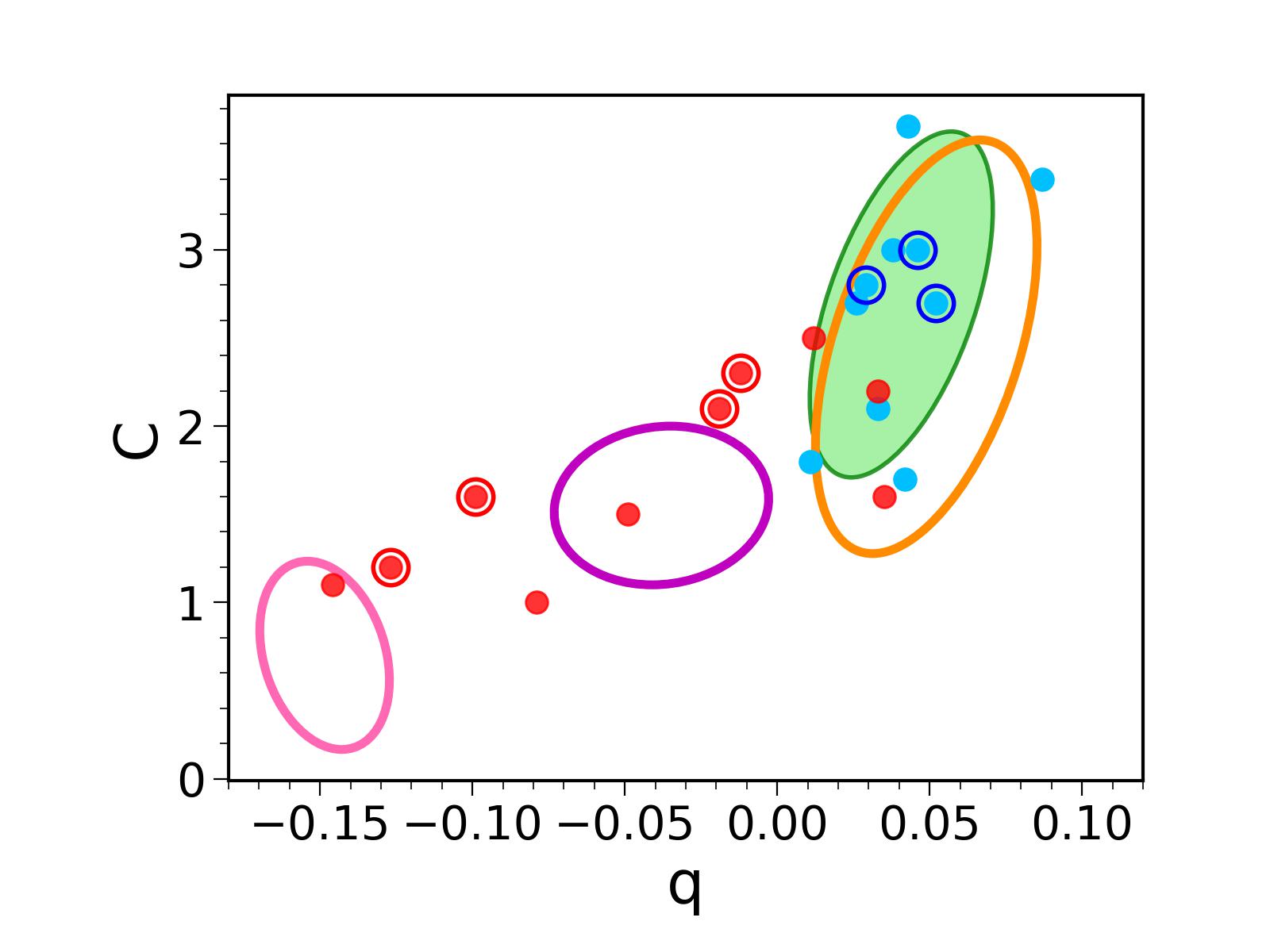}
 
  \caption{Maps of the curves of C spectral type in the $p$, $C$, $A$, $W$, $\Delta$ and $\varphi_{\rm{min}}$ vs $q$ planes. Group C1 is shown as blue full circles, the other curves as red full circles. Open red circles are for high effective temperatures and low mass-loss-rates and the open blue circles for low effective temperatures and high mass-loss-rates. Open ellipses show the regions covered by the groups S1 (orange), S2 (purple) and S3 (pink). The full green ellipse shows the C1 region. }
  \label{fig20}
\end{figure*}

  Figure \ref{fig20} displays maps of the carbon-rich curves projected on planes having the $q$ axis as abscissa. Clear correlations are observed between $q$ and $W$, $C$, $A$ and $\Delta$. In Paper II, we commented about a sample of 15 carbon-rich stars for which \citet{Bergeat2005} are quoting values of the effective temperatures and mass-loss-rates. Of these, a triplet (V Cyg, T Dra and R For) have the lower effective temperatures (1880, 1850 and 2000 K, respectively) and the larger mass-loss-rates (6.3, 8.2 and 5.0 $\times$10$^{-6}$ M$_\odot$ yr$^{-1}$, respectively). They are circled in blue in the maps of Figure \ref{fig20}. And a quartet (U Cyg, RZ Peg, RV Cen and RS Cyg) have the higher effective temperatures (2650, 2925, 2855 and 3100 K, respectively) and the smaller mass-loss rates (10, 1.0, 7.8 and 2.0$\times$10$^{-7}$ M$_{\odot}$ yr$^{-1}$, respectively). They are circled in red in the maps of Figure \ref{fig20}. They suggest that when carbon stars evolve at the end of the AGB, the light pulses become narrower with steeper ascending branches ($W$ decreases, $q$ and $\varphi_{\rm{min}}$ increase). Moreover, the colour index increases, the pulsations become less regular and their amplitude increases. The periods remain essentially constant. In order to guide the eye, we outline in Figure \ref{fig20} the region C1 covered by the cooler C stars ($p$$>$$-$0.5 and $q$$>$0.).

Figure \ref{fig20} displays in addition the location of the S curves in parameter space. A number of features are revealed:

- The S curves have oscillation amplitudes typically two magnitudes larger than the C curves. This is probably related, at last in part, to the different nature of the molecules on the photosphere and in the lower atmosphere;

- The regions covered by the S1 curves largely overlap with C1. Apart from the oscillation amplitudes, the only exception is the $W$ vs $q$ map that shows the former being close to the latter but having lower values of $W$, typically 0.1 smaller. As commented in the previous subsection, the shapes of the S1 curves are returning to a form closer to that of curves of less evolved stars after having evolved away from them when they were of Myes spectral type. This may suggest at least four possible scenarios for the stars having S1 light curves; one is that they keep evolving as S types and never become carbon rich \citep{Guandalini2008}, in which case their light curves would remain in the S1 region for the rest of their AGB life as long as they keep pulsating; a second scenario is that they evolve smoothly to carbon rich stars similar to the stars having curves in the C1 region, pulsating with progressively lower amplitudes and $q$ and $W$ values experiencing modest changes; a third (and fourth) scenario is that they stop pulsating altogether, as considered in Paper II, their light pulses becoming so brief that they would eventually fade away; they might then become carbon rich stars pulsating with light curves having lower q values and progressively cool down to the C1 region; or they may never clearly pulsate again and spend the rest of their AGB life as carbon rich stars having light curves classified as semi-regular or irregular. In most cases, we are unable to choose between these scenarios, illustrating how poor is our understanding of the underlying dynamics.

The triplet S1/S2/S3 spans the whole range of $q$, as do the curves of the carbon stars, from the cooler to the warmer. This may suggest, in similarity with the possible S1 to C1 transition discussed above, that stars having curves in the S3 group would progressively evolve to carbon stars having curves of low $q$ values, while stars having curves in the S2 group would progressively evolve to carbon stars having curves of intermediate $q$ values.  However, here again, this is ignoring the possible transition from Mira to so-called semi-regular or irregular; such a transition is probably likely given the very large population of non-Mira carbon rich stars.

The mass loss rates, $^{12}$C/$^{13}$C isotopic ratios and effective temperatures of many of the carbon-rich stars of our sample have been measured, however with large uncertainties. We find that they are generally correlated with the value of the $q$ parameter. Negative $q$ values are associated with large effective temperatures, typically $\sim$2800 K, low mass loss rates, typically below 10$^{-6}$ M$_\odot$ yr$^{-1}$ and low isotopic ratios. In contrast, positive $q$ values are associated with low effective temperatures, typically $\sim$2000 K, large mass loss rates, well in excess of 10$^{-6}$ M$_\odot$ yr$^{-1}$ and large isotopic ratios. Many peculiarities complicate the general picture: LX Cyg \citep{Uttenthaler2016} and BH Cru \citep{Uttenthaler2011, VanEck2017} have become C stars only recently; V CrB \citep{Feast2006, Abia2020} has a low metallicity and may be an extragalactic interloper; U Cyg \citep{Hinkle2016} is likely a J- or R-type C star in view of its low $^{12}$C/$^{13}$C isotopic ratio and other features; V Cyg \citep{Neufeld2010} is surrounded by a shell of very large water abundance possibly produced by the vaporization of orbiting comets or dwarf planets; T Dra \citep{AlonsoHernandez2024, AlonsoHernandez2025} is an X ray emitter.

\section{Stars excluded from the Table 1 sample}
\begin{figure*}
  \centering
  \includegraphics[width=0.75\textwidth,trim={0cm 0 0cm 0},clip]{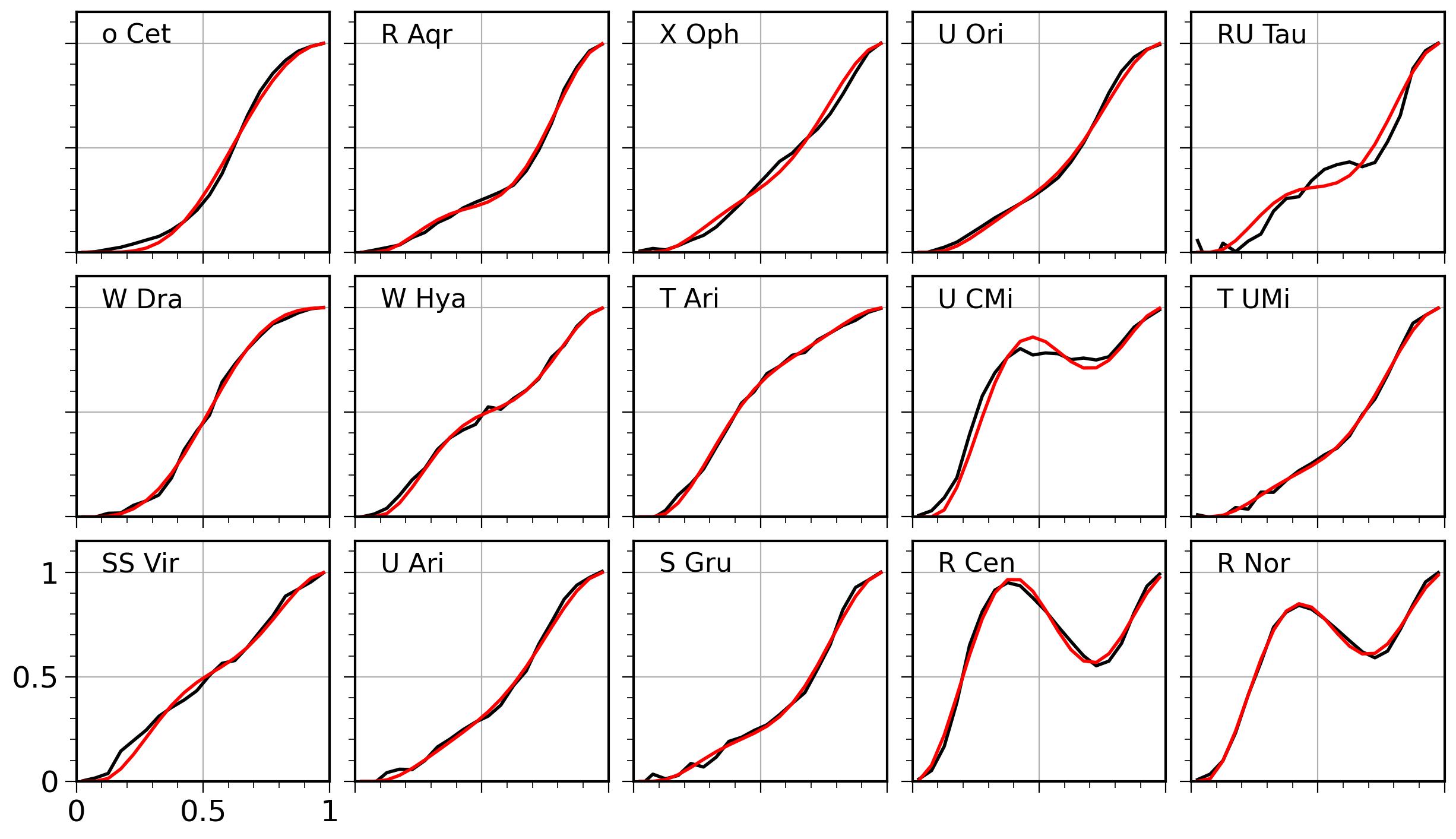}
  \caption{Normalized profiles of the ascending branches of the light curves mentioned in Section 7.}
  \label{fig21}
\end{figure*}

In the present section we review briefly the case of ten stars, which were given a special treatment in Paper II and were accordingly excluded from the sample listed in Table \ref{tab1}. They include five stars of Mno spectral type having light curves presenting peculiarities (W Dra, W Hya, T Ari, U CMi and T UMi), three binaries ($o$ Ceti, R Aqr and X Oph) and two stars presenting interesting features, U Ori and RU Tau. Their parameters are listed in Table \ref{tab6} and their normalized ascending branch profiles are displayed in Figure \ref{fig21}. In addition, we include the cases of SS Vir, a carbon-rich star excluded from the Merch\'{a}n Ben\'{i}tez sample as being semi-regular, and of U Ari and S Gru, both of Myes spectral type, which were excluded from the Paper II sample as having insufficient density of observations.

The curve of Mno star W Dra was said in Paper II to be only marginally different from those of the selected sample, in which it could be included. The present evaluation of its $p$ and $q$ parameters reveals indeed that it is clearly a member of the group of Mno outliers.

The curve of Mno star W Hya was instead found to be very different from those of the selected sample and to closely resemble curves of what is now referred to as group My2, such as that of R Hya, suggesting that the star has experienced TDU events. The present analysis supports this finding, the parameters of both curve and star being typically only 2$\sigma$ away from the centre of the My2 group.

The curves of T Ari and U CMi, both of Mno spectral type, were observed in Paper II to be very atypical, with a very large value of the $W$ parameter. The present analysis confirms this result. They share negative values of $q$, $-$0.09 and $-$0.16, respectively, which locate them between My2 and S3, close to the warmer C stars, far away from the regions covered by Mno stars. Indeed, they are closer to S star FF Cyg and Myes star T Cas than to any other star of the sample. Moreover, the parameters of the U CMi curve are remarkably close to those of the Hot Bottom Burners R Cen and R Nor, suggesting that U CMi might be one of these. Unfortunately, as remarked in Paper II, too little is known about these stars to suggest plausible explanations and the cases must remain unexplained.

T UMi experienced a sudden period change a few decades ago, usually interpreted as the effect of the proximity of a helium shell flash. The parameters listed in Table \ref{tab6} of Paper II were accordingly evaluated for cycles preceding year 1985. The present evaluation of their relation with the $p$ and $q$ parameters gives evidence for a good match with the region covered by Mno curves illustrated in Figure \ref{fig10} and Table \ref{tab3} (excluding Mno outliers); however, $C$ and $\Delta$ have slightly lower values.

$o$ Cet and R Aqr were excluded from the analysis of Paper II as being members of binaries. The present parameters match those of group My1 with the exception of too low an amplitude of oscillation for R Aqr (4.2 instead of 6.2$\pm$0.5 mag) and too large a value of $p$ for $o$ Cet (+0.16 instead of $-$0.46$\pm$0.28).

Several parameters of X Oph are well off the linear fits obeyed by the Mno curves (Table \ref{tab3}): $\Delta$=0.22 instead of 0.52$\pm$0.10, $W$=0.35 instead of 0.25$\pm$0.02, $\varphi_{\rm{min}}$=0.49 instead of 0.55$\pm$0.02 and, most importantly, $A$=1.7 instead of 4.9$\pm$0.5. This supports the explanation proposed in Paper II, the presence of a bright unresolved companion being the source of the light detected at the minima of the light curve.

The spectral type of U Ori is listed as M6-9.5 by both the SIMBAD data base and \citet{MerchanBenitez2023}, with a possible presence of technetium in the spectrum. It is a star that has been abundantly studied, in particular for maser emissions surrounding it, and its light curve has been measured with a high density of observations over the past century; however, as its period is very close to 1 year (371 days), the curve suffers important seasonal gaps which caused its exclusion from the analyses of Papers I and II, except for a mention in Table 6 of Paper II. The present analysis identifies it as a clear member of the My1 group, not far away from the region covered by stars of Mno spectral type.

The light curve of RU Tau was mentioned in Paper II as displaying occasional double maxima. The star, of spectral type M3.5, is quoted by \citet{MerchanBenitez2023} as being the only oxygen-rich Mira variable sharing with Hot Bottom Burners (HBB) R Cen and R Nor a long period (589 days), a low colour index (1.9) and a large width parameter (0.41). However, the analysis of its light curve did not support an HBB assignment. The present analysis gives $q$ values near 0.1 and $p$ values near $-$1.0 but the low density of observations prevents a reliable evaluation of the precise location in parameter space.

SS Vir is a semi-regular carbon-rich star mentioned by \citet{Bergeat2005} as having an effective temperature of 2560 K and a mass loss rate of 3.6 10$^{-7}$ M$_{\odot}$ yr$^{-1}$ and by \citet{Richichi2006} as having an effective temperature of 2445$\pm$41 K and a mass loss rate of 2 10$^{-7}$ M$_{\odot}$ yr$^{-1}$. Its light curve parameters listed in Table \ref{tab6} are consistent with these evaluations and make SS Vir a likely successor of the S2 group.

The light curves of U Ari and S Gru are typical of the My1 group, their parameters are listed in Table \ref{tab6}.

\begin{figure*}
  \centering
  \includegraphics[height=15cm,trim={0cm 0 0cm 0},clip]{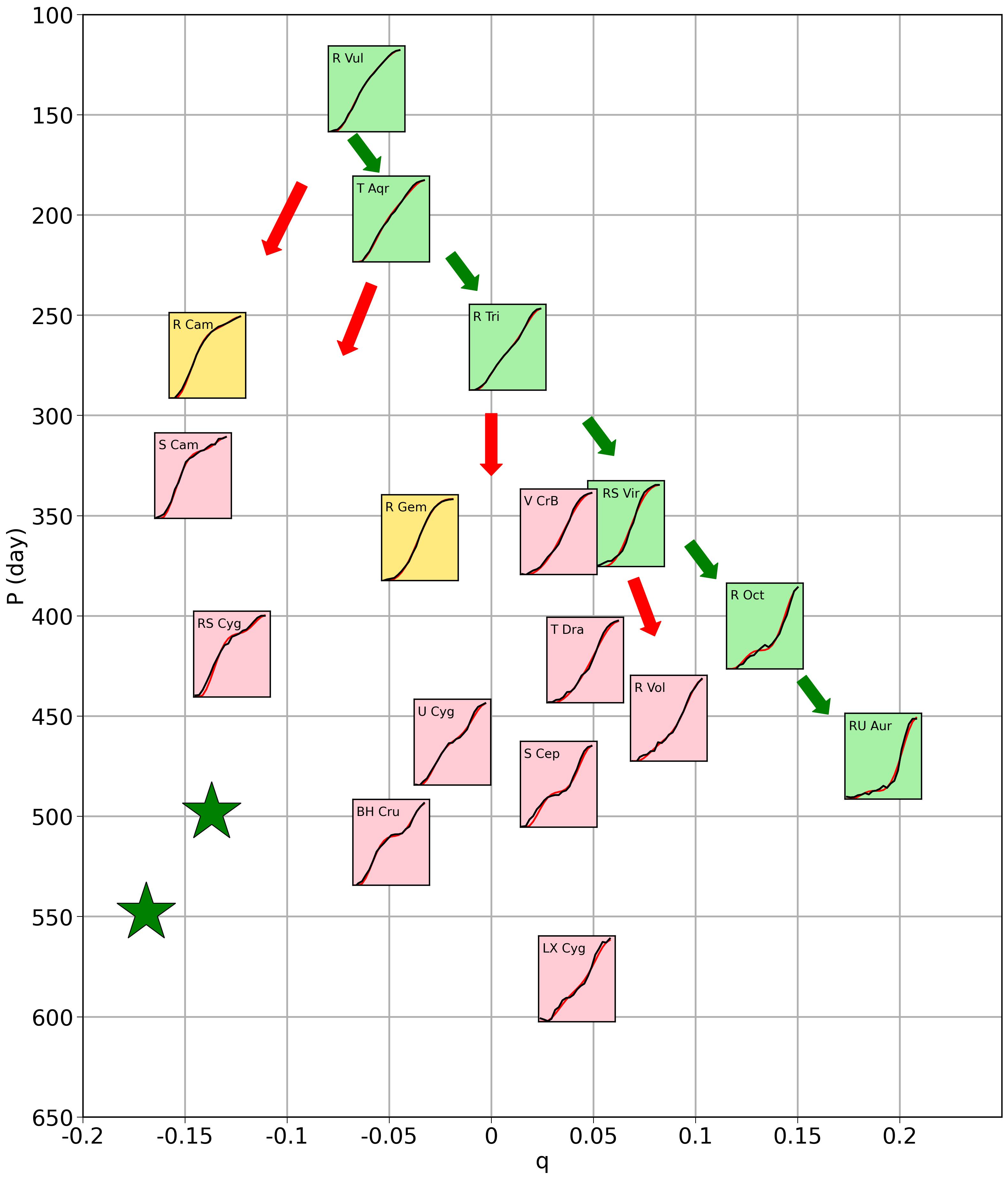}
   \includegraphics[height=14.9cm,trim={0cm -1.5cm 0cm 0cm},clip]{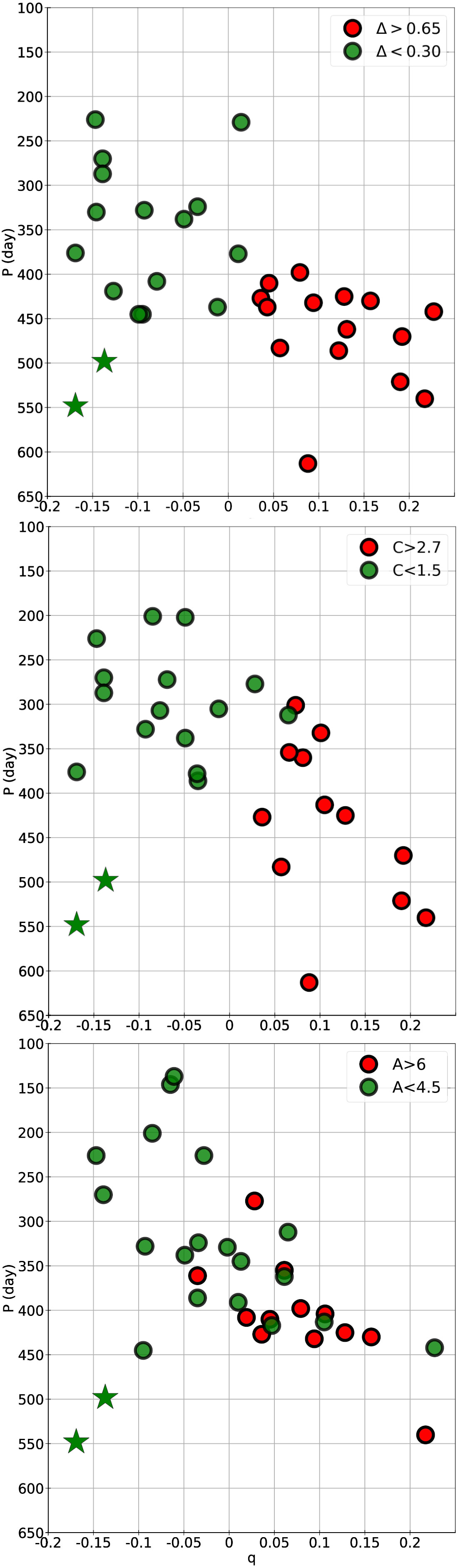}
  \caption{Left panel: the dependence of the ascending branch profiles on their location in the period vs $q$ plane is illustrated with typical examples (Mno in green, S in yellow and C in red). Arrows indicate schematically the evolution of the light curves associated with that of the stars on the AGB. Right column: three examples of the evolution of the curve parameters in the period vs $q$ plane for $\Delta$ (up, red dots for $\Delta$$>$0.65, green dots for $\Delta$$<$0.30), $C$ (centre, red dots for $C$$>$2.7, green dots for $C$$<$1.5, excluding carbon stars) and $A$ (down, red dots for $A$$>$6, green dots for $A$$<$4.5, excluding carbon stars). The green stars show the locations of the Hot Bottom Burners R Cen and R Nor.}
  \label{fig22}
\end{figure*}

In addition to the above ten stars, we also consider the cases of R Cen and R Nor, two well-established Hot Bottom Burners, which were mentioned in Paper II. Hot Bottom Burners have initial masses larger than $\sim$4 solar masses and their spectra include no technetium but strong lithium lines. R Cen and R Nor are accordingly of the Mno spectral type. As R Cen displays major changes of amplitude starting in the late sixties of the past century, probably associated with a recent thermal pulse, we consider only the part of the light curve covering between January 1919 and May 1966. For both stars, we use the second maximum as origin of the phase in order to preserve the universality of a featureless descending branch, implying a doubly-peaked ascending branch as displayed in Figure \ref{fig21}. We made fits to the normalised ascending branch using various constraints on the values of the $\varepsilon$ parameters, and obtained consistent results; the values quoted in Table \ref{tab6} correspond to the best fit obtained by leaving $\varepsilon_1$ and $\varepsilon_2$ free to vary, ($\varepsilon_1$,$\varepsilon_2$)=($-$0.01,+0.02) and ($-$0.05,+0.01), respectively. As illustrated in Figures \ref{fig17} and \ref{fig22}, the specificity of this pair of stars is remarkably revealed by the values taken by the parameters of the normalized ascending branches of their light curves.

\begin{deluxetable*}{lcc ccc ccc ccc c}
\tablenum{6}
\tablecaption{Parameters of the light curves mentioned in Section 7. \label{tab6}}
\tablewidth{0pt}
\tablehead{
 \colhead{Name}&\colhead{$P$}&\colhead{Spectral type}&\colhead{Tc}&\colhead{$^{12}$C/$^{13}$C}&\colhead{$C$}&\colhead{$A$}&\colhead{$W$}&\colhead{$\Delta$}&\colhead{$\varphi_{\rm{min}}$}&
 \colhead{$p$$\times$10$^3$}&\colhead{$q$$\times$10$^3$}&\colhead{$\chi^2$$\times$10$^4$}\\
  \colhead{}&\colhead{(day)}&\colhead{}&\colhead{}&\colhead{}&\colhead{(mag)}&\colhead{(mag)}&\colhead{}&\colhead{(mag)}&\colhead{}& \colhead{}&\colhead{}&\colhead{}
}
\startdata
W Dra&	279&	M3&	no&	$-$&	2.7&	4.8&	0.35&	0.5&	0.59&	262&	$-$16&	47\\
W Hya&	391&	M7.5&	no&	16&	0.4&	3.2&	0.37&	0.34&	0.48&	$-$714&	$-$3&	65\\
T Ari&	320&	M6&	no&	15&	1.6&	2.1&	0.5&	0.22&	0.55&	$-$397&	$-$91&	39\\
U CMi&	409&	M4&	no&	$-$&	2&	4&	0.47&	0.34&	0.39&	$-$1495&	$-$164&	442\\
T UMi&	317&	M6&	no&	$-$&	1.4&	4.9&	0.21&	0.31&	0.53&	$-$461&	104&	42\\
$o$ Cet&	333&	M5&	yes&	10&	1.5&	5.6&	0.24&	0.62&	0.62&	158&	80&	126\\
R Aqr&	386&	M6.5&	yes&	16&	1.6&	4.2&	0.23&	0.6&	0.57&	$-$662&	126&	45\\
X Oph&	334&	M3&	no&	12&	2&	1.7&	0.35&	0.22&	0.49&	$-$425&	79&	187\\
U Ori&	372&	M6&	poss&	25&	2.6&	5.5&	0.24&	0.55&	0.61&	$-$370&	82&	77\\
RU Tau&	589&	M3.5&	$-$&	$-$&	1.9&	3.7&	0.41&	$-$&	$-$&	$-$864&	91&	756\\
SS Vir&	357&	C4.5&	$-$&	$-$&	$-$&	2.1&	0.44&	0.3&	0.47&	$-$574&	$-$10&	199\\
U Ari&	372&	M4.5/4&	yes&	$-$&	2.1&	6.4&	0.22&	0.63&	0.57&	$-$331&	77&	82\\
S Gru&	401&	M8/5&	yes&	$-$&	1.9&	6.7&	0.2&	0.62&	0.56&	$-$521&	114&	96\\
R Cen&	548&	M4&	no&	$-$&	2.1&	5.2&	0.44&	0.25&	0.39&	$-$2249&	$-$169&	170\\
R Nor&	498&	M3&	no&	$-$&	2.3&	5.6&	0.35&	0.38&	0.43&	$-$1817&	$-$137&	51\\
\enddata
\end{deluxetable*}

\section{Summary}
In the wake of Papers I and II and of an abundant earlier literature, the aim of the present work is to progress in our understanding of the relation between the physics governing the pulsation of Mira variables and their evolution along the AGB. To this end, it seeks possible correlations between the stellar parameters describing the location of a star in the Hertzsprung-Russell diagram and parameters describing the shape and main features of its light curve. Specifically, it focuses on the ascending branch of the light cycle, which had been shown in Papers I and II to display a broad range of different shapes.  To this effect, it uses only two parameters, $p$ and $q$, measuring respectively the value and the slope of the normalized magnitude of the ascending branch at half-time between minimal and maximal light (Figure \ref{fig3}). This parameterization sheds new light on the problem, giving evidence for strong correlations between parameter $q$ and the stellar parameters of the star. In contrast, parameter $p$ does not suggest simple interpretations and several examples have been spotted of curves having values of $p$ lying clearly out of the mean trend while the state of the star displays no particular feature. 

The arguments developed throughout the article use as guideline the assumption that when a pulsating star evolves along the AGB, the continuity of the evolution of the stellar parameters (excluding thermal pulse episodes) is associated with a similar continuity of the parameters describing the shape of its light curve. There are good reasons in favour of the validity of such an assumption, both from what we know of the physics governing pulsations and from the correlations observed in the earlier literature, including Papers I and II, as well as in the present work. Yet, this is only an assumption, which needs to be constantly and critically challenged, preventing many of the suggestions made in the present work to be reliably ascertained. Accordingly, we spent considerable time and effort spotting cases displaying features not in conformity with the general picture. This implied a detailed scrutiny of such cases, significantly lengthening and complicating the article, but we believe that it is only at this price that a serious argumentation could be presented.
    
The presence of humps on the ascending branch is remarkably well described by the parameterization, describing them as expected episodes of a continuous evolution, a result reminiscent of that obtained for the Hertzsprung progression of Cepheids.
  
There is clear evidence for a group of stars that have not experienced strong enough TDU events for technetium to be detectable in their spectrum (referred to as Mno) to have light curves that evolve in parameter space as schematically illustrated in Figures \ref{fig10} and \ref{fig22}. Strong correlations are observed between $q$ and other shape-sensitive parameters, such as $W$ and $\varphi_{\rm{min}}$, but also stellar parameters sensitive to the state of the star, such as the period, the regularity of the pulsation regime, the colour index and the effective temperature. This result strengthens those obtained in Paper II, suggesting that the Mno stars of this group evolve along the AGB with their light curves having increasing values of $q$. 
  
Stars of M spectral type having experienced strong enough TDU events for the presence of technetium to be detectable in their spectrum (referred to as Myes) have curves that cover a broad region of the curve parameter space. Some (My1 and S1) can be reasonably assumed to be successors of the Mno stars described in the preceding paragraph, some very close to the compact group of warm Mno stars, some farther away. Other curves (My2 and S2) have $q$ values similar to those of warm Mno stars independently from their evolution as oxygen-rich, Mno, Myes or S, suggesting that they also are the progeny of the warm Mno stars. A common feature of the transition from Mno to Myes spectral types is a strong increase of the amplitude of oscillation, by over one magnitude. Moreover, it is also in this stage that the period experiences the stronger increase, staying nearly constant on average in the transitions from Myes to S and S to C spectral types. Abundant illustration of these features is given in Figures \ref{fig13} to \ref{fig17}, inviting to draw an idealised, but probably oversimplified picture of these observations (Figure \ref{fig22}), showing continuity between curves switching from Mno to Myes and S spectral types with their $q$ parameter experiencing only modest changes.
  
There is clear evidence for another group of stars of S spectral types (S3) to occupy low $q$ regions of the parameter space. Their existence had been noted in the earlier literature, and in particular in Papers I and II. It is reasonable to assume that these stars are the progeny of some of the stars of group My2, but this cannot be ascertained. Their case raises many questions, in particular concerning the low values of their colour indices. Their possible continuity with the stars of groups My1, S1 and S2 constitutes an attractive scenario that would offer a unified picture of the whole sample, with different evolutions depending essentially on the initial mass of the star (Figure \ref{fig22}). 

The transition to carbon-rich stars, when it happens, goes together with sudden changes of the amplitude of oscillations, decreasing by $\sim$2 mag, and of the width parameter $W$, increasing by $\sim$0.1, the other parameters, including the effective temperature, experiencing only small changes. Some of the warmer C stars may have precursors of insufficiently stable pulsation regimes to be accepted as Mira variables rather than being the progeny of stars of group S3.

As was already mentioned in Paper II, a major difficulty that has been recurrently met during the study is the lack of a clear picture of the physics governing the transition from a stable to an unstable regime of pulsation and conversely, namely of the nature of the relevant instability region in the Hertzsprung-Russell diagram. 

The specificity of Hot Bottom Burners R Cen and R Nor is remarkably revealed by the values taken by their light curve parameters.

\section*{acknowledgments}
  
To the extent that the analyses presented in the present article may have contributed some progress, the credit belongs to the innumerable observers around the world and to the AAVSO, without whom none of the results could have been obtained. We are accordingly deeply indebted to them: to the many observers for the high quality of their observations and to the AAVSO for the outstanding handling and reduction of the data that makes their use particularly easy and efficient. We thank Mai Nhat Tan for contributions to the early phase of the analyses presented in the article. This research has made use of the SIMBAD database, CDS, Strasbourg Astronomical Observatory, France. It is funded by the Vietnam National Foundation for Science and Technology Development (NAFOSTED) under grant number 103.99-2024.36. Financial support from the World Laboratory, the Rencontres du Vietnam is gratefully acknowledged.

\bibliography{lightcurves3}{}
\bibliographystyle{aasjournalv7}



\end{document}